\documentclass[a4paper,11pt]{article}

\usepackage{amsmath}
\usepackage{amssymb}
\usepackage{mathtools}

\usepackage[linktocpage]{hyperref}
\usepackage[dvipsnames,table]{xcolor}
\hypersetup{
    colorlinks,
    linkcolor={RoyalBlue},
    citecolor={RoyalBlue},
    urlcolor={RoyalBlue}
}

\def\gtwid{\mathrel{\raise.3ex\hbox{$>$\kern-.75em\lower1ex\hbox{$\sim$}}}}
\def\ltwid{\mathrel{\raise.3ex\hbox{$<$\kern-.75em\lower1ex\hbox{$\sim$}}}}
\def\square{\kern1pt\vbox{\hrule height 1.2pt\hbox{\vrule width 1.2pt\hskip 3pt
   \vbox{\vskip 6pt}\hskip 3pt\vrule width 0.6pt}\hrule height 0.6pt}\kern1pt}

\usepackage[margin=2.5cm]{geometry}

\numberwithin{equation}{section}

\usepackage{graphicx}
\graphicspath{ {./figures/} }

\allowdisplaybreaks

\usepackage{cite}

\usepackage[labelfont=bf]{caption}

\usepackage{orcidlink}

\begin{document}

\hfill {UFIFT-QG-26-03}

\

\begin{center}

\renewcommand{\thefootnote}{\fnsymbol{footnote}}

{\Large \bf 
Cancellation of one-parameter graviton gauge dependence
\vskip+1mm
in the effective scalar field equation in de Sitter
}

\setcounter{footnote}{0} 

\bigskip

Dra\v{z}en Glavan\,\orcidlink{0000-0002-1983-0448},${}^{\,a,}$\footnote[1]{email: 
	\href{mailto:glavan@fzu.cz}{\tt glavan@fzu.cz}}
Shun-Pei Miao\,\orcidlink{0000-0002-0754-6888},${}^{\,b,}$\footnote[2]{email:
	\href{mailto:spmiao5@mail.ncku.edu.tw}{\tt spmiao5@mail.ncku.edu.tw}}
Tomislav Prokopec\,\orcidlink{0000-0003-0391-5743},${}^{\,c,}$\footnote[3]{email: 
	\href{mailto:t.prokopec@uu.nl}{\tt t.prokopec@uu.nl}}
Richard P.~Woodard\,\orcidlink{0000-0003-0830-1396}${}^{\,d,}$\footnote[4]{email: 
	\href{mailto:woodard@phys.ufl.edu}{\tt woodard@phys.ufl.edu}}

\bigskip
\bigskip

{\it 
${}^a$\,CEICO, Institute of Physics of the Czech Academy of Sciences (FZU),
\\
Na Slovance 1992/2, 182 21 Prague 8, Czech Republic

\medskip

${}^b$\,Department of Physics, National Cheng Kung University,
\\
No.~1 University Road, Tainan City, 70101, Taiwan

\medskip

${}^c$\,Institute for Theoretical Physics, Spinoza Institute \& EMME$\Phi$,
Utrecht University,
\\
Postbus 80.195, 3508 TD Utrecht, The Netherlands

\medskip

${}^d$\,Department of Physics, University of Florida, Gainesville, FL 32611, U.S.A.
}

\

\

\

\parbox{0.9\linewidth}{
We investigate gauge dependence of one-graviton-loop corrections to the effective field equation of the
massless, minimally coupled scalar in de Sitter, obtained by including source and observer corrections to 
the effective self-mass correcting the equation. Using the $\Delta\alpha$ variation of the de Sitter-breaking 
graviton propagator in a one-parameter family of gauges, we compute the gauge-dependent contributions 
to the effective self-mass of a massless minimally coupled scalar mediating interactions between heavy scalars. 
We show that gauge dependence cancels provided the contributions from all diagram classes are collected, 
including one-loop corrections to external mode functions, which play a qualitatively new role relative to flat 
space. The resulting cancellation supports the construction of graviton gauge-independent cosmological 
quantum-gravitational observables from quantum-corrected effective equations.
}

\end{center}

\

\

\hrule

\tableofcontents

\

\hrule

\

\

\

\section{Introduction}

The rapid accelerated expansion of space during primordial inflation provides conditions for efficient 
gravitational particle production~\cite{Parker:1968mv}, experienced by fields non-conformally coupled to 
gravity. It is by this linear quantum field-theoretic mechanism that inflation produces a vast ensemble of 
long-wavelength scalar~\cite{Mukhanov:1981xt} and tensor~\cite{Starobinsky:1979ty} cosmological 
perturbations, i.e.~inflationary scalars and gravitons, respectively. These inflationary quanta—more precisely, 
the remnants they leave on the cosmic microwave background—have become the main probes of inflationary 
physics~\cite{Planck:2018jri}.

It should be noted that gravity couples universally, and that the vast ensemble of inflationary gravitons in 
principle must interact with all matter fields, and also with itself. In addition, the relatively high energy scale 
of inflation, potentially as high as the GUT scale of $10^{16}\, \mathrm{GeV}$, opens the possibility of 
inflationary quantum-gravitational effects that are not prohibitively small, while still being describable by 
the effective field theory of gravity~\cite{Donoghue:1994dn,Donoghue:1995cz,Burgess:2003jk,Donoghue:2012zc,Donoghue:2017ovt,Donoghue:2022eay}. Moreover, gravitons can mediate the effects of the rapid 
expansion to conformally coupled fields (such as electromagnetism), thereby breaking conformal invariance 
and potentially leaving observable imprints. Investigating this possibility is of our primary interest.

The naive dimensional-analysis estimate for the size of quantum-gravitational effects during inflation 
is~$\kappa^2H^2 \sim 10^{-11}$--$10^{-10}$, where $H$ is the almost-constant inflationary Hubble rate, 
and~$\kappa=\sqrt{16\pi G_{N}}$ is the loop-counting parameter of quantum gravity. However, explicit 
computations have shown that these corrections can be enhanced by large logarithms, either temporal or 
spatial. These computations are performed in rigid de Sitter space with constant Hubble rate, which is a good 
approximation for the purpose at hand and offers considerable technical simplifications. Thus far, six systems 
that exhibit logarithmic enhancement of loop corrections have been identified:\footnote{Temporal growth 
of the dynamical massless, conformally coupled scalar has also been reported~\cite{Boran:2017cfj}, but this 
result did not persist after further scrutiny~\cite{Glavan:2020ccz}.}
\begin{itemize}
\item Temporal growth of the dynamical fermion field strength~\cite{Miao:2006gj};
\item Temporal and spatial growth of the electric force~\cite{Glavan:2013jca};
\item Temporal growth of the dynamical photon field strength~\cite{Wang:2014tza};
\item Temporal growth of the dynamical graviton field strength~\cite{Tan:2021lza};
\item Spatial suppression of the force mediated by a massless minimally coupled scalar~\cite{Glavan:2021adm};
\item Temporal suppression of the gravitational force~\cite{Tan:2022xpn}.
\end{itemize}
The origin of large secular logarithms can be explained by a combination of stochastic
formalism and renormalization group~\cite{Miao:2021gic,Glavan:2023tet}.

However, a serious question pertains to these results: the {\it gauge issue}. All results above have been 
worked out in the simplest graviton gauge~\cite{Tsamis:1992xa,Woodard:2004ut}. The procedure was to 
compute the one-graviton-loop correction to the relevant 1-particle-irreducible (1PI) 2-point function 
(self-mass/self-energy/vacuum polarization)~\cite{Tsamis:1996qk,Miao:2005am,Kahya:2007bc,Miao:2012bj,Leonard:2013xsa,Glavan:2020gal}, and then use it to quantum-correct the effective field equations. 
The concern is that this procedure need not yield gauge-independent physical predictions, in particular for 
the logarithmically enhanced terms, and the explicit computations~\cite{Miao:2006gj,Glavan:2013jca,Wang:2014tza,Tan:2021lza,Glavan:2021adm,Tan:2022xpn} have been criticized on that account~\cite{Higuchi:2011vw,Miao:2011ng,Morrison:2013rqa,Miao:2013isa}.

A computation of the graviton loop correction to dynamical photons performed in the one-parameter family 
of exact covariant gauges~\cite{Mora:2012zi} brings this issue into focus. The vacuum polarization computed 
in this gauge~\cite{Glavan:2015ura} has been used to infer the correction to the field strength of a dynamical 
photon~\cite{Glavan:2016bvp}. This correction does not depend on the gauge-fixing parameter of the exact 
covariant gauge and does show secular enhancement, but its amplitude differs from the noncovariant gauge 
computation~\cite{Wang:2014tza}. This underscores the need to address the {\it gauge issue} of 
quantum-gravitational loop corrections in cosmological quantum field theory, i.e.~to purge gauge dependence 
from physical observables. In this work we report progress toward this goal by testing the de Sitter 
generalization of the proposal of~\cite{Miao:2017feh}, designed to produce gauge-independent 
one-loop-corrected effective field equations, by including the analogues of source and observer corrections 
to self-masses. 

The gauge dependence of 1PI two-point functions is well established in flat space, and using them directly 
to quantum-correct field equations is not in general the correct procedure for extracting physical gauge-
independent predictions (see e.g.~\cite{Leonard:2012fs}). In flat space this problem is solved for asymptotic 
particle scattering by the construction of the S-matrix, in gauge theories and in quantum gravity. In curved 
space the S-matrix does not generally exist because of the absence of asymptotic states, and in any case 
cosmological applications are inherently initial-value problems. This motivates constructing appropriate 
one-loop quantum-gravitational observables for inflation.

One-graviton-loop corrections to 1PI two-point functions in de Sitter space must inherit at least the flat-space 
gauge dependence. The relevant question is how much of this gauge dependence contaminates the 
logarithmically enhanced terms, and how to purge it. In flat space the S-matrix achieves gauge independence 
by accounting for source and observer corrections to the scattering process. One can apply inverse-scattering 
methods to the S-matrix in order to infer quantum-corrected long-range potentials~\cite{Donoghue:1994dn,Donoghue:1993eb,Bjerrum-Bohr:2002aqa,Bjerrum-Bohr:2002gqz}. 
Moreover, the relevant long-range contributions can be isolated more directly by analyzing the non-analytic 
behaviour of momentum-space amplitudes.

The S-matrix achieves gauge-independence by putting the amputated connected~$n$-point functions on 
shell by attaching to them asymptotic mode functions. In flat space this form is obtained by the 
Lehmann--Symanzik--Zimmermann (LSZ) reduction formula~\cite{Lehmann:1954rq}. A generalization of this 
procedure was proposed in~\cite{Miao:2017feh} that dispenses with taking asymptotic time limits, allowing 
generalizations to curved spacetimes.\footnote{Although the S-matrix does exist for massive scalars on de 
Sitter~\cite{Marolf:2012kh} (see also~\cite{Melville:2023kgd,Melville:2024ove}), the construction is likely
problematic for massless, minimally coupled scalars and gravitons precisely because of the secular growth 
that makes these fields so interesting.} In particular, one considers an amputated 4-point function in position 
space and puts it on shell by attaching mode functions at finite time. This object includes source and observer 
corrections much like the S-matrix. One then employs integration by parts, and the Donoghue 
Identities~\cite{Donoghue:1993eb,Bjerrum-Bohr:2002aqa,Bjerrum-Bohr:2002gqz,Donoghue:1996mt} to 
reduce the one-loop diagrams to the topology of the self-mass diagram, from which the effective 
gauge-independent self-mass can be read off. This effective self-mass can then be used to quantum-correct 
effective field equations, thus purging gauge dependence from them. This program has been demonstrated 
to work for long-range corrections to potentials in flat space for the scalar model~\cite{Miao:2017feh}, and 
for electromagnetism~\cite{Katuwal:2021thy}, and here we investigate whether gauge independence is 
maintained in its de Sitter space generalization.

In this work we report on the computation of all the diagrams contributing to the gauge-independent 
effective self-mass by using the gauge-dependent part of the graviton propagator in the simple 
one-parameter gauge~\cite{Glavan:2019msf,Glavan:2025azq}, that we refer to as the~$\Delta\alpha$ 
variation of the graviton propagator. We demonstrate that the gauge dependence drops out from the final 
answer by showing that all contributions from the $\Delta\alpha$ variation sum to zero. Thus, in principle,
we establish the gauge independence of the entire construction. Furthermore, we find that the previous 
computation~\cite{Glavan:2024elz}, that we performed using the simplest gauge graviton 
propagator~\cite{Tsamis:1992xa,Woodard:2004ut} corresponding to~$\Delta\alpha\!=\!0$, should be 
supplemented by contributions from additional diagrams we identify here in order to yield the full 
gauge-independent result.

We find that no Donoghue Identities are necessary for the present computation. On the one hand, this does 
not allow us to test the de Sitter versions of the Donoghue Identities used in~\cite{Glavan:2024elz}. On the 
other hand, it allows an unambiguous test of which diagrams and contributions must be included to ensure 
gauge independence. We find that, unlike in flat space, additional diagrams and contributions are necessary, 
including the one-loop corrections to the external mode functions, that are crucial for obtaining gauge 
independence. This is not seen in flat space, where such corrections can at most be absorbed into a constant 
multiplicative field-strength. In curved space, secular corrections to mode functions are possible; they combine 
with one-loop corrections to the amputated four-point function to yield a gauge-independent result. 

\medskip

We should also mention the approach to eliminate gauge dependence by taking the expectation values of 
gauge invariant operators. This is straightforward for linearized gravity, and was accomplished long ago
for cosmological perturbations~\cite{Bardeen:1980kt}. However, it is quite nontrivial when interactions are 
included because one must invariantly fix the points at which observations are made and the frames in which 
any tensor indices are reported~\cite{Tsamis:1989yu,Carlip:2001wq,Abramo:2001db,Geshnizjani:2002wp,Giddings:2005id,Tanaka:2013caa,Donnelly:2015hta,Khavkine:2015fwa,Marolf:2015jha,Frob:2017lnt,Frob:2017gyj}. 
Devising a reasonable way of accomplishing this is crucial because there is no local way to specify the 
coordinate system and frame fields~\cite{Torre:1993fq}, but permitting arbitrary nonlocal additions to the 
operator under study allows one to completely cancel any correction to the linearized operator from 
interaction vertices~\cite{Miao:2012xc}, including what are known to be real effects in flat space scattering. 
The problem has both a technical and a physical aspect. On the technical level, nonlinear additions 
to the operator under study inevitably involve new sorts of 
divergences~\cite{Tsamis:1989yu,Miao:2017vly,Frob:2017apy}, which are not removed by the BPHZ 
(Bogoliubov and Parasiuk~\cite{Bogoliubov:1957gp}, Hepp~\cite{Hepp:1966eg}
and Zimmermann~\cite{Zimmermann:1968mu,Zimmermann:1969jj}) renormalizations that suffice for 
noncoincident 1PI $n$-point functions. Removing these divergences requires the more complicated 
procedure of composite operator 
renormalization~\cite{Itzykson:1980rh,Weinberg:1996kr,Korchemsky:1987wg}. The physics issue is that 
the finite residues of the new divergences can change scaling and renormalization group flows. For example, 
flat space scattering amplitudes defined using invariant Green's functions at fixed geodesic separations 
produce an extra high energy logarithm at one loop order~\cite{Frob:2017apy}. 

\medskip
The paper is organized as follows. Section~\ref{sec: The problem the goal} outlines in more 
detail the main problem we address and the goal of constructing gauge-independent 
observables. Section~\ref{sec: Reduction strategy} gives the general strategy for the reduction 
to self-mass diagrams. Section~\ref{sec: Feynman diagrams} lists the Feynman rules for the 
model and all classes of diagrams necessary for the computation, including simplifications 
obtained by consolidating some classes. Section~\ref{sec: Propagators} collects the relevant 
properties of scalar and graviton propagators. The main computation is presented in 
Sections~\ref{sec: Reducing 4-point diagrams} and~\ref{sec: Mode function corrections}, where 
contributions from the four-point diagrams and from mode-function corrections are presented, 
respectively. The concluding Section~\ref{sec: Discussion} collects all contributions to show 
that they cancel and provides a discussion. Mathematical details and the discussion of 
additional diagrams are relegated to the appendices.

\section{The problem and the goal}
\label{sec: The problem the goal}

In order to investigate the problem of gauge dependence of the quantum-gravitationally-corrected effective field equations in de Sitter space, we consider a specific minimal model
where explicit computations are tractable. We consider a massive scalar~$\Psi$ with
$m\!\gg\!H$,  interacting via the massless, minimally coupled (MMC) scalar~$\varphi$ in de Sitter space. The covariant 
action that describes this system is given by
\begin{align}
S[g_{\mu\nu},\varphi,\Psi]
	={}&
	\int\! d^{D\!} x \, \sqrt{-g} \,
	\biggl[
	\frac{R - (D\!-\!2)\Lambda}{\kappa^2}
	-
	\frac{1}{2} g^{\mu\nu} (\partial_\mu \varphi) (\partial_\nu \varphi)
\nonumber \\
&	\hspace{3cm}
	-
	\frac{1}{2} g^{\mu\nu} (\partial_\mu \Psi) (\partial_\nu \Psi)
	-
	\frac{1}{2} \bigl( m^2 \!+\! \lambda \varphi \bigr) \Psi^2
	\biggr]
	\, ,
\label{action}
\end{align}
where~$\kappa^2 \!=\! 16\pi G$ is the rescaled Newton constant,~$\Lambda$
is the positive cosmological constant,~$R$ is the Ricci scalar,\footnote{Our conventions
are~$\eta_{\mu\nu} \!=\! (-1,1,\dots,1)$ for the Minkowski metric, 
and~$R_{\mu\nu} \!=\! \partial_\alpha \Gamma^\alpha_{\mu\nu} \!-\! 
	\partial_\nu \Gamma^\alpha_{\mu\alpha}
	\!+\! \Gamma^\alpha_{\mu\nu}\Gamma^{\beta}_{\beta\alpha}
	\!-\! \Gamma^\alpha_{\mu\beta}\Gamma^{\beta}_{\nu\alpha}$ for the Ricci tensor.}
and~$\lambda$ is a dimensionful coupling constant.

We are interested in investigating this two-scalar system on a spatially-flat de Sitter 
space background,
for which the metric is conformal to the Minkowski one,~$g_{\mu\nu} \!=\! a^2(\eta) \eta_{\mu\nu}$, when written in conformal time coordinate~$\eta$
that takes values on a negative semi-axis,~$-\infty\!<\!\eta\!<\!0$. The scale factor
of de Sitter space grows with conformal time,~$a(\eta)\!=\! - 1/(H\eta)$, while the Hubble
rate remains constant,~$H\!=\!\sqrt{\Lambda/3}$. In particular, we are interested in 
one-graviton loop corrections to the interaction between heavy scalars~$\Psi$ that is
mediated by the MMC scalar~$\varphi$. In this setup the fluctuations of the metric
field, that we define as
\begin{equation}
g_{\mu\nu} = a^2 \big( \eta_{\mu\nu} + \kappa h_{\mu\nu} \bigr) \, ,
\end{equation}
are quantized semiclassically, and the effects of their interactions are captured by perturbative quantum field 
theory. The interaction vertices encoded in the action~(\ref{action}) that we will need for our computation are 
collected in Table~\ref{OriginalVertices}. In the remainder of this section we outline what is the problem that 
arises in the study of quantum-gravitational corrections to the system defined in~(\ref{action}), while the 
following section is devoted to the outline of our strategy for addressing this problem.
\begin{table}[h!]
\vskip+5mm
\centering
\begin{tabular}{c c l}
\hline
	A
	&
	$\vcenter{\hbox{\includegraphics{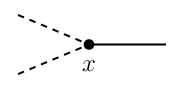}}}$
	&
	$- i \lambda a_x^D$
\\
\hline
	B
	&
	$\vcenter{\hbox{\includegraphics{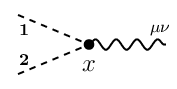}}}$
	&
	$
	- i \kappa a_x^{D-2} \Bigl[ - \partial{}_1^{(\mu} \partial{}_2^{\nu)}
		+ \tfrac{1}{2} \eta^{\mu\nu}
		\bigl( \partial{}_1 \!\cdot\! \partial{}_2 \!+\! a_x^2 m^2 \bigr) 
		\Bigr]
	$
\\
\hline
	C
	&
	$\vcenter{\hbox{\includegraphics{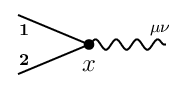}}}$
	&
	$
	- i \kappa a_x^{D-2} \Bigl[ - \partial{}_1^{(\mu} \partial{}_2^{\nu)}
		+ \tfrac{1}{2} \eta^{\mu\nu} \partial{}_1 \!\cdot\! \partial{}_2 
		\Bigr]
	$
\\
\hline
	D
	&
	$\vcenter{\hbox{\includegraphics{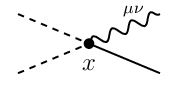}}}$
	&
	$
	- \tfrac{1}{2} i \kappa \lambda a_x^D \eta^{\mu\nu}
	$
\\
\hline
	E
	&
	$\vcenter{\hbox{\includegraphics{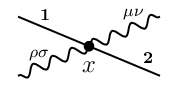}}}$
	&
	$\vcenter{\hbox{
	$\begin{aligned}[t]
	&- i \kappa^2 a_x^{D-2}
	\Bigl[
	\tfrac{1}{2} \partial{}_1^{(\mu} \eta^{\nu)(\rho} \partial{}_2^{\sigma)}
	+
	\tfrac{1}{2} \partial{}_1^{(\rho} \eta^{\sigma)(\mu} \partial{}_2^{\nu)}
	\!-\!
	\tfrac{1}{4} \partial{}_1^{(\mu} \partial{}_2^{\nu)} \eta^{\rho\sigma}
	\!-\!
	\tfrac{1}{4} \partial{}_1^{(\rho} \partial{}_2^{\sigma)} \eta^{\mu\nu} \\
	&\hspace{2.5cm}
	- \tfrac{1}{8} \bigl( 2 \eta^{\mu(\rho} \eta^{\sigma)\nu} 
	\!-\! \eta^{\mu\nu} \eta^{\rho\sigma} \bigr)
			\partial{}_1 \!\cdot\! \partial{}_2
			\Bigr]
			\end{aligned}$
			}}$
\\
\hline
	F
	&
	$\vcenter{\hbox{\includegraphics{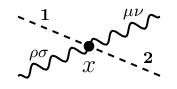}}}$
	&
	$\vcenter{\hbox{
	$\begin{aligned}[t]
	&
	- i \kappa^2 a_x^{D-2}
	\Bigl[
	\tfrac{1}{2} \partial{}_1^{(\mu} \eta^{\nu)(\rho} \partial{}_2^{\sigma)}
	\!+\!
	\tfrac{1}{2} \partial{}_1^{(\rho} \eta^{\sigma)(\mu} \partial{}_2^{\nu)}
	\!-\!
	\tfrac{1}{4} \partial{}_1^{(\mu} \partial{}_2^{\nu)} \eta^{\rho\sigma}
	\!-\!
	\tfrac{1}{4} \partial{}_1^{(\rho} \partial{}_2^{\sigma)} \eta^{\mu\nu}
	\\
	&
	\hspace{2.5cm}
	- \tfrac{1}{8} \bigl( 2 \eta^{\mu(\rho} \eta^{\sigma)\nu}  \!
	\!-\!
	\eta^{\mu\nu} \eta^{\rho\sigma} \bigr)
	\bigl( \partial{}_1 \!\cdot\! \partial{}_2 + a_x^2 m^2 \bigr)
	\Bigr]
	\end{aligned}$
	}}$
\\
\hline
	G
	&
	$\vcenter{\hbox{\includegraphics{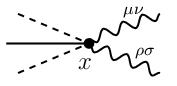}}}$
	&
	$
	- i \kappa^2 \lambda a_x^D \Bigl[ \tfrac{1}{8} \eta^{\mu\nu} \eta^{\rho\sigma}
		- \tfrac{1}{4} \eta^{\mu(\rho} \eta^{\sigma)\nu}
	\Bigr]
	$
\\
\hline
\end{tabular}
\vskip-1mm
\caption{List of 3-point, 4-point, and 5-point interaction vertices 
encoded in the action~(\ref{action}), that are relevant for the computation of the 
one-graviton-loop correction to the scalar exchange potential. Solid lines correspond to
the MMC scalar, dashed lines to the massive scalar, and wavy lines to the graviton.
 }
\label{OriginalVertices}
\vskip+5mm
\end{table}

The MMC scalar interacting with inflationary gravitons is the simplest system in 
de Sitter space that exhibits a large secular logarithm in the one graviton loop 
correction~\cite{Glavan:2021adm}.
The tree-level point-source potential mediated by the MMC scalar is given 
by~\cite{Akhmedov:2010ah,Glavan:2019yfc}
\begin{equation}
\varphi_0(r) = - \frac{K}{4\pi a r}
	- \frac{K H}{4\pi} \ln\Bigl[ \frac{a}{1+aHr} \Bigr]
	\xrightarrow{aHr \gg 1}
	\frac{K H}{4\pi} \ln(Hr)
	\, .
\label{TreeLevelPotential}
\end{equation}
The first term in the middle equality is just the de Sitter generalization of the flat space 
potential, while the second tail term, dominant for large physical distances,
is the pure de Sitter space contribution.
The tree-level potential in~(\ref{TreeLevelPotential}) is a solution to the classical field 
equation with a point source,
\begin{equation}
\mathcal{D} \varphi(x) = K a \delta^3(\vec{x}) \, ,
\label{FieldEq}
\end{equation}
where~$\mathcal{D} \!=\! \partial^\mu a^{D-2} \partial_\mu$, and~$K$ is the
scalar charge of the point particle. 

The long-range interaction in~(\ref{TreeLevelPotential}) receives quantum gravitational 
corrections from interactions with an ensemble of inflationary gravitons. These corrections
are inferred by computing the 
one-loop MMC scalar self-mass,~$-i\mathcal{M}^2$, with a single graviton running in the loop, 
and using it to correct the classical equation of motion~(\ref{FieldEq}),
\begin{equation}
\mathcal{D} \varphi(x)
	- \int\! d^4x' \, \mathcal{M}^2(x;x') \varphi(x')
	 = K a \delta^3(\vec{x}) \, .
\label{CorrectedFieldEq}
\end{equation}
Of course, it is the retarded self-mass component obtained in the
Schwinger-Keldysh formalism for nonequilibrium quantum field 
theory~\cite{Schwinger:1960qe,Mahanthappa:1962ex,Bakshi:1962dv,Bakshi:1963bn,Keldysh:1964ud,Chou:1984es,Jordan:1986ug,Ford:2004wc} that corrects this equation.
This self-mass has been computed in the simple graviton gauge~\cite{Glavan:2021adm}, 
and leads to the
following result for the one-loop corrected potential
\begin{equation}
\varphi(x) \xrightarrow{aHr \gg 1} \varphi_0(r) \times 
	\bigg[ 1 - \frac{\kappa^2 H^2}{8\pi^2} \ln(Hr) \bigg]
	\, ,
\label{PhiCorrection}
\end{equation}
that exhibits a large suppression of the tree-level potential for large physical distances.

The one-loop quantum gravitational correction in~(\ref{PhiCorrection}) enhanced by
the spatially-dependent logarithm is exciting, as it signals effects considerably larger
than expected by naive dimensional analysis. 
However, it is not clear that~(\ref{PhiCorrection}) is the full physical result, or whether a part
of it is a gauge artifact. This is because the
self-mass used to quantum correct the field equation~(\ref{CorrectedFieldEq}) is known to be a 
gauge-dependent object. This gauge dependence is not manifest because the result was 
obtained in a gauge without free gauge parameters, but we know that at least the flat
space gauge dependence must survive the de Sitter generalization.
For this reason, the main objective of our program is purging gauge dependence 
from 1PI 2-point functions that correct the field equations, in particular here from 
the MMC scalar self-mass.

Our intention is to construct an effective gauge-independent self-mass~$-i\mathcal{M}^2(x;x')$, that  can be used to correct the field equation~(\ref{FieldEq}).
In order to accomplish this, we appeal to the lessons from
the flat space inverse scattering method, where the long-range potential can be extracted
from the~$t$-channel $2\!\to\!2$ scattering of massive source and observer particles. 
The quantum-field-theoretic object corresponding to this process is the~$t$-channel
amputated 4-point function of the massive scalar, that is put on shell by attaching 
the asymptotic plane-wave mode functions.

We propose to consider the analogous object
in de Sitter space, with the massive scalar~$\Psi$ playing the same role of source
and observer, and consider its amputated 4-point function that is now put on shell
at finite times, by attaching mode functions that themselves satisfy quantum corrected
equations of motion. It is possible to reduce all one-loop diagrams contributing to
this object to the topology of the~$t$-channel self-mass diagram, put on shell by
the tree-level mode functions. This self-mass should capture the infrared effects
that are behind large logarithms in de Sitter space corrections. We then read off this 
effective self-mass and use it to 
quantum correct the field equations as in~(\ref{CorrectedFieldEq}). We conjecture that
this effective self-mass should be gauge-independent, and this work is devoted to 
showing the independence on one of the two gauge-fixing parameters
of the graviton propagator in the tractable non-covariant 
gauge~\cite{Glavan:2019msf,Glavan:2025azq}.
The general reduction procedure we follow to extract the one-loop
effective self-mass is outlined in the following section.

\section{Reduction strategy}
\label{sec: Reduction strategy}

\subsection{General reduction}
\label{subsec: General reduction}

Our starting point is the amputated connected four-point function for the massive scalar 
field~$\Psi$. We put this object on shell by attaching to it mode functions~$U(x)$ of the 
massive  scalar field,
\begin{equation}
\vcenter{\hbox{\includegraphics[width=2.cm]{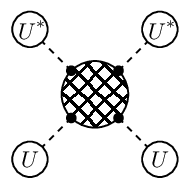}}}
	\equiv
	\int\! d^{D\!}x\, U(x) 
	\int\! d^{D\!}y \, U^*(y) 
	\int\! d^{D\!}x' \, U(x')
	\int\! d^{D\!}y' \, U^*(y')
	\left(
\vcenter{\hbox{\includegraphics[width=1.cm]{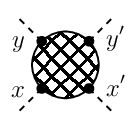}}}
\right)
\, .
\label{MainObject}
\end{equation}
It is crucial that the attached mode functions satisfy the quantum-corrected equation 
of motion,
\begin{equation}
\big( \mathcal{D}_x \!-\! a_x^Dm^2 \big) U(x)
	-
	\int\! d^4x' \, \mathcal{M}_m^2(x;x') U(x') = 0 \, ,
\label{Ueq}
\end{equation}
where~$-i\mathcal{M}_m^2$ is the self-mass of the massive scalar field~$\Psi$. In the limit of flat space, and 
upon specializing the mode functions to asymptotic plane waves, this expression reduces to the LSZ reduction 
formula that is known to yield the gauge-independent S-matrix. We find~(\ref{MainObject}) to be the 
appropriate curved space generalization of the LSZ reduction formula, and in this work we show its gauge 
independence.

The ultimate goal is to capture the physical part of the large logarithms in graviton loop corrections in de 
Sitter space, by isolating the gauge-independent part of the scalar self-mass. To this end we reduce the 
diagrams contributing to~(\ref{MainObject}) to the following form in the limit~$m/H \!\gg\! 1$,
\begin{equation}
\vcenter{\hbox{\includegraphics[width=2.4cm]{4ptFunction}}}
\hspace{-0.3cm}
\longrightarrow
\vcenter{\hbox{\includegraphics[width=2.4cm]{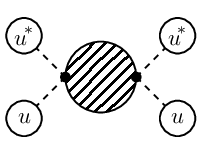}}}
\hspace{-0.cm}+\hspace{-0.cm}
\vcenter{\hbox{\includegraphics[width=3.cm]{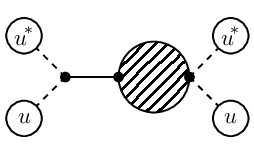}}}
\hspace{-0.cm}+\hspace{-0.cm}
\vcenter{\hbox{\includegraphics[width=3.cm]{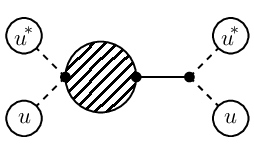}}}
\hspace{-0.cm}+\hspace{-0.cm}
\vcenter{\hbox{\includegraphics[width=3.5cm]{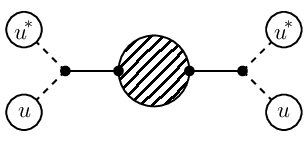}}}
\label{4ReducedDiagrams}
\end{equation}
This is accomplished by (i) taking advantage of the derivative nature of gravitational 
interactions and judiciously  integrating by parts to create favorable combinations of 
derivatives that ``pinch'' certain propagators, and
(ii) by applying Donoghue Identities~\cite{Donoghue:1994dn,Donoghue:1993eb,Bjerrum-Bohr:2002aqa,Donoghue:1996mt} that simplify diagrams while capturing the relevant infrared 
contributions from the loops. In this work we will not need any Donoghue Identities,
since only the manipulation of derivatives in interaction vertices and the
$\Delta\alpha$ variation of the graviton
propagator will be sufficient to demonstrate the vanishing of the gauge variation.

Essentially all identities for manipulating derivatives that we use pertain to 3-vertices,
such as the one below in Eq.~(\ref{VertexIdentity1}). The main vertex identity 
that we use could be considered a de Sitter space generalization of the 
energy-momentum conservation,
\begin{equation}
\vcenter{\hbox{\includegraphics[width=2.9cm]{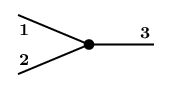}}}
\qquad \Longrightarrow\qquad
a^{D-2} \Big[ \partial_\mu^1 + \partial_\mu^2 + \partial_\mu^3 
	+ \delta_\mu^0 (D\!-\!2)aH \Big] 
	\longrightarrow 0 \, ,
\label{VertexIdentity1}
\end{equation}
where numbers on derivatives designate the leg that the derivative acts on;
generalization to vertices with more attached legs should be obvious.
From this identity follows the main vertex contraction identity we use frequently,
\begin{equation}
a^{D-2} \partial_1 \!\cdot\! \partial_2
	\longrightarrow
	\tfrac{1}{2} \big( \mathcal{D}_3 - \mathcal{D}_1 - \mathcal{D}_2 \big)
	\, .
\label{VertexIdentity2}
\end{equation}
These two fundamental identities are supplemented with another vertex identity
with a different power of the scale factor,
\begin{equation}
H a^{D-1} \bigl( \partial_1 \!\cdot\! \partial_2 \bigr)
	\longrightarrow
	\frac{H a}{2} \bigl( \mathcal{D}_3 \!-\! \mathcal{D}_1 \!-\! \mathcal{D}_2 \bigr)
	-
	H^2a^D \partial^3_0
	+
	\mathcal{O}(H^3)
	\, ,
\label{VertexIdentity3}
\end{equation}
and by the two commutation identities,
\begin{subequations}
\begin{align}
&
\mathcal{D} \partial_\mu
	=
	\Bigl[ \partial_\mu \!-\! \delta_\mu^0 (D \!-\! 2) H a \Bigr] \mathcal{D} 
	+
	\delta_\mu^0
	(D\!-\!2) H^2 a^{D} \partial_0
	\, ,
\\
&
\mathcal{D} Ha
	=
	Ha
	\bigl( \mathcal{D}
	\!-\!
	2H a^{D-1} \partial_0 \bigr)
	-
	D H^3 a^{D+1}
	\, .
\end{align}
\label{CommutationIdentities}%
\end{subequations}

\medskip

From the resulting reduced diagrams in~(\ref{4ReducedDiagrams}), only the last one is in the form that can 
be interpreted as self-mass. For the first three it is necessary to introduce one or two legs, which is 
accomplished easily by using the MMC scalar propagator equation of motion,
\begin{equation}
\mathcal{D}_x \, i \Delta_A(x;x') = i \delta^D(x\!-\!x') \, .
\label{Aeom}
\end{equation}
This way all the resulting reduced diagrams are written in the following form,
\begin{equation}
\vcenter{\hbox{\includegraphics[width=2.6cm]{4ptFunction}}}
\longrightarrow	\hspace{0.5cm}
\vcenter{\hbox{\includegraphics[width=4.2cm]{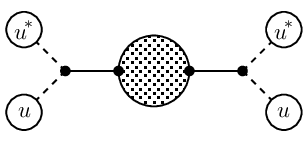}}}
\quad ,
\label{FinalDiagRed}
\end{equation}
from which we can read off the desired effective self-mass,
\begin{equation}
-i\mathcal{M}^2_{\rm eff}(x;x') =  \
\vcenter{\hbox{\includegraphics[width=2.cm]{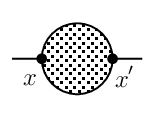}}}
\quad .
\end{equation}
%

\subsection{One-loop reduction}

Applying the outlined procedure at one-loop level in flat space is straightforward~\cite{Miao:2017feh}. 
It requires accounting for 6 classes of nonvanishing diagrams contributing to the 1PI 4-point function. 
Derivative interactions, together with the position space Donoghue identities are sufficient to accomplish 
the task. However, in de Sitter space the situation complicates significantly. In our previous work we have 
identified additional 4-point diagrams that contribute in de Sitter, but which vanish in flat space. In fact, in 
this work we find that the situation is even more complicated: we need to consider {\it all} one-loop 4-point 
functions, {\it and} the one-loop corrections to the attached mode functions. Only after accounting for all 
these contributions do we get that gauge dependence cancels. It therefore behooves us to review the 
reduction formalism and recast it in the language appropriate for our applications, which we do in this 
subsection.

\medskip

The amputated 4-point function is best written as a skeleton expansion in terms of 
1PI $n$-point functions, and resummed two-point functions,
\begin{equation}
\vcenter{\hbox{\includegraphics[width=2.6cm]{4ptFunction}}}
=
\vcenter{\hbox{\includegraphics[width=4.5cm]{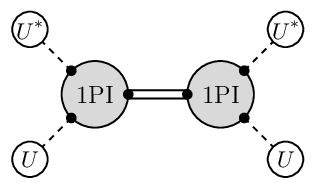}}}
+
\vcenter{\hbox{\includegraphics[width=2.6cm]{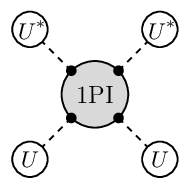}}}
\ .
\label{SkeletonExpansion}
\end{equation}
Here the resummed MMC scalar propagator, denoted by the double solid line, satisfies the 
Dyson-Schwinger equation
\begin{equation}
\vcenter{\hbox{\includegraphics[width=1.75cm]{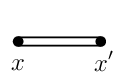}}}
	\ = \
	\vcenter{\hbox{\includegraphics[width=1.75cm]{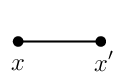}}}
	\ + 
	\vcenter{\hbox{\includegraphics[width=3.5cm]{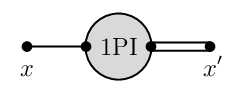}}}
	\quad ,
\label{MasslessDSeq}
\end{equation}
where the single-line term on the right-hand side corresponds to the tree-level propagator.

The diagram corresponding to the tree-level exchange 
potential~(\ref{TreeLevelPotential}) is
\begin{equation}
\vcenter{\hbox{\includegraphics[width=3.2cm]{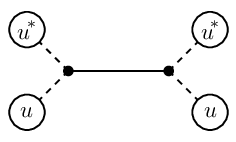}}}
\ ,
\label{TreeLevelDiagram}
\end{equation}
where the tree-level mode function~$u$ satisfies
\begin{equation}
\big( \mathcal{D}_x \!-\! a_x^Dm^2 \big) u(x) = 0 \, .
\label{tree-levelUeom}
\end{equation}
The one-loop corrections to the exchange potential will be captured by the following 
diagrams:
\begin{align}
&
\vcenter{\hbox{\includegraphics[width=4.5cm]{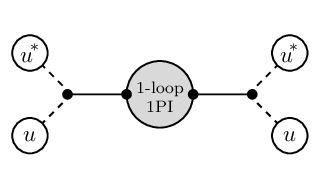}}}
+
\vcenter{\hbox{\includegraphics[width=3.5cm]{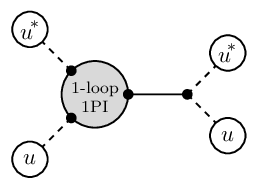}}}
+
\vcenter{\hbox{\includegraphics[width=3.5cm]{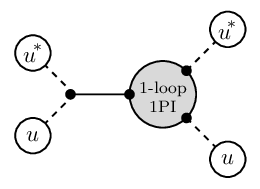}}}
+
\vcenter{\hbox{\includegraphics[width=2.6cm]{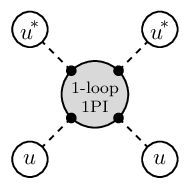}}}
\nonumber \\
&
+
\vcenter{\hbox{\includegraphics[width=3.1cm]{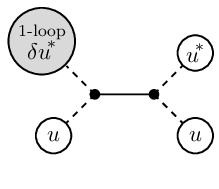}}}
+
\vcenter{\hbox{\includegraphics[width=3.1cm]{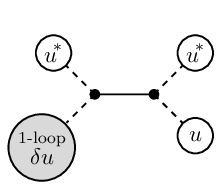}}}
+
\vcenter{\hbox{\includegraphics[width=3.1cm]{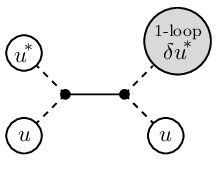}}}
+
\vcenter{\hbox{\includegraphics[width=3.1cm]{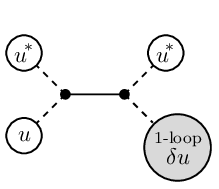}}}
\label{General1loopDiagrams}
\end{align}
The first diagram above descends from the first diagram on the right-hand side
in~(\ref{SkeletonExpansion}) and comes from correcting the massless propagator by 
solving~(\ref{MasslessDSeq}) to one-loop order. The two middle diagrams in the
first line also descend from the first diagram in~(\ref{SkeletonExpansion}),
by truncating the 1PI 3-vertex to one-loop order. Analogously, the last diagram in the
first line above descends from the second diagram on the right-hand side 
in~(\ref{SkeletonExpansion}) by truncating the 1PI 4-vertex to one-loop order. 
On the other hand, all the diagrams of the second line
above come from correcting the massive mode function,~$U \!=\! u \!+\! \delta u$, 
by solving~(\ref{Ueq}) for the one-loop correction,
\begin{equation}
\big( \mathcal{D}_x \!-\! a_x^Dm^2 \big) \delta u(x)
	=
	\int\! d^4x' \, \mathcal{M}_m^2(x;x') u(x') \, ,
\label{oneloopUeq}
\end{equation}
where~$\mathcal{M}_m^2(x;x')$ is the one-loop self-mass of the massive scalar.

\section{Feynman diagrams}
\label{sec: Feynman diagrams}

\begin{figure}[h!]
\center
\hspace{3.cm}
\includegraphics[width=4.2cm]{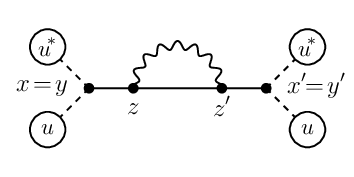}
\hfill
\includegraphics[width=4.cm]{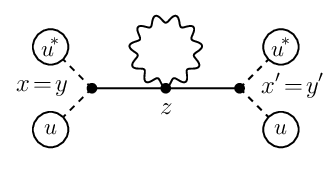}
\hspace{3.cm}
\vskip-6mm
\hspace{4.8cm} $\boldsymbol{0a}$ \hfill$\boldsymbol{0b}$ \hspace{4.8cm}
\vskip+6mm
\hspace{0.7cm}
\includegraphics[width=4.5cm]{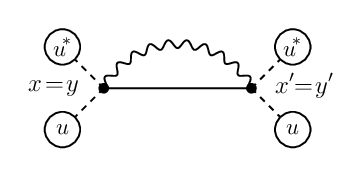}
\hfill
\includegraphics[width=4.5cm]{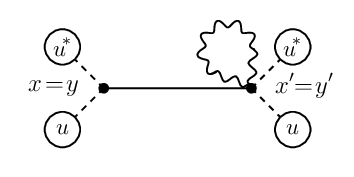}
\hfill
\includegraphics[width=4.5cm]{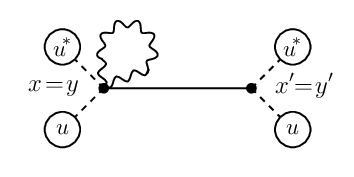}
\hspace{0.7cm}
\vskip-7mm
\hspace{2.7cm} $\boldsymbol{1a}$ \hfill$\boldsymbol{1b}$\hfill $\boldsymbol{1c}$ \hspace{2.7cm}
\vskip+6mm
\includegraphics[width=3.5cm]{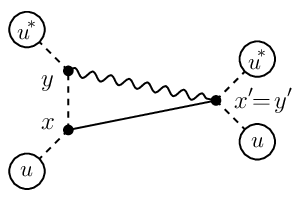}
\hfill
\includegraphics[width=3.5cm]{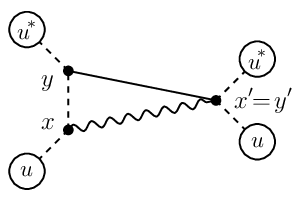}
\hfill
\includegraphics[width=3.5cm]{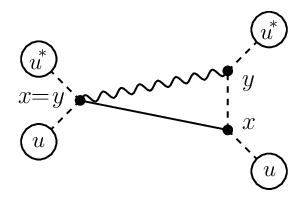}
\hfill
\includegraphics[width=3.5cm]{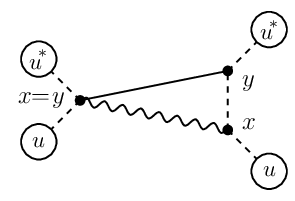}
\vskip-6mm
\hspace{1.4cm} $\boldsymbol{2a}$ \hfill$\boldsymbol{2b}$\hfill $\boldsymbol{2c}$
\hfill $\boldsymbol{2d}$ \hspace{1.4cm}
\vskip+6mm
\hspace{3cm}
\includegraphics[width=3.8cm]{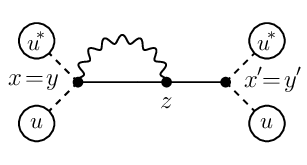}
\hfill
\includegraphics[width=3.8cm]{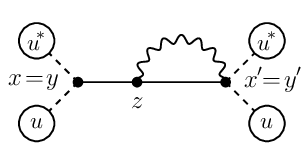}
\hspace{3cm}
\vskip-5mm
\hspace{4.7cm} $\boldsymbol{3a}$ \hfill$\boldsymbol{3b}$ \hspace{4.7cm}
\vskip+6mm
\includegraphics[width=3.5cm]{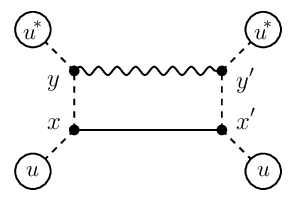}
\hfill
\includegraphics[width=3.5cm]{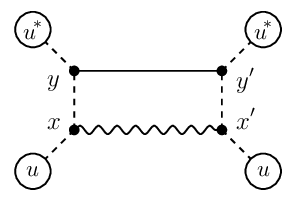}
\hfill
\includegraphics[width=3.5cm]{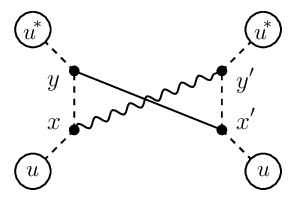}
\hfill
\includegraphics[width=3.5cm]{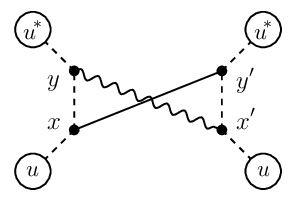}
\vskip-5mm
\hspace{1.4cm} $\boldsymbol{4a}$ \hfill$\boldsymbol{4b}$\hfill $\boldsymbol{4c}$
\hfill $\boldsymbol{4d}$ \hspace{1.4cm}
\vskip+6mm
\includegraphics[width=3.5cm]{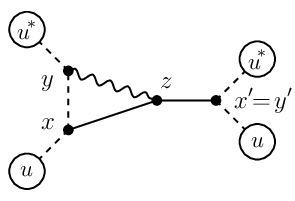}
\hfill
\includegraphics[width=3.5cm]{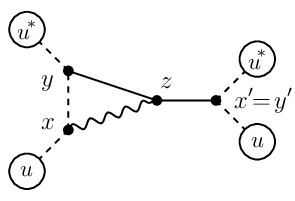}
\hfill
\includegraphics[width=3.5cm]{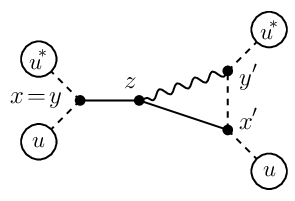}
\hfill
\includegraphics[width=3.5cm]{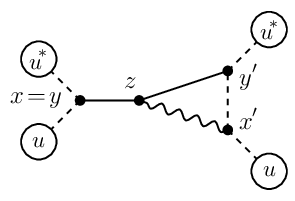}
\vskip-6mm
\hspace{1.4cm} $\boldsymbol{5a}$ \hfill$\boldsymbol{5b}$\hfill $\boldsymbol{5c}$
\hfill $\boldsymbol{5d}$ \hspace{1.4cm}
\vskip+6mm
\includegraphics[width=3.5cm]{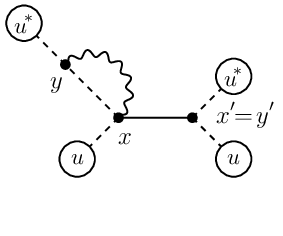}
\hfill
\includegraphics[width=3.5cm]{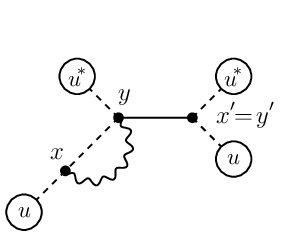}
\hfill
\includegraphics[width=3.5cm]{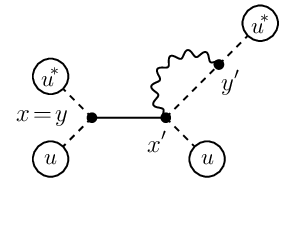}
\hfill
\includegraphics[width=3.5cm]{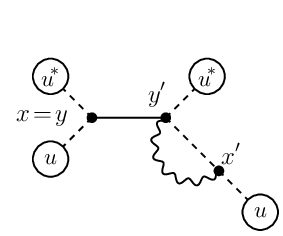}
\vskip-6mm
\hspace{1.6cm} $\boldsymbol{6a}$ \hfill$\boldsymbol{6b}$\hfill $\boldsymbol{6c}$ \ \
\hfill $\boldsymbol{6d}$ \hspace{1.5cm}
\vskip+6mm
\hspace{3cm}
\includegraphics[width=3.5cm]{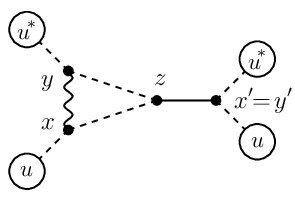}
\hfill
\includegraphics[width=3.5cm]{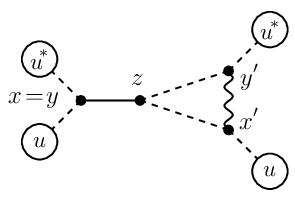}
\hspace{3cm}
\vskip-6mm
\hspace{4.7cm} $\boldsymbol{7a}$ \hfill$\boldsymbol{7b}$ \hspace{4.7cm}
\vskip+1mm
\caption{Eight classes of one-loop diagrams contributing to the $t$-channel
of the connected four-point function, that is put on shell by the attached mode functions.}
\label{4ptDiagrams}
\end{figure}

In this section we give expressions for all the relevant one-loop diagrams in the specific model given 
in~(\ref{action}). These diagrams are constructed from interaction vertices given in 
Table~\ref{OriginalVertices}. The diagrams coming from the amputated 4-point function, corresponding to 
the first line in~(\ref{General1loopDiagrams}), are depicted in Fig.~\ref{4ptDiagrams}, while diagrams 
descending from the mode function corrections, corresponding to the second line 
in~(\ref{General1loopDiagrams}) are given in subsection~\ref{subsec: One-loop diagrams for mode function corrections} later on in the section.

We should comment on the form in which we give expressions corresponding to 
diagrams in Fig.~\ref{4ptDiagrams}. It is unnecessary to
explicitly write some parts of diagrams that are common for all of them. For this reason 
we do not explicitly write the external mode functions, nor the integrals that attach
them to the main part of the diagram. Instead, we give expressions for
4-point functions~$-iV(x;y;x';y')$ that are integrated over to obtain the diagrams,
\begin{equation}
\big( \text{Diag.~}n \big)
	= \int\! d^{D\!}x \, u(x)
	\int\! d^{D\!}y \, u^*(y)
	\int\! d^{D\!}x' \, u(x')
	\int\! d^{D\!}y \, u^*(y') \times
	\Big[ -iV_n(x;y;x';y') \Big]
	\, ,
\end{equation}
where $n\!=\!0,\dots,7$.
When manipulating the 4-point functions that appear in the integrand above,
we immediately drop the terms from~$-iV$ that annihilate the external mode function
via its equation of motion~(\ref{tree-levelUeom}).

We should also comment on the condensed notation for derivatives that we employ 
henceforth, that we find tremendously useful when performing computations.
We found it very convenient to indicate where derivatives act by putting
accents on them, rather than this being indicated by their placement. Given the
preponderance of indices, this strategy saves a lot of time and effort, and makes for 
far more transparent expressions when one acclimates to reading it. Derivatives acting 
on the graviton propagator, or on the integrated propagators contained inside of it
are always denoted by a tilde accent. Derivatives acting on either the scalar propagator in 
the loop, be it massive or massless, are denoted without accents, and derivatives acting on
the scalar propagator outside the loop (again massive or massless) are denoted by a bar.
There are some cases where these conventions lead to some ambiguity, which is removed
by explicitly indicating the coordinate on the derivative, that can be matched with the
corresponding vertex, and consequently the corresponding propagator. For example,
the following expression
\begin{align}
\MoveEqLeft[5]
\Big\{
	\partial_z^\mu i \Delta_A(x;z) 
	\Big\}
	\!\times\!
	\Big\{ \partial_z^\nu 
	i \big[ {}_{\mu\nu} \Delta_{\rho\sigma} \big](z;z') \Big\}
	\!\times\!
	\Big\{
	\partial_{z'}^\rho i \Delta_A(z;z')
	\Big\}
	\!\times\!
	\Big\{ \partial_{z'}^\sigma i \Delta_A(z';x') \Big\}
\nonumber \\
&
	\longrightarrow
	\overline{\partial}{}_z^\mu 
	\widetilde{\partial}{}_z^\nu
	\partial{}_{z'}^\rho
	\overline{\partial}{}_{z'}^\sigma
	\times
	i \Delta_A(x;z)
	\begin{pmatrix}
	i \big[ {}_{\mu\nu}\Delta_{\rho\sigma} \big](z;z')
	\\
	i \Delta_A(z;z')
	\end{pmatrix}
	i \Delta_A(z';x')
	\, .
\end{align}
Furthermore, wherever possible, we try to write propagators in the loops in parentheses
in order to visually resemble the diagram itself. In places where we unpack the condensed 
notation this is usually mentioned explicitly, and otherwise should be clear from the
context.

\subsection{One-loop diagrams for 4-point functions}
\label{subsec: One-loop diagrams for 4-point functions}

In what follows we list the eight classes of diagrams that we consider, which we denote 
by $0, 1, \dots, 7$, that are diagrammatically given in Fig.~\ref{4ptDiagrams}.

\paragraph{Class 0.} The self-mass diagrams are comprised
of two diagrams depicted in the first line of Fig.~\ref{4ptDiagrams}.
These 3-vertex and 4-vertex diagrams, respectively, are given by the following expressions:
\begin{align}
- i V_{0a}
	={}&
	(\kappa\lambda)^2
	\delta^D(x\!-\!y) \delta^D(x'\!-\!y')
	(a_xa_{x'})^D \int\! d^{D\!}z \, d^{D\!}z' \,
	(a_za_{z'})^{D-2}
	\Bigl[
		\overline{\partial}{}_z^\mu \partial_z^\nu
		-
		\tfrac{1}{2} \eta^{\mu\nu} \bigl( \overline{\partial}_z \!\cdot\! \partial_z \bigr)
		\Bigr]
\nonumber \\
&	\hspace{2cm}
	\times\!
	\Bigl[
		\overline{\partial}{}_{z'}^\rho \partial_{z'}^\sigma
		-
		\tfrac{1}{2} \eta^{\rho\sigma} 
			\bigl( \overline{\partial}_{z'} \!\cdot\! \partial_{z'} \bigr)
		\Bigr]
	i \Delta_A(x;z)
	\left(\begin{matrix}
	i \bigl[ {}_{\mu\nu}\Delta_{\rho\sigma} \bigr](z;z')
	\\
	i \Delta_A(z;z')
	\end{matrix}\right)
	i \Delta_A(z';x')
	\, ,
\label{defV0a}
\\
- i V_{0b} ={}&
	i (\kappa\lambda)^2 \delta^D(x\!-\!y) \delta^D(x'\!-\!y')
	(a_x a_{x'})^D
	\int\! d^{D\!}z \, d^{D\!}z' \, a_z^{D-2} \,
		\theta^{\omega\lambda\mu\nu\rho\sigma}\,
\nonumber \\
&	\hspace{4cm}
	\times
	\overline{\partial}{}^z_\omega \overline{\partial}{}^{z'}_\lambda
	i \Delta_A(x;z)
	\left(\begin{matrix}
	i \bigl[ {}_{\mu\nu} \Delta_{\rho\sigma} \bigr](z;z)
	\\
	\delta^D(z \!-\! z')
	\end{matrix}\right)
	i \Delta_A(z';x')
	\, ,
\label{defV0b}
\end{align}
where the constant tensor structure in the second expression is
\begin{equation}
\theta^{\omega\lambda\mu\nu\rho\sigma}
	=
	\eta^{\omega(\mu} \eta^{\nu)(\rho} \eta^{\sigma)\lambda}
	-
	\tfrac{1}{4} \eta^{\omega(\mu} \eta^{\nu)\lambda}\eta^{\rho\sigma}
	-
	\tfrac{1}{4} \eta^{\omega(\rho} \eta^{\sigma)\lambda}\eta^{\mu\nu}
	-
	\tfrac{1}{4} \eta^{\omega\lambda} \eta^{\mu(\rho} \eta^{\sigma)\nu}
	+
	\tfrac{1}{8} \eta^{\omega\lambda} \eta^{\mu\nu} \eta^{\rho\sigma}
	\, .
\label{ThetaStructure}
\end{equation}
The full computation would require to include the counterterm diagram as well, as was done 
in~\cite{Glavan:2024elz}. However, since the~$\Delta\alpha$ variation will produce a gauge independent 
primitive result, we do not need to consider renormalization to prove gauge independence.

\paragraph{Class 1.} This class of diagrams corresponds to the second line in 
Fig.~\ref{4ptDiagrams}. It contains the diagram representing correlation between vertices,
\begin{equation}
- i V_{1a} =
	- \tfrac{1}{4}(\kappa\lambda)^2
	\delta^D(x\!-\!y) \delta^D(x'\!-\!y')
	(a_x a_{x'})^D
	\eta^{\mu\nu} \eta^{\rho\sigma}
	i \bigl[ {}_{\mu\nu} \Delta_{\rho\sigma} \bigr] (x;x')
	i \Delta_A(x;x') 
	\, ,
\label{defV1a}
\end{equation}
as well as the local graviton corrections to 3-vertices,
\begin{align}
- i V_{1b} ={}&
	\tfrac{1}{4} (\kappa\lambda)^2
	\delta^D(x\!-\!y) \delta^D(x' \!-\! y')
	(a_x a_{x'})^D
\nonumber \\
&	\hspace{3.5cm}
	\times
	\Bigl[
	\eta^{\mu(\rho} \eta^{\sigma)\nu}
	-
	\tfrac{1}{2} \eta^{\mu\nu} \eta^{\rho\sigma}
	\Bigr]
	i \bigl[ {}_{\mu\nu} \Delta_{\rho\sigma} \bigr] (x;x)
	i \Delta_A(x;x') 
	\, ,
\label{defV1b}
\\
- i V_{1c} ={}&
	\tfrac{1}{4} (\kappa\lambda)^2
	\delta^D(x\!-\!y) \delta^D(x' \!-\! y')
	(a_x a_{x'})^D
\nonumber \\
&	\hspace{3.5cm}
	\times
	\Bigl[
	\eta^{\mu(\rho} \eta^{\sigma)\nu}
	-
	\tfrac{1}{2} \eta^{\mu\nu} \eta^{\rho\sigma}
	\Bigr]
	i \Delta_A(x;x') 
	i \bigl[ {}_{\mu\nu} \Delta_{\rho\sigma} \bigr] (x';x')
	\, .
\label{defV1c}
\end{align}
%

\paragraph{Class 2.}
This class of diagrams corresponds to vertex-source and vertex-observer correlations,
that are depicted in the third row of Fig.~\ref{4ptDiagrams}. The expressions corresponding
to these diagrams are:
\begin{align}
- i V_{2a} ={}&
	- \tfrac{i}{2}(\kappa\lambda)^2 \delta^D(x'\!-\!y') (a_xa_{x'})^D a_y^{D-2}
\nonumber \\
&	\hspace{1cm}
	\times\!
	\Bigl[
		\overline{\partial}{}_{y}^\mu \partial_{y}^\nu
        	-
		\tfrac{1}{2} \eta^{\mu\nu} 
			\bigl( \overline{\partial}_{y} \!\cdot\! \partial_{y} \!+\!a_y^2m^2 \bigr)
		\Bigr]
	i \Delta_m(x;y)
	\begin{pmatrix}
	i \bigl[ {}_{\mu\nu} \Delta^{\rho}{}_{\rho} \bigr] (y;x')
	\\
	i \Delta_A(x;x') 
	\end{pmatrix}
	\, ,
\label{defV2a}
\\
- i V_{2b} ={}&
	- \tfrac{i}{2}(\kappa\lambda)^2\delta^D(x'\!-\!y') (a_ya_{x'})^D a_x^{D-2}
\nonumber \\
&	\hspace{1cm}
	\times\!
	\Bigl[
		\overline{\partial}{}_{x}^\mu \partial_{x}^\nu
		-
		\tfrac{1}{2} \eta^{\mu\nu} 
			\bigl( \overline{\partial}_{x} \!\cdot\! \partial_{x}\!+\!a_x^2m^2 \bigr)
		\Bigr]
	i \Delta_m(x;y)
	\begin{pmatrix}
	i \Delta_A(y;x') 
	\\
	i \bigl[ {}_{\mu\nu} \Delta^{\rho}{}_{\rho} \bigr] (x;x')
	\end{pmatrix}
	\, ,
\label{defV2b}
\\
- i V_{2c} ={}&
	- \tfrac{i}{2}(\kappa\lambda)^2 \delta^D(x\!-\!y)(a_xa_{x'})^D a_{y'}^{D-2}
\nonumber \\
&	\hspace{1cm}
	\times\!
	\Bigl[
		\overline{\partial}{}_{y'}^\rho \partial_{y'}^\sigma
		-
		\tfrac{1}{2} \eta^{\rho\sigma} 
			\bigl( \overline{\partial}_{y'} \!\cdot\! \partial_{y'}\!+\!a_{y'}^2m^2 \bigr)
		\Bigr]
	\begin{pmatrix}
	i \bigl[ {}^{\mu}{}_{\mu} \Delta_{\rho\sigma} \bigr] (x;y')
	\\
	i \Delta_A(x;x') 
	\end{pmatrix}
	i \Delta_m(x';y')
	\, ,
\label{defV2c}
\\
- i V_{2d} ={}&
	- \tfrac{i}{2}(\kappa\lambda)^2\delta^D(x\!-\!y) (a_xa_{y'})^D a_{x'}^{D-2}
\nonumber \\
&	\hspace{1cm}
	\times\!
	\Bigl[
		\overline{\partial}{}_{x'}^\rho \partial_{x'}^\sigma
		-
		\tfrac{1}{2} \eta^{\rho\sigma} 
			\bigl( \overline{\partial}_{x'} \!\cdot\! \partial_{x'}\!+\!a_{x'}^2m^2 \bigr)
		\Bigr]
	\begin{pmatrix}
	i \Delta_A(x;y') 
	\\
	i \bigl[ {}^{\mu}{}_{\mu} \Delta_{\rho\sigma} \bigr] (x;x')
	\end{pmatrix}
	i \Delta_m(x';y')
	\, .
\label{defV2d}
\end{align}
%

\paragraph{Class 3.}
The fourth row of Fig.~\ref{4ptDiagrams} two diagrams comprising this class, that accounts for correlations 
between vertices and the force carrier,
\begin{align}
- i V_{3a} ={}&
	\!
	- \tfrac{i}{2}(\kappa\lambda)^2 \delta^D(x'\!-\!y') (a_xa_{x'})^D
	\! \int \! d^{D\!}z\, a_z^{D-2}
\nonumber \\
&	\hspace{2cm}
\times\!
	\Bigl[
		\overline{\partial}{}_{z}^\rho \partial_{z}^\sigma
		-
		\tfrac{1}{2} \eta^{\rho\sigma} 
			\overline{\partial}_{z} \!\cdot\! \partial_{z}
		\Bigr]
	\begin{pmatrix}
	i \bigl[ {}^{\mu}{}_{\mu}\Delta_{\rho\sigma} \bigr] (x;z)
	\\
	i \Delta_A(x;z)
	\end{pmatrix}
	i \Delta_A(z;x') 
	\, ,
\label{defV3a}
\\
- i V_{3b} ={}&
	-\tfrac{i}{2}(\kappa\lambda)^2\delta^D(x\!-\!y) 
	(a_xa_{x'})^D
	\! \int \! d^{D\!} z\, a_z^{D-2}
\nonumber \\
&	\hspace{2cm}
	\times \!
	\Bigl[
		\overline{\partial}{}_{z}^\mu \partial_{z}^\nu
		-
		\tfrac{1}{2} \eta^{\mu\nu} 
			\overline{\partial}_{z} \!\cdot\! \partial_{z}
		\Bigr]	i \Delta_A(x;z) 
	\begin{pmatrix}
	i \bigl[ {}_{\mu\nu}\Delta^{\rho}{}_{\rho} \bigr] (z;x')
	\\
	i \Delta_A (z;x')
	\end{pmatrix}
	\, .
\label{defV3b}
\end{align}
%

\paragraph{Class 4.}
The four diagrams comprising the class 4 of (fifth row of 
Fig.~\ref{4ptDiagrams}) represent source-observer correlations,
\begin{align}
- i V_{4a} ={}&
	(\kappa\lambda)^2 (a_x a_{x'})^D
	(a_y a_{y'})^{D-2}
	\Bigl[
		\overline{\partial}{}_y^\mu \partial_y^\nu
		-
		\tfrac{1}{2} \eta^{\mu\nu}
		\bigl( \overline{\partial}{}_y \!\cdot\! \partial_y
			+
			a_y^2m^2 \bigr)
		\Bigr]
\nonumber \\
&
	\times \!
	\Bigl[
		\overline{\partial}{}_{y'}^\rho \partial{}_{y'}^\sigma
		-
		\tfrac{1}{2} \eta^{\rho\sigma}
		\bigl( \overline{\partial}{}_{y'} \!\cdot\! \partial_{y'}
		+
		a_{y'}^2m^2 \bigr)
		\Bigr]
	i \Delta_m(x;y)
	\begin{pmatrix}
	i \bigl[ {}_{\mu\nu}\Delta_{\rho\sigma} \bigr](y;y')
	\\
	i \Delta_A(x;x')
	\end{pmatrix}
	i \Delta_m(x';y')
	\, ,
\label{defV4a}
\\
- i V_{4b} ={}&
	(\kappa\lambda)^2 (a_y a_{y'})^D
	(a_x a_{x'})^{D-2}
	\Bigl[
		\overline{\partial}{}_x^\mu \partial_x^\nu
		-
		\tfrac{1}{2} \eta^{\mu\nu}
		\bigl( \overline{\partial}{}_x \!\cdot\! \partial_x
			+
			a_x^2 m^2 \bigr)
		\Bigr]
\nonumber \\
	&
	\times \!
	\Bigl[
		\overline{\partial}{}_{x'}^\rho \partial_{x'}^\sigma
		-
		\tfrac{1}{2} \eta^{\rho\sigma}
		\bigl( \overline{\partial}{}_{x'} \!\cdot\! \partial_{x'}
			+
			a_{x'}^2m^2 \bigr)
		\Bigr]
	i \Delta_m(x;y)
	\begin{pmatrix}
	i \Delta_A(y;y')
	\\
	i \bigl[ {}_{\mu\nu}\Delta_{\rho\sigma} \bigr](x;x')
	\end{pmatrix}
	i \Delta_m(x';y')
	\, ,
\label{defV4b}
\\
- i V_{4c} ={}&
	(\kappa\lambda)^2 (a_y a_{x'})^D
	(a_x a_{y'})^{D-2}
	\Bigl[
		\overline{\partial}{}_x^\mu \partial_x^\nu
		-
		\tfrac{1}{2} \eta^{\mu\nu}
		\bigl( \overline{\partial}{}_x \!\cdot\! \partial_x
			+
			a_x^2m^2 \bigr)
		\Bigr]
\nonumber \\
&
	\times \!
	\Bigl[
		\overline{\partial}{}_{y'}^\rho \partial_{y'}^\sigma
		-
		\tfrac{1}{2} \eta^{\rho\sigma}
		\bigl( \overline{\partial}{}_{y'} \!\cdot\! \partial_{y'}
			+
			a_{y'}^2m^2 \bigr)
		\Bigr]
	i \Delta_m(x;y)
	\begin{pmatrix}
	i \Delta_A(y;x')
	\\
	i \bigl[ {}_{\mu\nu}\Delta_{\rho\sigma} \bigr](x;y')
	\end{pmatrix}
	i \Delta_m(x';y')
	\, ,
\label{defV4c}
\\
- i V_{4d} ={}&
	(\kappa\lambda)^2 (a_x a_{y'})^D
	(a_y a_{x'})^{D-2}
	\Bigl[
		\overline{\partial}{}_y^\mu \partial_y^\nu
		-
		\tfrac{1}{2} \eta^{\mu\nu}
		\bigl( \overline{\partial}{}_y \!\cdot\! \partial_y
			+
			a_y^2 m^2 \bigr)
		\Bigr]
\nonumber \\
&
	\times \!
	\Bigl[
		\overline{\partial}{}_{x'}^\rho \partial_{x'}^\sigma
		-
		\tfrac{1}{2} \eta^{\rho\sigma}
		\bigl( \overline{\partial}{}_{x'} \!\cdot\! \partial_{x'}
			+
			a_{x'}^2m^2 \bigr)
		\Bigr]
	i \Delta_m(x;y)
	\begin{pmatrix}
	i \bigl[ {}_{\mu\nu}\Delta_{\rho\sigma} \bigr](y;x')
	\\
	i \Delta_A(x;y')
	\end{pmatrix}
	i \Delta_m(x';y')
	\, .
\label{defV4d}
\end{align}
%

\paragraph{Class 5.}
Correlations of the force carrier with the source and the observer are captured
by diagrams in this class. They are given in the sixth line of Fig.~\ref{4ptDiagrams},
with expressions corresponding to them being:
\begin{align}
 - i V_{5a} ={}&
 	(\kappa\lambda)^2 \delta^D(x'\!-\!y')
 	(a_x a_{x'})^D \int\! d^{D\!}z\,
 	(a_y a_{z})^{D-2}
 	\Bigl[
		\overline{\partial}{}_y^\mu \partial_y^\nu
		-
		\tfrac{1}{2} \eta^{\mu\nu} \bigl( \overline{\partial}{}_y \!\cdot\! \partial_y
		+
		a_y^2m^2 \bigr)
		\Bigr]
\nonumber \\
&
	\times\!
	\Bigl[
		\overline{\partial}{}_{z}^\rho \partial_{z}^\sigma
			-
			\tfrac{1}{2} \eta^{\rho\sigma} \overline{\partial}{}_{z} \!\cdot\! \partial_{z}
		\Bigr]
	i \Delta_m(x;y)
	\left(\begin{matrix}
	i \bigl[ {}_{\mu\nu}\Delta_{\rho\sigma} \bigr](y;z)
	\\
	i \Delta_A(x;z)
	\end{matrix}\right)
	i \Delta_A(z;x')
	\, ,
\label{defV5a}
\\
 - i V_{5b} ={}&
 	(\kappa\lambda)^2 \delta^D(x'\!-\!y') 
 	(a_y a_{x'})^D \! \int\! d^{D\!}z\,
 	(a_x a_{z})^{D-2}
  	\Bigl[
		\overline{\partial}{}_x^\mu \partial_x^\nu
		-
		\tfrac{1}{2} \eta^{\mu\nu}
			\bigl( \overline{\partial}{}_x \!\cdot\! \partial_x
			+
			a_x^2m^2 \bigr)
		\Bigr]
\nonumber \\
&
	\times\!
	\Bigl[
		\overline{\partial}{}_{z}^\rho \partial_{z}^\sigma
			-
			\tfrac{1}{2} \eta^{\rho\sigma}
		\overline{\partial}{}_{z} \!\cdot\! \partial_{z}
		\Bigr]
	i \Delta_m(x;y)
	\begin{pmatrix}
	i \Delta_A(y;z)
	\\
	i \bigl[ {}_{\mu\nu}\Delta_{\rho\sigma} \bigr](x;z)
	\end{pmatrix}
	i \Delta_A(z;x')
	\, ,
\label{defV5b}
\\
- i V_{5c} ={}&
	(\kappa\lambda)^2 \delta^D(x\!-\!y) 
	(a_x a_{x'})^D \! \int\! d^{D\!} z \, (a_{z}a_{y'})^{D-2}
	\Bigl[
		\overline{\partial}{}_z^\mu \partial_z^\nu
		-
		\tfrac{1}{2} \eta^{\mu\nu}
		\overline{\partial}{}_z \!\cdot\! \partial_z
		\Bigr]
\nonumber \\
&
	\times\!
	\Bigl[
		\overline{\partial}{}_{y'}^\rho \partial_{y'}^\sigma
		-
		\tfrac{1}{2} \eta^{\rho\sigma}
		\bigl( \overline{\partial}{}_{y'} \!\cdot\! \partial_{y'} + a_{y'}^2m^2 \bigr)
		\Bigr]
	i \Delta_A(x;z)
	\left(\begin{matrix}
	i \bigl[ {}_{\mu\nu}\Delta_{\rho\sigma} \bigr](z;y')
	\\
	i \Delta_A(z;x')
	\end{matrix}\right)
	i \Delta_m(x';y')
	\, ,
\label{defV5c}
\\
- i V_{5d} ={}&
	(\kappa\lambda)^2 
	(a_x a_{y'})^D \delta^D(x\!-\!y)
	\! \int\! d^{D\!} z\,
	(a_{z}a_{x'})^{D-2}
	\Bigl[
		\overline{\partial}{}_z^\mu \partial_z^\nu
		-
		\tfrac{1}{2} \eta^{\mu\nu} \overline{\partial}{}_z \!\cdot\! \partial_z
		\Bigr]
\nonumber \\
&
\times \!
	\Bigl[
		\overline{\partial}{}_{x'}^\rho \partial_{x'}^\sigma
		-
		\tfrac{1}{2} \eta^{\rho\sigma}
		\bigl( \overline{\partial}{}_{x'} \!\cdot\! \partial_{x'} + a_{x'}^2m^2 \bigr)
		\Bigr]
		i \Delta_A(x;z)
		\begin{pmatrix}
		i \Delta_A(z;y')
		\\
		i \bigl[ {}_{\mu\nu}\Delta_{\rho\sigma} \bigr](z;x')
		\end{pmatrix}
		i \Delta_m(x';y')
		\, .
\label{defV5d}
\end{align}
%

\paragraph{Class 6.}
This class of diagrams, depicted in the seventh row in Fig.~\ref{4ptDiagrams},
was not considered previously in~\cite{Glavan:2024elz}. However, it is necessary to
include it to prove gauge independence. These diagrams represent vertex-observer
and vertex-source self-correlations,
\begin{align}
- i V_{6a} ={}&
	-
	\tfrac{i}{2} (\kappa \lambda)^2
	\delta^D(x'\!-\!y')
	(a_xa_{x'})^D a_y^{D-2}
\nonumber \\
&	\hspace{1cm}
	\times\!
	\Bigl[
		\overline{\partial}_y^{(\rho} \partial_y^{\sigma)}
		-
		\tfrac{1}{2} \eta^{\rho\sigma} 
			\bigl( \overline{\partial}_y \!\cdot\! \partial_y \!+\! a_y^2 m^2 \bigr) 
		\Bigr]
		\eta^{\mu\nu}
	\begin{pmatrix}
	i \bigl[ {}_{\mu\nu}\Delta_{\rho\sigma} \bigr](x;y)
	\\
	i \Delta_m(x;y) 
	\end{pmatrix}
	i \Delta_A(x;x')
	\, ,
\label{defV6a}
\\
- i V_{6b} ={}&
	-
	\tfrac{i}{2} (\kappa \lambda)^2
	\delta^D(x'\!-\!y') (a_ya_{x'})^D
	a_x^{D-2}
\nonumber \\
&	\hspace{1cm}
	\times\!
	\Bigl[ \overline{\partial}_x^{(\mu} \partial_x^{\nu)}
		-
		\tfrac{1}{2} \eta^{\mu\nu} 
			\bigl( \overline{\partial}_x \!\cdot\! \partial_x \!+\! a_x^2 m^2 \bigr) 
		\Bigr]
		\eta^{\rho\sigma}
	\begin{pmatrix}
	i \Delta_m(x;y) 
	\\
	i \bigl[ {}_{\mu\nu}\Delta_{\rho\sigma} \bigr](x;y)
	\end{pmatrix}
	i \Delta_A(y;x')
	\, ,
\label{defV6b}
\\
- i V_{6c} ={}&
	-
	\tfrac{i}{2} (\kappa \lambda)^2
	\delta^D(x\!-\!y)
	(a_xa_{x'})^D
	a_{y'}^{D-2}
\nonumber \\
&	\hspace{1cm}
	\times\!
	\Bigl[ \overline{\partial}_{y'}^{(\rho} \partial_{y'}^{\sigma)}
		-
		\tfrac{1}{2} \eta^{\rho\sigma} 
			\bigl( \overline{\partial}_{y'} \!\cdot\! \partial_{y'} \!+\! a_{y'}^2 m^2 \bigr) 
		\Bigr]
		\eta^{\mu\nu}
		i \Delta_A(x;x')
	\begin{pmatrix}
	i \bigl[ {}_{\mu\nu}\Delta_{\rho\sigma} \bigr](x';y')
	\\
	i \Delta_m(x';y')
	\end{pmatrix}
	\, ,
\label{defV6c}
\\
- i V_{6d} ={}&
	-
	\tfrac{i}{2} (\kappa \lambda)^2
	\delta^D(x\!-\!y)
	(a_x a_{y'})^D a_{x'}^{D-2}
\nonumber \\
&	\hspace{1cm}
	\times\!
	\Bigl[ \overline{\partial}_{x'}^{(\mu} \partial_{x'}^{\nu)}
		-
		\tfrac{1}{2} \eta^{\mu\nu} 
			\bigl( \overline{\partial}_{x'} \!\cdot\! \partial_{x'} \!+\! a_{x'}^2 m^2 \bigr) 
		\Bigr]
		\eta^{\rho\sigma}
	i \Delta_A(x;y')
	\begin{pmatrix}
	i \Delta_m(x';y') 
	\\
	i \bigl[ {}_{\mu\nu}\Delta_{\rho\sigma} \bigr](x';y')
	\end{pmatrix}
	\, .
\label{defV6d}
\end{align}
%

\paragraph{Class 7.}
The last class of 4-point diagrams, corresponding to the last line in Fig.~\ref{4ptDiagrams}, were also not 
considered previously in~\cite{Glavan:2024elz}, but we now find it to be necessary. The two diagrams 
comprising it represent source-source and observer-observer self-correlations,
\begin{align}
- i V_{7a} ={}&
	(\kappa \lambda)^2
	\delta^D(x'\!-\!y')
	(a_xa_y)^{D-2} \Bigl[ \overline{\partial}_x^{(\mu} \partial_x^{\nu)}
		- \tfrac{1}{2} \eta^{\mu\nu} 
			\bigl( \overline{\partial}_x \!\cdot\! \partial_x \!+\! a_x^2 m^2 \bigr) 
		\Bigr]
\label{defV7a}
\\
&	\hspace{-0.7cm}
    \times \!
	\Bigl[ \overline{\partial}_y^{(\rho} \partial_y^{\sigma)}
		- \tfrac{1}{2} \eta^{\rho\sigma} 
			\bigl( \overline{\partial}_y \!\cdot\! \partial_y \!+\! a_y^2 m^2 \bigr) 
		\Bigr]
	\int\! d^{D\!}z \, (a_za_{x'})^D
	i \bigl[ {}_{\mu\nu}\Delta_{\rho\sigma} \bigr] (x;y)
	\begin{pmatrix}
	i \Delta_m(y;z)
	\\
	i \Delta_m(x;z) 
	\end{pmatrix}
	i \Delta_A(z;x')
	\, ,
\nonumber \\
- i V_{7b} ={}&
	(\kappa \lambda)^2
	\delta^D(x\!-\!y)
	(a_x'a_y')^{D-2} \Bigl[ \overline{\partial}_{x'}^{(\mu} \partial_{x'}^{\nu)}
		- \tfrac{1}{2} \eta^{\mu\nu} 
			\bigl( \overline{\partial}_{x'} \!\cdot\! \partial_{x'} \!+\! a_{x'}^2 m^2 \bigr) 
		\Bigr]
\label{defV7b}
\\
&	\hspace{-0.7cm}
	\times \!
	\Bigl[ \overline{\partial}_{y'}^{(\rho} \partial_{y'}^{\sigma)}
			- \tfrac{1}{2} \eta^{\rho\sigma} 
			\bigl( \overline{\partial}_{y'} \!\cdot\! \partial_{y'} \!+\! a_{y'}^2 m^2 \bigr) 
		\Bigr]
	\int\! d^{D\!}z \, (a_{x}a_z)^D
	i \Delta_A(x;z)
	\begin{pmatrix}
	i \Delta_m(z;y')
	\\
	i \Delta_m(z;x') 
	\end{pmatrix}
	i \bigl[ {}_{\mu\nu}\Delta_{\rho\sigma} \bigr] (x';y')
	\, .
\nonumber
\end{align}
%

\subsection{Consolidated one-loop diagrams for the 4-point function}

\begin{figure}[h!]
\vskip+3mm
\center
\includegraphics[width=3.5cm]{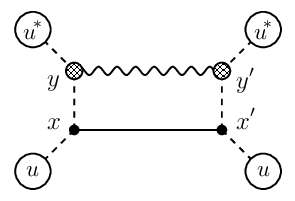}
\hfill
\includegraphics[width=3.5cm]{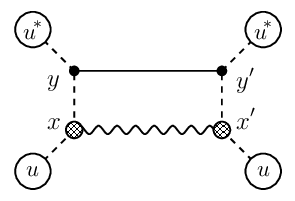}
\hfill
\includegraphics[width=3.5cm]{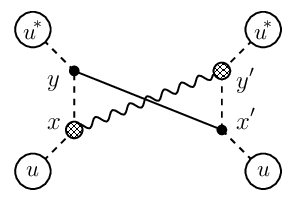}
\hfill
\includegraphics[width=3.5cm]{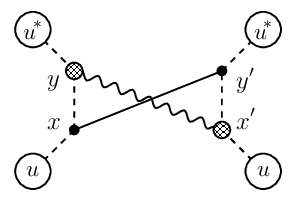}
\vskip-5mm
\hspace{1.4cm} $\boldsymbol{\overline{\overline{4a}}}$ \hfill$\boldsymbol{\overline{\overline{4b}}}$\hfill $\boldsymbol{\overline{\overline{4c}}}$
\hfill $\boldsymbol{\overline{\overline{4d}}}$ \hspace{1.4cm}
\vskip+6mm
\includegraphics[width=3.5cm]{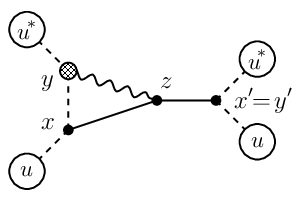}
\hfill
\includegraphics[width=3.5cm]{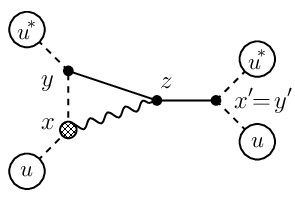}
\hfill
\includegraphics[width=3.5cm]{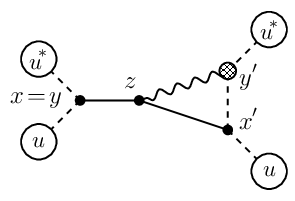}
\hfill
\includegraphics[width=3.5cm]{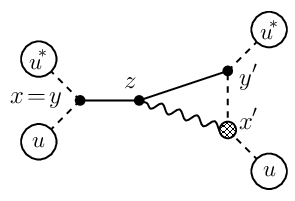}
\vskip-5mm
\hspace{1.4cm} $\boldsymbol{\overline{\overline{4a}}}$ \hfill$\boldsymbol{\overline{\overline{4b}}}$\hfill $\boldsymbol{\overline{\overline{4c}}}$
\hfill $\boldsymbol{\overline{\overline{4d}}}$ \hspace{1.4cm}
\caption{{\it First line:} diagram classe~$\overline{\overline{4}}$ obtained by consolidating
diagram classes in~(\ref{defV4a})--(\ref{defV4d}), and class 2 in (\ref{defV2a})--(\ref{defV2d}),
together with diagram~$1a$ in~(\ref{defV1a}). {\it Second line:} diagram class~$\overline{5}$
obtained by consolidating diagrams from class 5 in~(\ref{defV5a})--(\ref{defV5d}), and
class 3 in~(\ref{defV3a})--(\ref{defV3b}). The consolidated vertex~$\overline{\text{B}}$,
represented as a hatched vertex node, is defined in Eq.~(\ref{ConsolidatedBvertex}).
}
\label{consolidated1PIdiagrams}
\end{figure}

A considerable simplification for the contributing diagrams in classes 1--5 was
realized in~\cite{Glavan:2024elz}. This simplification applies regardless of the graviton
gauge, and is attained by the consolidation of some classes of diagrams, that is
effectively captured by eliminating vertex D and modifying vertex B,
\begin{equation}
\text{vertex B} \, \longrightarrow \, \text{vertex $\overline{\rm B}$} \, ,
\qquad \quad
\text{vertex D} \, \longrightarrow \, 0 \, ,
\label{VertexConsolidation1}
\end{equation}
where the new consolidated vertex is given by:
\begin{equation}
\overline{\rm B} \qquad
\vcenter{\hbox{\includegraphics{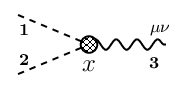}}}
\qquad
- i \kappa \Bigl[ - a_x^{D-2} \partial_1^{(\mu} \partial_2^{\nu)}
		+ \tfrac{1}{4} \eta^{\mu\nu} \mathcal{D}_3
		\Bigr]
	\, .
\label{ConsolidatedBvertex}
\end{equation}
This consolidation collapses (i) diagram classes 2 and 4, together with diagram $1a$, into a consolidated 
diagram class~$\overline{\overline{4}}$, given in the first row of Fig.~\ref{consolidated1PIdiagrams}, and 
(ii) diagram classes 3 and 5 into a consolidated diagram class $\overline{5}$, given in the second row of 
Fig~\ref{consolidated1PIdiagrams}. The two remaining diagrams in class 1 are henceforth referred to as 
class~$\overline{1}$.

\paragraph{Class $\boldsymbol{\overline{\overline{4}}}$.}
The diagrams from classes 2 and 4, and the diagram~$1a$ combine into four diagrams 
that make this new consolidated class:
\begin{align}
- i V_{\overline{\overline{4a}}}
	={}&
	(\kappa\lambda)^2 (a_x a_{x'})^D
	\Bigl[
		a_y^{D-2} \overline{\partial}{}_y^\mu \partial_y^\nu
			-
			\tfrac{1}{4} \eta^{\mu\nu} \widetilde{\mathcal{D}}_y
		\Bigr]
	\Bigl[
		a_{y'}^{D-2} \overline{\partial}{}_{y'}^\rho \partial_{y'}^\sigma
			-
			\tfrac{1}{4} \eta^{\rho\sigma} \widetilde{\mathcal{D}}_{y'}
		\Bigr]
\nonumber \\
&	\hspace{1cm}
	\times
	i \Delta_m(x;y)
	\begin{pmatrix}
	i \bigl[ {}_{\mu\nu}\Delta_{\rho\sigma} \bigr](y;y')
	\\
	i \Delta_A(x;x')
	\end{pmatrix}
	i \Delta_m(x';y')
	\, ,
\label{4barbara}
\\
 - i V_{\overline{\overline{4b}}}
 	={}&
	(\kappa\lambda)^2 (a_y a_{y'})^D
	\Bigl[
		a_x^{D-2} \overline{\partial}{}_x^\mu \partial_x^\nu
			-
			\tfrac{1}{4} \eta^{\mu\nu} \widetilde{\mathcal{D}}_x
		\Bigr]
	\Bigl[
		a_{x'}^{D-2} \overline{\partial}{}_{x'}^\rho \partial_{x'}^\sigma
			-
			\tfrac{1}{4} \eta^{\rho\sigma} \widetilde{\mathcal{D}}_{x'}
		\Bigr]
\nonumber \\
&	\hspace{1cm}
	\times
	i \Delta_m(x;y)
	\begin{pmatrix}
	i \Delta_A(y;y')
	\\
	i \bigl[ {}_{\mu\nu}\Delta_{\rho\sigma} \bigr](x;x')
	\end{pmatrix}
	i \Delta_m(x';y')
	\, ,
\label{4barbarb}
\\
- i V_{\overline{\overline{4c}}}
	={}&
	(\kappa\lambda)^2 (a_y a_{x'})^D
	\Bigl[
		a_x^{D-2} \overline{\partial}{}_x^\mu \partial_x^\nu
			-
			\tfrac{1}{4} \eta^{\mu\nu} \widetilde{\mathcal{D}}_x
		\Bigr]
	\Bigl[
		a_{y'}^{D-2} \overline{\partial}{}_{y'}^\rho \partial_{y'}^\sigma
			-
			\tfrac{1}{4} \eta^{\rho\sigma} \widetilde{\mathcal{D}}_{y'}
		\Bigr]
\nonumber \\
&	\hspace{1cm}
	\times
	i \Delta_m(x;y)
	\begin{pmatrix}
	i \Delta_A(y;x')
	\\
	i \bigl[ {}_{\mu\nu}\Delta_{\rho\sigma} \bigr](x;y')
	\end{pmatrix}
	i \Delta_m(x';y')
	\, ,
\label{4barbarc}
\\
- i V_{\overline{\overline{4d}}}
	={}&
	(\kappa\lambda)^2 (a_x a_{y'})^D
	\Bigl[
		a_y^{D-2} \overline{\partial}{}_y^\mu \partial_y^\nu
			-
			\tfrac{1}{4} \eta^{\mu\nu} \widetilde{\mathcal{D}}_y
		\Bigr]
	\Bigl[
		a_{x'}^{D-2} \overline{\partial}{}_{x'}^\rho \partial_{x'}^\sigma
			-
			\tfrac{1}{4} \eta^{\rho\sigma} \widetilde{\mathcal{D}}_{x'}
		\Bigr]
\nonumber \\
&	\hspace{1cm}
	\times
	i \Delta_m(x;y)
	\begin{pmatrix}
	i \bigl[ {}_{\mu\nu}\Delta_{\rho\sigma} \bigr](y;x')
	\\
	i \Delta_A(x;y')
	\end{pmatrix}
	i \Delta_m(x';y')
	\, .
\label{4barbard}
\end{align}
%

\paragraph{Class $\boldsymbol{\overline{5}}$.}
This consolidated class is made up of four diagrams resulting from combining diagrams
of classes 3 and 5:
\begin{align}
- i V_{\overline{5a}}
	={}&
	(\kappa\lambda)^2
	\delta^D(x'\!-\!y')
	(a_x a_{x'})^D
	\Bigl[
		a_y^{D-2} \overline{\partial}{}_y^\mu \partial_y^\nu
			-
			\tfrac{1}{4} \eta^{\mu\nu} \widetilde{\mathcal{D}}_y
		\Bigr]
	\int\! d^{D\!}z \, a_{z}^{D-2}
\nonumber \\
&	\hspace{1cm}
	\times \!
	\Bigl[ \overline{\partial}_{z}^\rho \partial_{z}^\sigma
			-
			\tfrac{1}{2} \eta^{\rho\sigma} \overline{\partial}_{z} \!\cdot\! \partial_{z}
		\Bigr]
	i \Delta_m(x;y)
	\begin{pmatrix}
	i \bigl[ {}_{\mu\nu} \Delta_{\rho\sigma} \bigr](y;z)
	\\
	i \Delta_A(x;z)
	\end{pmatrix}
	i \Delta_A(z;x')
	\, ,
\label{defV5bara}
\\
- i V_{\overline{5b}}
	={}&
	(\kappa\lambda)^2
	\delta^D(x'\!-\!y')
	(a_y a_{x'})^D
	\Bigl[
		a_x^{D-2} \overline{\partial}{}_x^\mu \partial_x^\nu
			-
			\tfrac{1}{4} \eta^{\mu\nu} \widetilde{\mathcal{D}}_x
		\Bigr]
	\int\! d^{D\!}z \, a_{z}^{D-2}
\nonumber \\
&	\hspace{1cm}
	\times \!
	\Bigl[ \overline{\partial}_{z}^\rho \partial_{z}^\sigma
			-
			\tfrac{1}{2} \eta^{\rho\sigma} \overline{\partial}_{z} \!\cdot\! \partial_{z}
		\Bigr]
	i \Delta_m(x;y)
	\begin{pmatrix}
	i \Delta_A(y;z)
	\\
	i \bigl[ {}_{\mu\nu} \Delta_{\rho\sigma} \bigr](x;z)
	\end{pmatrix}
	i \Delta_A(z;x')
	\, ,
\label{defV5barb}
\\
- i V_{\overline{5c}}
	={}&
	(\kappa\lambda)^2
	\delta^D(x\!-\!y)
	(a_x a_{x'})^D
	\Bigl[
		a_{y'}^{D-2} \overline{\partial}{}_{y'}^\mu \partial_{y'}^\nu
			-
			\tfrac{1}{4} \eta^{\mu\nu} \widetilde{\mathcal{D}}_{y'}
		\Bigr]
	\int\! d^{D\!}z \, a_z^{D-2}
\nonumber \\
&	\hspace{1cm}
	\times \!
	\Bigl[
		\overline{\partial}{}_z^\rho \partial_z^\sigma
			-
			\tfrac{1}{2} \eta^{\rho\sigma} \overline{\partial}_z \!\cdot\! \partial_z
		\Bigr]
	i \Delta_A(x;z)
	\begin{pmatrix}
	i \bigl[ {}_{\rho\sigma}\Delta_{\mu\nu} \bigr](z;y')
	\\
	i \Delta_A(z;x')
	\end{pmatrix}
	i \Delta_m(x';y')
	\, ,
\label{defV5barc}
\\
- i V_{\overline{5d}}
	={}&
	(\kappa\lambda)^2
	\delta^D(x\!-\!y)
	(a_x a_{y'})^D
	\Bigl[
		a_{x'}^{D-2} \overline{\partial}{}_{x'}^\mu \partial_{x'}^\nu
			-
			\tfrac{1}{4} \eta^{\mu\nu} \widetilde{\mathcal{D}}_{x'}
		\Bigr]
	\int\! d^{D\!}z \, a_z^{D-2}
\nonumber \\
&	\hspace{1cm}
	\times \!
	\Bigl[
		\overline{\partial}{}_z^\rho \partial_z^\sigma
			-
			\tfrac{1}{2} \eta^{\rho\sigma} \overline{\partial}_z \!\cdot\! \partial_z
		\Bigr]
	i \Delta_A(x;z)
	\begin{pmatrix}
	i \Delta_A(z;y')
	\\
	i \bigl[ {}_{\rho\sigma}\Delta_{\mu\nu} \bigr](z;x')
	\end{pmatrix}
	i \Delta_m(x';y')
	\, .
\label{defV5barad}
\end{align}
%

\subsection{One-loop diagrams for mode function corrections}
\label{subsec: One-loop diagrams for mode function corrections}

The one-loop diagrams corresponding to the mode function corrections in the second line 
of~(\ref{General1loopDiagrams}) are given by, respectively,
\begin{align}
\big( \text{Diag.~}\delta u,a \big)
	={}&
	-\lambda^2
	\int\! d^{D\!}x \, u(x) \delta u^*(x)
	\int\! d^{D\!}x' \, u(x') \, u^*(x') \times
	i\Delta_A(x;x')
	\, ,
\label{DiagDeltaUa}
\\
\big( \text{Diag.~}\delta u,b \big)
	={}&
	-\lambda^2
	\int\! d^{D\!}x \, \delta u(x) u^*(x)
	\int\! d^{D\!}x' \, u(x') \, u^*(x') \times
	i\Delta_A(x;x')
	\, ,
\label{DiagDeltaUb}
\\
\big( \text{Diag.~}\delta u,c \big)
	={}&
	-\lambda^2
	\int\! d^{D\!}x \, u(x) u^*(x)
	\int\! d^{D\!}x' \, u(x') \, \delta u^*(x') \times
	i\Delta_A(x;x')
	\, ,
\label{DiagDeltaUc}
\\
\big( \text{Diag.~}\delta u,d \big)
	={}&
	-\lambda^2
	\int\! d^{D\!}x \, u(x) u^*(x)
	\int\! d^{D\!}x' \, \delta u(x') \, u^*(x') \times
	i\Delta_A(x;x')
	\, ,
\label{DiagDeltaUd}
\end{align}
with~$\delta u$ being the one-loop correction to the mode function coming
from solving~(\ref{oneloopUeq}).

\begin{figure}[h!]
\vskip+5mm
\hspace{3.7cm}
$I$
\hspace{5.1cm}
$I\!I$
\vskip-4mm
\centering
\hspace{4.cm}
\includegraphics[height=1.7cm]{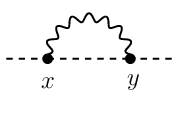}
\hfill
\includegraphics[height=1.7cm]{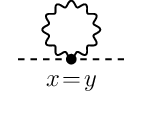}
\hspace{4.cm}
\vskip-6mm
\caption{One-loop diagrams representing contributions to the self-mass of the massive 
scalar field: the 3-vertex diagram ($I$) and the 4-vertex diagram~($I\!I$).}
\label{MassiveSelfMassDiagrams}
\end{figure}

The mode function equation is corrected by the one-loop self-mass for the massive
scalar. We split this self-mass into two parts, one made up of 3-vertices, and another
made up of a 4-vertex, diagrammatically depicted in Fig.~\ref{MassiveSelfMassDiagrams}. 
The expressions corresponding to the two self-mass contributions are given by
\begin{align}
-i\mathcal{M}_I^2(x;y)
	={}&
	- \kappa^2 
	(a_xa_{y})^{D-2} 
	 \Bigl[ \overline{\partial}{}_x^{(\mu} \partial{}_x^{\nu)}
		- \tfrac{1}{2} \eta^{\mu\nu}
		\bigl( \overline{\partial}{}_x \!\cdot\! \partial{}_x \!+\! a_x^2 m^2 \bigr) 
		\Bigr]
\nonumber \\
&	\hspace{1.5cm}
	\times\!
	 \Bigl[ \overline{\partial}{}_y^{(\rho} \partial{}_y^{\sigma)}
		- \tfrac{1}{2} \eta^{\rho\sigma}
		\bigl( \overline{\partial}{}_y \!\cdot\! \partial{}_y \!+\! a_y^2 m^2 \bigr) 
		\Bigr]
	\begin{pmatrix}
	i \big[ {}_{\mu\nu} \Delta_{\rho\sigma} \big](x;y)
	\\
	i \Delta_m(x;y)
	\end{pmatrix}
	\, ,
\label{defMI}
\\
-i\mathcal{M}_{I\!I}^2(x;y)
	={}&
	-i\kappa^2
	\delta^D(x\!-\!y)
	a_x^{D-2}
	\Bigl[
	\tfrac{1}{2}
	\overline{\partial}{}_x^{(\mu} \eta^{\nu)(\rho} \overline{\partial}{}_{y}^{\sigma)}
	+
	\tfrac{1}{2}
	\overline{\partial}{}_x^{(\rho} \eta^{\sigma)(\mu} \overline{\partial}{}_{y}^{\nu)}
	-
	\tfrac{1}{4} \overline{\partial}{}_x^{(\mu} \overline{\partial}{}_{y}^{\nu)} \eta^{\rho\sigma}
	-
	\tfrac{1}{4} \overline{\partial}{}_x^{(\rho} \overline{\partial}{}_{y}^{\sigma)} 
		\eta^{\mu\nu}
\nonumber \\
&	\hspace{1.5cm}
	- \tfrac{1}{8} \bigl( 2 \eta^{\mu(\rho} \eta^{\sigma)\nu}  \!
	\!-\!
	\eta^{\mu\nu} \eta^{\rho\sigma} \bigr)
	\bigl( \overline{\partial}{}_x \!\cdot\! \overline{\partial}{}_{y} \!+\! a_x^2 m^2 \bigr)
	\Bigr]
	i \big[ {}_{\mu\nu} \Delta_{\rho\sigma} \big](x;y)
	\, ,
\label{defMII}
\end{align}
where we interpret barred derivatives as acting outside on whatever attaches to the self-mass, in line with 
the conventions in the rest of the paper. Accordingly, we split the corrections to mode functions into two 
classes as well,~$\delta u \!=\! \delta u_I \!+\! \delta u_{I\!I}$, each sourced by the corresponding contribution 
to the self-mass,
\begin{equation}
\big( \mathcal{D}_x \!-\! a_x^Dm^2 \big) \delta u_N(x)
	=
	\int\! d^Dy \, \mathcal{M}_N^2(x;y) \, u(y) \, ,
\qquad\quad
N = I , I\!I \, ,
\label{deltaUNeq}
\end{equation}
where in this equation all external $x$-derivatives in the self-mass are reflected to act on the self-mass.

\section{Propagators}
\label{sec: Propagators}

The diagrammatics of the preceding subsection are independent of the choice of
gauge for the graviton propagator. Here we collect expressions and results for the 
scalar propagators and the graviton propagator in the simple one-parameter gauge,
that we will use to evaluate the diagrams.

The propagator for a scalar field of mass~$M$ in de Sitter space satisfies the following 
equation of motion
\begin{equation}
\bigl( \mathcal{D}_x - a_x^DM^2 \bigr) i \Delta_M(x;x')
	=
	i \delta^D(x\!-\!x')
	\, .
\label{ScalarEOM}
\end{equation}
The solutions to this equation are de Sitter invariant 
for~$M^2\!>\!0$~\cite{Chernikov:1968zm}, but break de Sitter invariance already 
in the massless limit~$M^2\!=\!0$~\cite{Allen:1987tz,Onemli:2002hr}, and also for
tachyonic masses~$M^2\!<\!0$~\cite{Janssen:2008px}.
Apart from the scalar propagator with a heavy mass~$m/H\!\gg\!1$, 
whose specific form we will
not need, we will need four scalar propagators with specific masses:
\begin{equation}
M_W^2 =
	- D H^2
	\, ,
\qquad
M_A^2 = 0
	\, ,
\qquad
M_B^2 =
	(D\!-\!2) H^2
	\, ,
\qquad
M_C^2 = 2(D\!-\!3) H^2
	\, .
\end{equation}
The first two will thus contain de Sitter breaking parts, while the last two will be completely
de Sitter invariant. Their power-series representations are~\cite{Miao:2011fc}:
\begin{align}
i \Delta_W(x;x') ={}&
	\frac{ H^{D-2} }{ (4\pi)^{ \frac{D}{2} } }
	\biggl\{
	\Gamma\bigl( \tfrac{D-2}{2} \bigr)
	\Bigl( \frac{4}{y} \Bigr)^{\! \frac{D-2}{2} }
	+
	\frac{ 4 \, \Gamma \bigl( \frac{D+2}{2} \bigr) }{ (D\!-\!4) (D\!-\!2) }
	\Bigl( \frac{4}{y} \Bigr)^{\! \frac{D-4}{2}}
	+
	\frac{2\,\Gamma\bigl( \frac{D+6}{2} \bigr)}{ (D\!-\!6) (D\!-\!4) }
	\Bigl( \frac{4}{y} \Bigr)^{\! \frac{D-6}{2}}
\nonumber \\
&
	+
	\frac{\Gamma(D)}{\Gamma(\frac{D}{2})}
	\biggl[
	\frac{D \!+\! 1}{2D}
	+
	(D\!-\!1) a_xa_{x'}
	-
	\frac{1}{D\!-\!1} \Bigl( \frac{a_x}{a_{x'}} + \frac{a_{x'}}{a_x} \Bigr)
	\biggr]
\nonumber \\
&
	+
	\frac{\Gamma(D)}{2\,\Gamma\bigl( \frac{D}{2} \bigr)}
		\biggl[
		\psi\bigl( - \tfrac{D}{2} \bigr)
		- \psi\bigl( \tfrac{D+1}{2} \bigr)
		- \psi(D\!+\!1)
		+ \gamma_{E}
		- \frac{ 2\ln(a_x a_{x'}) }{D \!+\! 1}
		\biggr] (y-2)
\nonumber \\
&
	+
	\sum_{n=2}^{\infty}
	\biggl[
	\frac{ \Gamma\bigl( \frac{D+4}{2} \!+\! n \bigr) }
		{ \bigl(n \!-\! \frac{D-4}{2} \bigr) \bigl( n \!-\! \frac{D-2}{2} \bigr) (n\!-\!1)! }
		\Bigl( \frac{y}{4} \Bigr)^{\! n - \frac{D-4}{2}}
		\!\! -
		\frac{ \Gamma(n\!+\!D) }{ n(n\!-\!1) \, \Gamma\bigl( n \!+\! \frac{D}{2} \bigr) }
		\Bigl( \frac{y}{4} \Bigr)^{\! n }
	\biggr]
	\biggr\}
	\, ,
\label{DeltaWdef}
\\
i \Delta_A(x;x') ={}&
	\frac{ H^{D-2} }{ (4\pi)^{ \frac{D}{2} } }
	\biggl\{
	\Gamma\bigl( \tfrac{D-2}{2} \bigr)
	\Bigl( \frac{4}{y} \Bigr)^{\! \frac{D-2}{2} }
	+
	\frac{ 2 \, \Gamma\bigl( \frac{D+2}{2} \bigr) }{ D\!-\!4 } 
		\Bigl( \frac{4}{y} \Bigr)^{\! \frac{D-4}{2} }
	+
	\frac{ \Gamma ( D \!-\! 1 ) }
		{ \Gamma\bigl( \frac{D}{2} \bigr) } 
	\Bigl[ 
	\ln(aa') + \Psi \Bigr]
\nonumber \\
&
	-
	\sum_{n=1}^{\infty}
	\biggl[
	\frac{ \Gamma\bigl( \frac{D+2}{2} \!+\! n \bigr)  }
		{ \bigl( \frac{4-D}{2} \!+\! n \bigr) \, (n\!+\!1)! } 
		\Bigl( \frac{y}{4} \Bigr)^{\! n - \frac{D-4}{2} }
	\!-
	\frac{ \Gamma( D \!-\! 1 \!+\! n )  }
		{ n \, \Gamma\bigl( \frac{D}{2} \!+\! n \bigr) }
		\Bigl( \frac{y}{4} \Bigr)^{\! n }
	\biggr]
	\biggr\}
	\, ,
\label{DeltaAdef}
\\
i \Delta_B(x;x') ={}&
\frac{ H^{D-2} }{ (4\pi)^{ \frac{D}{2} } }
	\biggl\{
	\Gamma\bigl( \tfrac{D-2}{2} \bigr)
	\Bigl( \frac{4}{y} \Bigr)^{\! \frac{D-2}{2} }
	+
	\sum_{n=0}^{\infty}
	\biggl[
	\frac{ \Gamma\bigl( \frac{D}{2} \!+\! n \bigr) }{ (n\!+\!1)! } 
		\Bigl( \frac{y}{4} \Bigr)^{\! n - \frac{D-4}{2} }
	\!-
	\frac{ \Gamma ( D \!-\! 2 \!+\! n ) }{ \Gamma\bigl( \frac{D}{2} \!+\! n \bigr) }
		\Bigl( \frac{y}{4} \Bigr)^{\! n }
	\biggr]
	\biggr\}
	\, ,
\label{DeltaBdef}
\\
i \Delta_C(x;x') ={}&
	\frac{ H^{D-2} }{ (4\pi)^{ \frac{D}{2} } }
	\biggl\{
	\Gamma\bigl( \tfrac{D-2}{2} \bigr)
	\Bigl( \frac{4}{y} \Bigr)^{\! \frac{D-2}{2} }
	-
	\sum_{n=0}^{\infty}
	\biggl[
	\frac{ \bigl( \frac{6-D}{2} \!+\! n \bigr) \, \Gamma\bigl( \frac{D-2}{2} \!+\! n \bigr) }
			{ (n\!+\!1)! } 
		\Bigl( \frac{y}{4} \Bigr)^{\! n - \frac{D-4}{2} }
\nonumber \\
&	\hspace{7cm}
	-
	\frac{ \Gamma( D \!-\! 3 \!+\! n ) \, (1+n)  }
		{ \Gamma\bigl( \frac{D}{2} \!+\! n \bigr) } 
		\Bigl( \frac{y}{4} \Bigr)^{\! n }
	\biggr]
	\biggr\}
	\, ,
\label{DeltaCdef}
\end{align}
where~$\Psi \!=\! - \psi( \tfrac{2-D}{2}) 
	+ \psi(\tfrac{D-1}{2})
	+ \psi(D\!-\!1)
	- \gamma_E$.

We will need a particular kind of integrated propagator
that satisfies equations sourced by scalar propagators,
\begin{equation}
\bigl( \mathcal{D}_x - a_x^DM_I^2 \bigr) K_{IJ}(x;x') = 
	a_x^{D-2} i \Delta_J(x;x') \, ,
\qquad
\bigl( \mathcal{D}_{x'} - a_{x'}^D M_J^2 \bigr) K_{IJ}(x;x') = 
	a_{x'}^{D-2} i \Delta_I(x;x') \, ,
\end{equation}
so that its solution can be written in an integral form~\cite{Glavan:2019msf},
\begin{equation}
K_{IJ}(x;x') = - i \int\! d^{D\!}z \, a_z^{D-2} \,
	i \Delta_I(x;z) \, i \Delta_J(z;x')
	\, .
\end{equation}
In particular we will need its two diagonal instances, that can be evaluated in terms of 
differences of ordinary scalar propagators~\cite{Glavan:2025azq},
\begin{equation}
K_{AA}(x;x') =
	\frac{ i \Delta_B(x;x') - i \Delta_W(x;x') }{2(D\!-\!1)H^2 a_xa_{x'} }
	\, ,
\qquad
K_{BB}(x;x') =
	\frac{ i \Delta_C(x;x') - i \Delta_A(x;x') }{2(D\!-\!3)H^2 a_xa_{x'} }
	\, .
\label{defKdiff}
\end{equation}

Using the results for the scalar propagators collected above we can write the
graviton propagator in the simple one-parameter 
gauge~\cite{Glavan:2019msf,Glavan:2025azq},
\begin{equation}
i \bigl[ {}_{\mu\nu}\Delta_{\rho\sigma} \bigr](x;x')
	=
	i \bigl[ {}_{\mu\nu} \Delta_{\rho\sigma} \bigr](x;x')
	\Big|_{\alpha=1}
	+
	\Delta \alpha \!\times
	i \bigl[ {}_{\mu\nu}\Theta_{\rho\sigma} \bigr](x;x')
	\, ,
\label{GravitonPropagator}
\end{equation}
where~$\Delta\alpha\!=\!\alpha \!-\! 1$ can take arbitrary finite values. 
The first term is the graviton propagator in the
simplest gauge~\cite{Tsamis:1992xa,Woodard:2004ut},
\begin{align}
\MoveEqLeft[5]
	i \bigl[ {}_{\mu\nu} \Delta_{\rho\sigma} \bigr](x;x')
	\Big|_{\alpha=1}
	=
	2\Bigl[ \overline{\eta}_{\rho(\mu} \overline{\eta}_{\nu)\sigma}
		- \frac{ \overline{\eta}_{\mu\nu} \overline{\eta}_{\rho\sigma} }{ D \!-\! 3 } \Bigr]
	i \Delta_A (x;x')
	-
	4 \delta^0_{(\mu} \overline{\eta}_{\nu)(\rho} \delta^0_{\sigma)} \,
	i \Delta_B (x;x')
\nonumber \\
&
	+
	\frac{2}{(D \!-\! 2)(D \!-\! 3)}
	\Bigl[ \overline{\eta}_{\mu\nu} \!+\! (D\!-\!3) \delta_\mu^0 \delta_\nu^0 \Bigr]
	\Bigl[ \overline{\eta}_{\rho\sigma} \!+\! (D\!-\!3) \delta_\rho^0 \delta_\sigma^0 \Bigr]
	i \Delta_C (x;x')
	\, ,
\end{align}
corresponding to the choice~$\alpha\!=\!1$ for the gauge-fixing parameter. The 
computation in that gauge has been reported in~\cite{Glavan:2024elz}. Here we are 
interested in computing the~$\Delta \alpha$ contribution in the one-parameter simple
gauge, and proving gauge independence. This requires us to recompute all the diagrams
using the second part of the graviton propagator in~(\ref{GravitonPropagator}),
\begin{equation}
i \bigl[ {}_{\mu\nu}\Theta_{\rho\sigma} \bigr](x;x')
	=
	-
	4\partial{}_{(\mu}^x \overline{\eta}_{\nu)(\rho} \partial{}^{x'}_{\sigma)} K_{AA}(x;x')
	+
	4\Bigl[ \delta{}^0_{(\mu} \partial{}_{\nu)}^x \!-\! \eta_{\mu\nu} Ha_x \Bigr]
	\Bigl[ \delta{}^0_{(\rho} \partial{}_{\sigma)}^{x'} \!-\! \eta_{\rho\sigma} Ha_{x'} \Bigr]
	K_{BB}(x;x')
	\, .
\label{DeltaAlphaGraviton}
\end{equation}
that we call the~$\Delta\alpha$ variation.

Note that the derivative operators acting on integrated propagators in the graviton 
propagator annihilate some of the terms in the power-series expansion of
specific propagators~(\ref{DeltaWdef})--(\ref{DeltaCdef}), so that we 
can drop them from the integrated propagators.
In this paper we will need only the dimensionally regulated 
coincidence limits of the~$AA$ integrated propagator 
and its first two derivatives,
\begin{align}
K_{AA}(x;x') \xrightarrow{x' \to x}{}&
	\frac{-k}{H^2a_x^2}
	\biggl[
	\ln(a_x) + \frac{ \Psi }{2} + \frac{2D\!-\!3}{2(D\!-\!2)}
	\biggr]
	\, ,
\label{KAAcoincidence}
\\
\partial_\mu^x K_{AA}(x;x') \xrightarrow{x' \to x}{}&
	\frac{k \delta_\mu^0}{Ha_x} \biggl[
	\ln(a_x) + \frac{\Psi \!-\! 1}{2} + \frac{2D \!-\! 3}{2(D \!-\! 2)}
	\biggr] \, ,
\label{dKAAcoincidence}
\\
\partial_\mu^x \partial_\nu^{x'}
	K_{AA}(x;x') \xrightarrow{x' \to x}{}&
	- k \biggl[
	\Bigl( \ln(a_x) + \frac{ \Psi \!+\! 1 }{2} \Bigr) \overline{\eta}_{\mu\nu} 
	- \frac{ (D\!-\!3) }{2(D\!-\!2)} \delta_\mu^0 \delta_\nu^0
	\biggr]
\label{ddKAAcoincidence}
	\, ,
\end{align}
and the same for the~$BB$ integrated propagator,
\begin{align}
K_{BB}(x;x') \xrightarrow{x' \to x}{}&
	\frac{-k}{(D\!-\!3) H^2a_x^2}
	\biggl[
	\ln(a_x) + \frac{ \Psi  }{2}- \frac{1}{2(D\!-\!2)(D\!-\!3)} 
	\biggr]
	\, ,
\label{KBBcoincidence}
\\
\partial_\mu^x K_{BB}(x;x') \xrightarrow{x' \to x}{}&
	\frac{k \delta_\mu^0}{(D\!-\!3)Ha_x} \biggl[
	\ln(a_x) + \frac{ \Psi \!-\! 1}{2} - \frac{1}{2(D\!-\!2)(D\!-\!3)}
	\biggr]
	\, ,
\label{dKBBcoincidence}
\\
\partial^x_\mu \partial^{x'}_\nu 
	K_{BB}(x;x') \xrightarrow{x' \to x}{}&
	\frac{-k}{(D\!-\!3)} \biggl[
	\Bigl( \ln(a_x) 
		\!+\! \frac{ \Psi \!-\! 3 }{2}
		\!+\! \frac{D \!-\! 4}{2(D\!-\!2) (D\!-\!3) }
		\Bigr) \delta_\mu^0\delta_\nu^0
	\!+\! \frac{D\!-\!3}{2(D\!-\!2)} \overline{\eta}_{\mu\nu}
	\biggr]
	\, .
\label{ddKBBcoincidence}
\end{align}
where we introduced a dimensionful constant
\begin{equation}
k = \frac{H^{D-2}}{(4\pi)^{\frac{D}{2}}}
	\frac{\Gamma(D\!-\!1)}{\Gamma\big( \frac{D}{2} \big)} \, .
\end{equation}
These limits are obtained from~(\ref{DeltaWdef})--(\ref{DeltaCdef}) and~(\ref{defKdiff}). We will only be 
using the logarithm parts of these coincidence limits, as only they are connected to the large logarithmic 
correction of the exchange potential that we are interested in. However, since the logarithm always comes 
together with the divergent constant~$\Psi$ in the same combination, our proof of the cancellation of gauge 
dependence for the $\Delta \alpha$ part also holds true for one-loop divergences.

\section{Reducing 4-point diagrams}
\label{sec: Reducing 4-point diagrams}

It is convenient to break up the computation of the 4-point diagrams in 
Fig.~\ref{4ptDiagrams} into two parts, one coming from the~$\Delta\alpha$ variation 
of the graviton propagator~(\ref{DeltaAlphaGraviton}) containing~$K_{AA}$, 
and the other part containing~$K_{BB}$; accordingly we refer to these parts 
as~$AA$ and~$BB$.

\subsection{Preliminary reduction of $AA$ parts}
\label{subsec: Preliminary reduction of AA parts}

In each of the diagrams from classes 0--7
we need to contract the tensor structure of the two vertices
(either $\overline{\rm B}$, C, or B) 
into the tensor structure of the~$AA$ part of the 
graviton propagator. There are three different contractions of a single vertex into 
the~$AA$ tensor structure. Contracting the C-vertex gives
\begin{align}
\MoveEqLeft[6]
a_z^{D-2} \Bigl[
	\overline{\partial}{}_z^\mu \partial_z^\nu
	-
	\tfrac{1}{2} \eta^{\mu\nu} \overline{\partial}_z \!\cdot\! \partial_z \Bigr]
	\!\times\!
	\Bigl[ - 4 \widetilde{\partial}{}^z_{(\mu} \overline{\eta}_{\nu)(\rho}
		\widetilde{\partial}{}^{z'}_{\sigma)} \Bigr]
	=
	-2 a_z^{D-2} \Bigl[
	\bigl( \overline{\partial}{}_z \!\cdot\! \widetilde{\partial}{}_z \bigr)
		\nabla^z_{(\rho}
		\widetilde{\partial}{}^{z'}_{\sigma)}
\nonumber \\
&
	+
	\bigl( \partial_z \!\cdot\! \widetilde{\partial}{}_z \bigr)
		\overline{\nabla}{}^z_{(\rho}
		\widetilde{\partial}{}^{z'}_{\sigma)} 
	-
	\bigl( \overline{\partial}_z \!\cdot\! \partial_z \bigr)
		\widetilde{\nabla}{}^z_{(\rho} \widetilde{\partial}{}^{z'}_{\sigma)}
	\Bigr]
\longrightarrow
	2D_z
		\overline{\nabla}{}^z_{(\rho}
		\widetilde{\partial}{}^{z'}_{\sigma)}
	+
	2\overline{D}_z
		\nabla^z_{(\rho}
		\widetilde{\partial}{}^{z'}_{\sigma)}
	\, .
\label{C-AA-contraction}
\end{align}
Here the first equality is obtained by just expanding out the product and contracting the indices.
Getting to the final expression required the use of the  second vertex identity~(\ref{VertexIdentity2}) for contracted derivatives, followed by the first vertex 
identity~(\ref{VertexIdentity1}) applied to spatial derivatives. 
Applying the same sequence
of operations we derive the contraction of the $\overline{\rm B}$-vertex into the
graviton~$AA$ tensor structure,
\begin{align}
\MoveEqLeft[6]
	\Bigl[
		a_z^{D-2} \overline{\partial}{}_z^\mu \partial_z^\nu
		- \tfrac{1}{4} \eta^{\mu\nu} \widetilde{\mathcal{D}}_z
		\Bigr]
	\!\times\!
	\Bigl[
	-
	4 \widetilde{\partial}{}^z_{(\mu} \overline{\eta}_{\nu)(\rho} 
		\widetilde{\partial}{}^{z'}_{\sigma)} 
	\Bigr]
	=
	-
	2 a_z^{D-2}
	\bigl( \overline{\partial}_z \!\cdot\! \widetilde{\partial}_z \bigr)
	\nabla^z_{(\rho} 
		\widetilde{\partial}{}^{z'}_{\sigma)} 
\nonumber \\
&
	-2 
	a_z^{D-2} \bigl( \partial_z \!\cdot\! \widetilde{\partial}_z \bigr)
	\overline{\nabla}{}^z_{(\rho} 
		\widetilde{\partial}{}^{z'}_{\sigma)} 
	+\widetilde{\mathcal{D}}_z
		\widetilde{\nabla}{}^z_{(\rho} 
		\widetilde{\partial}{}^{z'}_{\sigma)} 
\longrightarrow
	\bigl( \mathcal{D}_z \!-\! \overline{\mathcal{D}}_z \bigr)
	\bigl( 2\overline{\nabla}{}^z_{(\rho} \!+\! \widetilde{\nabla}{}^z_{(\rho} \bigr)
		\widetilde{\partial}{}^{z'}_{\sigma)} 
		\, .
\label{B-AA-contraction}
\end{align}
and also the contraction of the B-vertex,
\begin{align}
\MoveEqLeft[4]
a_z^{D-2}
	\Bigl[
		\overline{\partial}{}_z^{(\mu} \partial_z^{\nu)}
		-
		\tfrac{1}{2} \eta^{\mu\nu} 
			\bigl( \overline{\partial}_z \!\cdot\! \partial_z \!+\! a_z^2 m^2 \bigr) 
		\Bigr]
	\!\times\!
	\Bigl[ - 4 \widetilde{\partial}{}^z_{(\mu} \overline{\eta}_{\nu)(\rho}
		\widetilde{\partial}{}^{z'}_{\sigma)} \Bigr]
\nonumber \\
&
	=
	-2 a_z^{D-2}
	\Bigl[
		\big( \overline{\partial}{}_z \!\cdot\! \widetilde{\partial}{}_z \big)
		\nabla^z_{(\rho}
		\widetilde{\partial}{}^{z'}_{\sigma)}
		+
		\big( \partial_z \!\cdot\! \widetilde{\partial}{}_z \big)
		\overline{\nabla}{}^z_{(\rho}
		\widetilde{\partial}{}^{z'}_{\sigma)}
		-
		\bigl( \overline{\partial}_z \!\cdot\! \partial_z \!+\! a_z^2 m^2 \bigr) 
		\widetilde{\nabla}{}^z_{(\rho} \widetilde{\partial}{}^{z'}_{\sigma)}
		\Bigr]
\nonumber \\
&
\longrightarrow
	2 \big( \mathcal{D}_z \!-\! a_z^Dm^2 \big)
		\overline{\nabla}{}_{(\rho}^z \widetilde{\partial}{}_{\sigma)}^{z'}
	+
	2 \big( \overline{\mathcal{D}}_z \!-\! a_z^Dm^2 \big)
		\nabla{}^z_{(\rho} \widetilde{\partial}{}_{\sigma)}^{z'}
	\, .
\label{contractionAA3}
\end{align}
We give contractions of the second vertex as needed for each diagram class.

\paragraph{Diagram class 0.}
There are two C-vertices in the~$0a$ diagram~(\ref{defV0a}), which requires contracting 
the second C-vertex into~(\ref{C-AA-contraction}) to obtain the necessary contraction,
\begin{align}
\MoveEqLeft[5]
a_z^{D-2} \Bigl[
	\overline{\partial}{}_z^\mu \partial_z^\nu
	\!-\!
	\tfrac{1}{2} \eta^{\mu\nu} \overline{\partial}_z \!\cdot\! \partial_z \Bigr]
	\!\times\!
	\Bigl[ - 4 \widetilde{\partial}{}^z_{(\mu} \overline{\eta}_{\nu)(\rho}
		\widetilde{\partial}{}^{z'}_{\sigma)} \Bigr]
	\!\times\!
	a_{z'}^{D-2} 
	\Bigl[
		\overline{\partial}{}_{z'}^\rho \partial_{z'}^\sigma
		\!-\!
		\tfrac{1}{2} \eta^{\rho\sigma} \overline{\partial}_{z'} \!\cdot\! \partial_{z'}
		\Bigr]
\nonumber \\
={}&
	a_{z'}^{D-2} 
	\Bigl[
	\overline{\mathcal{D}}_z
		\bigl( \nabla_z \!\cdot\! \overline{\nabla}_{z'} \bigr)
		\bigl( \widetilde{\partial}{}_{z'} \!\cdot\! \partial_{z'} \bigr)
	+
	\overline{\mathcal{D}}_z
		\bigl( \nabla_z \!\cdot\! \nabla_{z'} \bigr)
		\bigl( \widetilde{\partial}{}_{z'} \!\cdot\! \overline{\partial}{}_{z'} \bigr)
		-
	\overline{\mathcal{D}}_z
		\bigl( \nabla_z \!\cdot\! \widetilde{\nabla}_{z'} \bigr)
		\bigl( \overline{\partial}_{z'} \!\cdot\! \partial_{z'} \bigr)
\nonumber \\
&
	+
	\mathcal{D}_z
		\bigl( \overline{\nabla}{}_z \!\cdot\! \overline{\nabla}_{z'} \bigr)
		\bigl( \widetilde{\partial}{}_{z'} \!\cdot\! \partial_{z'} \bigr)
	+
	\mathcal{D}_z
		\bigl( \overline{\nabla}{}_z \!\cdot\! \nabla_{z'} \bigr)
		\bigl( \widetilde{\partial}{}_{z'} \!\cdot\! \overline{\partial}{}_{z'} \bigr)
	-
	\mathcal{D}_z
		\bigl( \overline{\nabla}{}_z \!\cdot\! \widetilde{\nabla}{}_{z'} \bigr)
		\bigl( \overline{\partial}_{z'} \!\cdot\! \partial_{z'} \bigr)
		\Bigr]
\nonumber \\
\longrightarrow{}&
	-
	\overline{\mathcal{D}}_z \bigl( \nabla_z \!\cdot\! \nabla_{z'} \bigr) 
		\overline{\mathcal{D}}_{z'}
	-
	\overline{\mathcal{D}}_z \bigl( \nabla_z \!\cdot\! \overline{\nabla}_{z'} \bigr) 
		\mathcal{D}_{z'}
	-
	\mathcal{D}_z \bigl( \overline{\nabla}{}_z \!\cdot\! \nabla_{z'}  \bigr) 
		\overline{\mathcal{D}}_{z'}
	-
	\mathcal{D}_z \bigl( \overline{\nabla}{}_z \!\cdot\! \overline{\nabla}_{z'} \bigr) 
		\mathcal{D}_{z'}
        \, .
\end{align}
Just as for the first set of contractions, we first expanded out the product,
and then used the second vertex identity, followed by the first vertex identity
applied to spatial derivatives to obtain the final result.
Substituting this contraction into diagram~(\ref{defV0a}), using the equation of motion for 
the~$A$-type scalar propagator~(\ref{ScalarEOM}),
and unpacking the condensed notation produces
\begin{align}
\Bigl[ - i V_{0a} \Bigr]_{AA}^{\Delta\alpha}
	={}&
	(\kappa\lambda)^2
	\delta^D(x\!-\!y) \delta^D(x'\!-\!y')
	(a_xa_{x'})^D
	\biggl\{
	K_{AA}(x;x')
	\!\times\! \nabla_x \!\cdot\! \nabla_{x'} \, i \Delta_A(x;x')
\nonumber \\
&
	+
	\nabla_{y} \!\cdot\! 
		\Bigl[ K_{AA} (x;y) \!\times\! \nabla_{y} i \Delta_A(y;x') \Bigr]
	+
	\nabla_{y'} \!\cdot\! 
		\Bigl[ \nabla_{y'} i \Delta_A(x;y') \!\times\! K_{AA} (y';x') \Bigr]
\nonumber \\
&
	-
	i \int\! d^{D\!}z \, d^{D\!}z' \,
	\bigl[ \nabla_z^\mu i \Delta_A(x;z) \bigr]
	\left(\begin{matrix}
	K_{AA} (z;z')
	\\
	\mathcal{D}_z \delta^D(z\!-\!z')
	\end{matrix}\right)
	\bigl[ \nabla^{z'}_\mu i \Delta_A(z';x') \bigr]
	\biggr\}
	\, .
\label{V0aAAresult}
\end{align}

Diagram $0b$ in~(\ref{defV0b}) requires the contraction of the constant tensor 
structure~(\ref{ThetaStructure}) and the two derivatives in~(\ref{defV0b}) into 
the~$AA$ graviton tensor structure. Upon performing that contraction, and after using 
the fact that 
in the coincident limit, $z'\to z$, spatial derivatives acting on the graviton propagator 
can be reflected as  $\widetilde{\nabla}_{z} \!\leftrightarrow\! -\widetilde{\nabla}_{z'}$, 
we obtain
\begin{align}
\MoveEqLeft[2]
\Bigl[ - i V_{0b} \Bigr]_{AA}^{\Delta\alpha} =
	i (\kappa\lambda)^2 \delta^D(x\!-\!y) \delta^D(x'\!-\!y')
	(a_x a_{x'})^D
	\int\! d^{D\!}z \, a_z^{D-2} \,
	\partial^z_\mu i \Delta_A(x;z)
\label{AA0bRES}
\\
&
	\times\!
    \Bigl\{
    \Bigl[
    \tfrac{1}{2} (D\!-\!1) \eta^{\mu\nu}
    \partial{}_z
    \!\cdot\!
    \partial{}_{z'}
    -
    (D\!-\!1)
    \partial_z^\mu \partial_{z'}^\nu 
    -
    \overline{\eta}^{\mu\nu}
    \partial{}_z
    \!\cdot\!
    \partial{}_{z'}
    \Bigr]
        K_{AA}(z;z')
    \Bigr\}_{z' \to z}
    \!\times
    	\partial^z_\nu i \Delta_A(z;x')
    \, ,
\nonumber 
\end{align}
where the condensed notation has been unpacked.

\paragraph{Diagram class~$\boldsymbol{\overline{1}}$.}
Diagrams~$1b$ and~$1c$ in~(\ref{defV1b}) and~(\ref{defV1c})
require the contraction of the simple constant tensor structure in
the G-vertex into the tensor structure of the~$AA$ part of the graviton propagator,
\begin{equation}
\Bigl[
	\eta^{\mu(\rho} \eta^{\sigma)\nu}
	-
	\tfrac{1}{2} \eta^{\mu\nu} \eta^{\rho\sigma}
	\Bigr]
	\!\times\!
	\Bigl[ - 4 \widetilde{\partial}{}^z_{(\mu} \overline{\eta}_{\nu)(\rho}
		\widetilde{\partial}{}^{z'}_{\sigma)} \Bigr]
	=
	-
	2 (D\!-\!1) ( \widetilde{\partial}{}_z \!\cdot\! \widetilde{\partial}{}_{z'} )
	\, .
\end{equation}
Applying this to the two diagrams then yields
\begin{align}
\Bigl[ - i V_{1b} \Bigr]_{AA}^{\Delta\alpha} ={}&
	- \tfrac{1}{2} (D\!-\!1) (\kappa\lambda)^2
	\delta^D(x\!-\!y) \delta^D(x' \!-\! y')
	(a_x a_{x'})^D
	\big[ \partial_x \!\cdot\! \partial_{y} K_{AA}(x;y) \big]
	\!\times\!
	i \Delta_A(x;x') 
	\, ,
\label{AA1bRES}
\\
\Bigl[ - i V_{1c} \Bigr]_{AA}^{\Delta\alpha} ={}&
	- \tfrac{1}{2} (D\!-\!1) (\kappa\lambda)^2
	\delta^D(x\!-\!y) \delta^D(x' \!-\! y')
	(a_x a_{x'})^D
	i \Delta_A(x;x') 
	\!\times\!
	\big[ \partial_{x'} \!\cdot\! \partial_{y'} K_{AA}(x';y') \big]
	\, .
\label{AA1cRES}
\end{align}
%

\paragraph{Diagram class~$\boldsymbol{\overline{\overline{4}}}$.}
The tensor contractions necessary for the~$\overline{\overline{4}}$ diagrams are 
obtained by contracting another~$\overline{\rm B}$-vertex into~(\ref{B-AA-contraction}).
\begin{align}
\MoveEqLeft[5]
\Bigl[
	a_z^{D-2} \overline{\partial}{}_z^\mu \partial_z^\nu
		\!-\! \tfrac{1}{4} \eta^{\mu\nu} \widetilde{\mathcal{D}}_z
		\Bigr]
	\!\times\!
	\Bigl[
	-
	4 \widetilde{\partial}{}^z_{(\mu} \overline{\eta}_{\nu)(\rho} 
		\widetilde{\partial}{}^{z'}_{\sigma)} 
	\Bigr]
	\!\times\!
	\Bigl[
	a_{z'}^{D-2} \overline{\partial}{}_{z'}^\rho \partial_{z'}^\sigma
		\!-\! \tfrac{1}{4} \eta^{\rho\sigma} \widetilde{\mathcal{D}}_{z'}
		\Bigr]
\nonumber \\
={}&
\tfrac{1}{2} 
	a_{z'}^{D-2} 
	\bigl( \mathcal{D}_z \!-\! \overline{\mathcal{D}}_z \bigr)
	\bigl( 2\overline{\nabla}{}_z \!+\! \widetilde{\nabla}_z \bigr) 
		\!\cdot\! 
		\Bigl[
	\overline{\nabla}{}_{z'}
		\bigl( \widetilde{\partial}{}_{z'} \!\cdot\! \partial_{z'} \bigr)
	+
	\nabla_{z'}
		\bigl( \widetilde{\partial}{}_{z'} \!\cdot\! \overline{\partial}{}_{z'} \bigr)
	-
	\tfrac{1}{2} \widetilde{\nabla}{}_{z'}
		\widetilde{\mathcal{D}}_{z'}
		\Bigr]
\nonumber \\
\longrightarrow{}&
	-
	\tfrac{1}{4} 
	\bigl( \mathcal{D}_z \!-\! \overline{\mathcal{D}}_z \bigr)
	\bigl( 2\overline{\nabla}{}_z \!+\! \widetilde{\nabla}_z \bigr) 
		\!\cdot\! 
	\bigl( 2\overline{\nabla}{}_{z'} \!+\! \widetilde{\nabla}_{z'} \bigr) 
		\bigl( \mathcal{D}_{z'} \!-\! \mathcal{\overline{D}}_{z'} \bigr)
	\, .
\label{AAcontractionFor4}
\end{align}
We arrive at this result by applying the second vertex identity~(\ref{VertexIdentity2}), followed by the first 
vertex identity~(\ref{VertexIdentity1}) applied to the spatial derivatives. Substituting this contraction into 
the diagram~$\overline{\overline{4a}}$ in~(\ref{4barbara}) we have
\begin{align}
\Bigl[ - i V_{\overline{\overline{4a}}} \Bigr]_{AA}^{\Delta\alpha}
	={}&
	-
	\frac{(\kappa\lambda)^2}{4} (a_x a_{x'})^D
	\bigl( 2\overline{\nabla}{}_y \!+\! \widetilde{\nabla}_y \bigr) 
		\!\cdot\! 
	\bigl( 2\overline{\nabla}{}_{y'} \!+\! \widetilde{\nabla}_{y'} \bigr) 
\nonumber \\
&
	\times
	\bigl( \mathcal{D}_y \!-\! \overline{\mathcal{D}}_y \bigr)
	i \Delta_m(x;y)
	\left(\begin{matrix}
	K_{AA}(y;y')
	\\
	i \Delta_A(x;x')
	\end{matrix}\right)
	\bigl( \mathcal{D}_{y'} \!-\! \mathcal{\overline{D}}_{y'} \bigr)
	i \Delta_m(x';y')
	\, .
\end{align}
There are two favorable combinations of derivatives, each of which pinches
one massive propagator, owing to~$\overline{\mathcal{D}}$ producing~$a^Dm^2$
when acting on the external mode function according to~(\ref{tree-levelUeom}),
\begin{equation}
\bigl( \mathcal{D}_y \!-\! \overline{\mathcal{D}}_y \bigr)
	i \Delta_m(x;y)
	\longrightarrow
	\bigl( \mathcal{D}_y \!-\! a_y^D m^2 \bigr)
	i \Delta_m(x;y)
	=
	i \delta^D(x\!-\!y)
	\, .
\end{equation}
This reduces the diagram to the topology of the first diagram on the right-hand side
of~(\ref{4ReducedDiagrams}) by eliminating massive propagators from it,
\begin{align}
\Bigl[ - i V_{\overline{\overline{4a}}} \Bigr]_{AA}^{\Delta\alpha}
	={}&
	\frac{(\kappa\lambda)^2}{4} (a_x a_{x'})^D
	\delta^D(x\!-\!y)
	\delta^D(x'\!-\!y')
	\bigl( 2\overline{\nabla}{}_y \!+\! \widetilde{\nabla}_y \bigr) 
		\!\cdot\! 
	\bigl( 2\overline{\nabla}{}_{y'} \!+\! \widetilde{\nabla}_{y'} \bigr) 
	\left(\begin{matrix}
	K_{AA}(y;y')
	\\
	i \Delta_A(x;x')
	\end{matrix}\right)
	\, .
\end{align}
However, this form is still not amenable to be interpreted as the effective self-mass 
contribution on account of the derivatives acting on external mode functions.
These derivatives can be moved away from external mode functions,
and into the loop only 
after the~$\overline{\overline{4a}}$ contribution is combined with the tree remaining 
ones~$\overline{\overline{4b}}$,~$\overline{\overline{4c}}$, and~$\overline{\overline{4d}}$,
that are obtained by reflecting the coordinates,
\begin{align}
\Bigl[ - i V_{\overline{\overline{4b}}} \Bigr]_{AA}^{\Delta\alpha}
	\! ={}&
	\frac{ (\kappa\lambda)^2 }{4} 
	\delta^D(x\!-\!y)
	\delta^D(x'\!-\!y')
	(a_y a_{y'})^D
	\bigl( 2\overline{\nabla}{}_x \!+\! \widetilde{\nabla}_x \bigr) 
		\!\cdot\! 
	\bigl( 2\overline{\nabla}{}_{x'} \!+\! \widetilde{\nabla}_{x'} \bigr) 
	\!
	\begin{pmatrix}
	i \Delta_A(y;y')
	\\
	K_{AA}(x;x')
	\end{pmatrix}
	,
\\
\Bigl[ - i V_{\overline{\overline{4c}}} \Bigr]_{AA}^{\Delta\alpha}
	\! ={}&
	\frac{(\kappa\lambda)^2}{4} 
	\delta^D(x\!-\!y)
	(a_y a_{x'})^D
	\delta^D(x'\!-\!y')
	\bigl( 2\overline{\nabla}{}_x \!+\! \widetilde{\nabla}_x \bigr) 
		\!\cdot\! 
	\bigl( 2\overline{\nabla}{}_{y'} \!+\! \widetilde{\nabla}_{y'} \bigr) 
	\!
	\begin{pmatrix}
	i \Delta_A(y;x')
	\\
	K_{AA}(x;y')
	\end{pmatrix}
	\, ,
\\
\Bigl[ - i V_{\overline{\overline{4d}}} \Bigr]_{AA}^{\Delta\alpha}
	\! ={}&
	\frac{ (\kappa\lambda)^2 }{4} 
	\delta^D(x\!-\!y)
	\delta^D(x'\!-\!y')
	(a_x a_{y'})^D
	\bigl( 2\overline{\nabla}{}_y \!+\! \widetilde{\nabla}_y \bigr) 
		\!\cdot\! 
	\bigl( 2\overline{\nabla}{}_{x'} \!+\! \widetilde{\nabla}_{x'} \bigr) 
	\!
	\begin{pmatrix}
	K_{AA}(y;x')
	\\
	i\Delta_A(x;y')
	\end{pmatrix}
	\, .
\end{align}
Adding these contributions together, and making use of the
delta functions in vertices produces a relatively simple expression,
\begin{align}
\Bigl[ - i V_{\overline{\overline{4a+b+c+d}}} \Bigr]_{AA}^{\Delta\alpha}
	={}&
	(\kappa\lambda)^2
	\delta^D(x\!-\!y)
	\delta^D(x' \!-\! y')
	(a_x a_{x'})^D
\nonumber \\
&	\hspace{1.7cm}
	\times
	\bigl( \overline{\nabla}{}_y \!+\! \overline{\nabla}{}_x 
		\!+\! \widetilde{\nabla}_y \bigr) 
		\!\cdot\! 
	\bigl( \overline{\nabla}{}_{y'} \!+\! \overline{\nabla}{}_{x'} 
		\!+\! \widetilde{\nabla}_{y'} \bigr) 
	\begin{pmatrix}
	K_{AA}(y;y')
	\\
	i \Delta_A(x;x')
	\end{pmatrix}
	\, .
\end{align}
Upon closer examination, we may use the first vertex identity~(\ref{VertexIdentity1})
to move spatial derivatives onto the scalar propagator, thus obtaining the
final expression that, after unpacking the notation, reads
\begin{equation}
\Bigl[ - i V_{\overline{\overline{4a+b+c+d}}} \Bigr]_{AA}^{\Delta\alpha}
	=
	(\kappa\lambda)^2
	\delta^D(x\!-\!y)
	\delta^D(x' \!-\! y')
	(a_x a_{x'})^D
	\!\times\!
	K_{AA}(x;x')
	\Bigl[ \nabla_x \!\cdot\! \nabla_{x'} \, i \Delta_A(x;x') \Bigr]
	\, .
\label{V4a+b+c+dAAresult}
\end{equation}
%

\paragraph{Diagram class~$\boldsymbol{\overline{5}}$.}
For this class of diagrams we need to contract 
the~$\overline{\text{B}}$-vertex
into~(\ref{B-AA-contraction}), which is done following the same sequence of operations
as for the other contractions in this section,
\begin{align}
\MoveEqLeft[5]
\Bigl[
	a_z^{D-2} \overline{\partial}{}_z^\mu \partial_z^\nu
		\!-\! \tfrac{1}{4} \eta^{\mu\nu} \widetilde{\mathcal{D}}_z
		\Bigr]
	\!\times\!
	\Bigl[
	-
	4 \widetilde{\partial}{}^z_{(\mu} \overline{\eta}_{\nu)(\rho} 
		\widetilde{\partial}{}^{z'}_{\sigma)} 
	\Bigr]
	\!\times\!
	a_{z'}^{D-2} 
	\Bigl[
		\overline{\partial}{}_{z'}^\rho \partial_{z'}^\sigma
		\!-\!
		\tfrac{1}{2} \eta^{\rho\sigma} \overline{\partial}_{z'} \!\cdot\! \partial_{z'}
		\Bigr]
\nonumber \\
={}&
	\tfrac{ 1}{2}
	a_{z'}^{D-2}
	\bigl( \mathcal{D}_z \!-\! \overline{\mathcal{D}}_z \bigr)
	\bigl( 2\overline{\nabla}{}_z \!+\! \widetilde{\nabla}_z \bigr)
	\!\cdot\!
	\Bigl[
		\overline{\nabla}{}_{z'}
		\bigl( \widetilde{\partial}{}_{z'} \!\cdot\! \partial_{z'} \bigr)
		+
		\nabla_{z'}
		\bigl( \widetilde{\partial}{}_{z'} \!\cdot\! \overline{\partial}{}_{z'} \bigr)
		-
		\widetilde{\nabla}{}_{z'}
		\bigl( \overline{\partial}_{z'} \!\cdot\! \partial_{z'} \bigr)
		\Bigr]
\nonumber \\
\longrightarrow{}&
	- \frac{1}{2}
	\bigl( \mathcal{D}_z \!-\! \overline{\mathcal{D}}_z \bigr)
		\bigl( 2\overline{\nabla}{}_z \!+\! \widetilde{\nabla}_z \bigr)
		\!\cdot\!
	\Bigl(
		\nabla_{z'} \overline{\mathcal{D}}_{z'}
		+
		\overline{\nabla}{}_{z'} \mathcal{D}_{z'}
		\Bigr)
		\, .
    \label{AAcontractionFor5}
\end{align}
Applying this contraction to diagram~$\overline{5a}$ in~(\ref{defV5a}) gives
\begin{align}
\MoveEqLeft[1]
\Bigl[ - i V_{\overline{5a}} \Bigr]_{AA}^{\Delta\alpha}
	=
	- \frac{(\kappa\lambda)^2}{2}
	\delta^D(x'\!-\!y')
	(a_x a_{x'})^D
	\int\! d^{D\!}z' \,
	\bigl( \mathcal{D}_y \!-\! \overline{\mathcal{D}}_y \bigr) i \Delta_m(x;y)
\nonumber \\
&
	\times\!
	\bigl( 2\overline{\nabla}{}_y \!+\! \widetilde{\nabla}_y \bigr)
		\!\cdot\!
	\biggl[
	\left( \begin{matrix}
	K_{AA}(y;z')
	\\
	\nabla_{z'} i \Delta_A(x;z')
	\end{matrix} \right)
	\overline{\mathcal{D}}_{z'} i \Delta_A(z';x')
	+
	\left( \begin{matrix}
	K_{AA}(y;z')
	\\
	\mathcal{D}_{z'} i \Delta_A(x;z')
	\end{matrix} \right)
	\overline{\nabla}{}_{z'} \,
	i \Delta_A(z';x')
	\biggr]
	\, ,
\end{align}
where accents on derivatives still denote where they act, but where we have in addition 
distributed the derivatives to make the following steps more obvious.
Pinching the massive propagator by the favorable combination of derivatives,
and also an additional~$A$-type scalar propagator in each term then gives
\begin{align}
\Bigl[ - i V_{\overline{5a}} \Bigr]_{AA}^{\Delta\alpha}
	={}&
	\frac{(\kappa\lambda)^2}{2}
	\delta^D(x\!-\!y)
	\delta^D(x'\!-\!y')
	(a_x a_{x'})^D
\nonumber \\
&
	\times\!
	\bigl( 2\overline{\nabla}{}_y \!+\! \widetilde{\nabla}_y \bigr)
		\!\cdot\!
	\Bigl[
	K_{AA}(y;x')
	\nabla_{x'} i \Delta_A(x;x')
	+
	K_{AA}(y;x)
	\nabla_{x}
	i \Delta_A(x;x')
	\Bigr]
	\, .
\end{align}
Here the barred derivative acts on the corresponding external leg, while the tilded 
derivative acts on the~$AA$ integrated propagator descended from the graviton 
propagator. 

The~$\overline{5b}$ diagram is obtained by interchanging the  coordinate
arguments~$x\!\leftrightarrow\!y$. However, due to the delta function identifying the
points~$x$ and~$y$ the two diagrams only differ in the derivative acting on the external
leg, and combine together into
\begin{align}
\MoveEqLeft[4]
\Bigl[ - i V_{\overline{5a+b}} \Bigr]^{\Delta\alpha}_{AA} 
	=
	(\kappa\lambda)^2
	\delta^D(x\!-\!y)
	\delta^D(x'\!-\!y')
	(a_x a_{x'})^D
\nonumber \\
&
	\times\!
	\bigl( \overline{\nabla}{}_x \!+\! \overline{\nabla}{}_y \!+\! \widetilde{\nabla}_x \bigr)
		\!\cdot\!
	\Bigl[
	K_{AA}(x;x')
	\nabla_{x'} i \Delta_A(y;x')
	+
	K_{AA}(x;y)
	\nabla_{y}
	i \Delta_A(y;x')
	\Bigr]
	\, .
\end{align}
We can now apply the first vertex identity~(\ref{VertexIdentity1}) for
the spatial derivatives in the~$x\!=\!y$ vertex; for the first term in the brackets
this is effectively a 5-point vertex, while for the second term this is effectively a 4-point
vertex,
\begin{align}
\Bigl[ - i V_{\overline{5a+b}} \Bigr]_{AA}^{\Delta\alpha}
	={}&
	-
	(\kappa\lambda)^2
	\delta^D(x\!-\!y)
	\delta^D(x'\!-\!y')
	(a_x a_{x'})^D
\nonumber \\
&
	\times\!
	\Bigl\{ 
	K_{AA}(x;x')
	\nabla_x \!\cdot\! \nabla_{x'} i \Delta_A(x;x')
	+
	\nabla_y \!\cdot\!
	\Bigl[
	K_{AA}(x;y)
	\!\times\!
	\nabla_{y}
	i \Delta_A(y;x')
	\Bigr]
	\Bigr\}
	\, .
\label{V5a+bAAresult}
\end{align}
The remaining two diagrams~$\overline{5c}$ and~$\overline{5d}$ are obtained 
from this result by simultaneously exchanging coordinate 
arguments~$(x,y)\!\leftrightarrow\!(x',y')$,
\begin{align}
\MoveEqLeft[5]
\Bigl[ - i V_{\overline{5c+d}} \Bigr]_{AA}^{\Delta\alpha}
	=
	-
	(\kappa\lambda)^2
	\delta^D(x\!-\!y)
	\delta^D(x'\!-\!y')
	(a_x a_{x'})^D
\nonumber \\
&
	\times\!
	\Bigl\{ 
	K_{AA}(x;x')
	\nabla_x \!\cdot\! \nabla_{x'} \, i \Delta_A(x;x')
	+
	\nabla_{y'} \!\cdot\!
	\Bigl[
	\nabla_{y'} \,
	i \Delta_A(x;y')
	\!\times\!
	K_{AA}(y';x')
	\Bigr]
	\Bigr\}
	\, .
\label{V5c+dAAresult}
\end{align}
%

\paragraph{Diagram class 6.}
For the four diagrams in this class we need to take the trace of the 
contraction~(\ref{contractionAA3}),
\begin{align}
\MoveEqLeft[6]
a_z^{D-2}
	\Bigl[
		\overline{\partial}{}_z^{(\mu} \partial_z^{\nu)}
		-
		\tfrac{1}{2} \eta^{\mu\nu} 
			\bigl( \overline{\partial}_z \!\cdot\! \partial_z \!+\! a_z^2 m^2 \bigr) 
		\Bigr]
	\!\times\!
	\Bigl[ - 4 \widetilde{\partial}{}^z_{(\mu} \overline{\eta}_{\nu)(\rho}
		\widetilde{\partial}{}^{z'}_{\sigma)} \Bigr]
		\eta^{\rho\sigma}
\nonumber \\
&
\longrightarrow
	2 \big( \mathcal{D}_z \!-\! a_z^Dm^2 \big)
		( \overline{\nabla}{}_z \!\cdot\! \widetilde{\nabla}{}_{z'} )
	+
	2 \big( \overline{\mathcal{D}}_z \!-\! a_z^Dm^2 \big)
		( \nabla{}_z \!\cdot\! \widetilde{\nabla}{}_{z'} )
	\, .
\end{align}
Substituting this into the diagram~(\ref{defV6a}) gives
\begin{equation}
\Big[ - i V_{6a} \Big]_{AA}^{\Delta\alpha} =
	\tfrac{1}{2} (\kappa \lambda)^2
	\delta^D(x\!-\!y) \delta^D(x'\!-\!y')
	(a_xa_{x'})^D \,
	\overline{\nabla}{}_x \!\cdot\! \nabla_x K_{AA}(x;x)
	\!\times\!
	i \Delta_A(x;x')
	\, ,
\label{AA6aRES}
\end{equation}
where we used that
\begin{equation}
\delta^D(x\!-\!y)\partial^x_\mu K_{AA}(x;y) \!=\! \tfrac{1}{2} 
	\delta^D(x\!-\!y)\partial^x_\mu K_{AA}(x;x) 
	\, .
\label{CoincidentDerivative}
\end{equation}
After adding to~(\ref{AA6aRES}) the~$6b$ diagram, obtained by reflecting~$x\!\leftrightarrow\!y$, we can 
get rid of derivatives acting on external mode functions by integrating by parts,
\begin{equation}
\Big[ - i V_{6a+b} \Big]_{AA}^{\Delta\alpha} =
	- \tfrac{1}{2} (\kappa \lambda)^2
	\delta^D(x\!-\!y) \delta^D(x'\!-\!y')
	(a_xa_{x'})^D
	\nabla_x \!\cdot\! \Big[ \nabla_x K_{AA}(x;x) \!\times\! i \Delta_A(x;x') \Big]
	\, .
\label{6a+bRED}
\end{equation}
The sum of the two remaining diagrams is then obtained by reflecting pairs of 
coordinates~$(x,y) \!\leftrightarrow\!(x',y')$ in the expression above,
\begin{equation}
\Big[ - i V_{6c+d} \Big]_{AA}^{\Delta\alpha} =
	- \tfrac{1}{2} (\kappa \lambda)^2
	\delta^D(x\!-\!y) \delta^D(x'\!-\!y')
	(a_xa_{x'})^D
	\nabla_{x'} \!\cdot\! \Big[ i \Delta_A(x;x') \!\times\! \nabla_{x'} K_{AA}(x';x') \Big]
	\, .
\label{6c+dRED}
\end{equation}
%

\paragraph{Diagram class 7.}
The last diagram class from Fig.~\ref{4ptDiagrams} requires contracting another B-vertex
into~(\ref{contractionAA3}),
\begin{align}
\MoveEqLeft[2]
a_z^{D-2}
	\Bigl[
		\overline{\partial}{}_z^{(\mu} \partial_z^{\nu)}
		-
		\tfrac{1}{2} \eta^{\mu\nu} 
			\bigl( \overline{\partial}_z \!\cdot\! \partial_z \!+\! a_z^2 m^2 \bigr) 
		\Bigr]
	\!\times\!
	\Bigl[ - 4 \widetilde{\partial}{}^z_{(\mu} \overline{\eta}_{\nu)(\rho}
		\widetilde{\partial}{}^{z'}_{\sigma)} \Bigr]
	\!\times\!
	a_{z'}^{D-2}
	\Bigl[
		\overline{\partial}{}_{z'}^{(\rho} \partial_{z'}^{\sigma)}
		-
		\tfrac{1}{2} \eta^{\rho\sigma}
			\bigl( \overline{\partial}_{z'} \!\cdot\! \partial_{z'} \!+\! a_{z'}^2 m^2 \bigr) 
		\Bigr]
\nonumber \\
={}&
	a_{z'}^{D-2}
	\big( \mathcal{D}_z \!-\! a_z^Dm^2 \big)
		\Big[
		( \overline{\nabla}{}_z \!\cdot\! \overline{\nabla}{}_{z'})
		( \widetilde{\partial}{}_{z'} \!\cdot\! \partial_{z'} )
		+
		( \overline{\nabla}_z \!\cdot\! \nabla_{z'})
		( \widetilde{\partial}{}_{z'} \!\cdot\! \overline{\partial}{}_{z'} )
		-
		(\overline{\nabla}{}_z \!\cdot\! \widetilde{\nabla}{}_{z'})
		\big( \overline{\partial}{}_{z'} \!\cdot\! \partial_{z'} \!+\! a_{z'}^D m^2 \big)
		\Big]
\nonumber \\
&
	+
	a_{z'}^{D-2}
	\big( \overline{\mathcal{D}}_z \!-\! a_z^Dm^2 \big)
		\Big[
		( \nabla_z \!\cdot\! \overline{\nabla}{}_{z'} )
		( \widetilde{\partial}{}_{z'} \!\cdot\! \partial_{z'} )
		+
		( \nabla_z \!\cdot\! \nabla_{z'} )
		( \widetilde{\partial}{}_{z'} \!\cdot\! \overline{\partial}{}_{z'} )
		-
		( \overline{\nabla}{}_z \!\cdot\! \widetilde{\nabla}{}_{z'} )
		\big( \overline{\partial}{}_{z'} \!\cdot\! \partial_{z'} \!+\! a_{z'}^Dm^2 \big)
		\Big]
\nonumber \\
\longrightarrow{}&
	-
	\big( \mathcal{D}_z \!-\! a_z^Dm^2 \big)
		( \overline{\nabla}{}_z \!\cdot\! \overline{\nabla}{}_{z'})
		\big( \mathcal{D}_{z'} \!-\! a_{z'}^Dm^2 \big)
	-
	\big( \mathcal{D}_z \!-\! a_z^Dm^2 \big)
		( \overline{\nabla}_z \!\cdot\! \nabla_{z'})
		\big( \overline{\mathcal{D}}_{z'} - a_{z'}^Dm^2 \big)
\nonumber \\
&
	-
	\big( \overline{\mathcal{D}}_z \!-\! a_z^Dm^2 \big)
		( \nabla_z \!\cdot\! \overline{\nabla}{}_{z'} )
		\big( \mathcal{D}_{z'} \!-\! a_{z'}^Dm^2 \big)
	-
	\big( \overline{\mathcal{D}}_z \!-\! a_z^Dm^2 \big)
		( \nabla_z \!\cdot\! \nabla_{z'} )
		\big( \overline{\mathcal{D}}_{z'} \!-\! a_{z'}^Dm^2 \big)
	\, .
\label{7AAcontraction}
\end{align}
This result is reached by first expanding the expression and contracting all the indices.
This is followed by applying the second vertex identity~(\ref{VertexIdentity2})
to contracted spacetime derivatives, and then the first vertex 
identity~(\ref{VertexIdentity1}) to spatial derivatives.
Using this contraction for the diagram~$7a$ in~(\ref{defV7a}), and using equations
of motion for the external mode functions and the massive propagator, we get
\begin{equation}
\Big[ - i V_{7a} \Big]_{AA}^{\Delta\alpha} =
	(\kappa \lambda)^2
	\delta^D(x\!-\!y) \delta^D(x'\!-\!y')
	(a_xa_{x'})^D
	\overline{\nabla}{}_x \!\cdot\! \overline{\nabla}{}_y
	\!\times\!
	K_{AA}(x;x)
	i \Delta_A(x;x')
	\, .
\label{AA7aRED}
\end{equation}
Even though this result is reduced to the topology of the third diagram on the right-hand side 
of~(\ref{4ReducedDiagrams}), it is not in the form that can be interpreted directly as contributing to the 
effective self-mass. This is because the two contracted derivatives still act on external mode functions.
The same is true for the other diagram of this class that is obtained by reflecting pairs of coordinates,
\begin{equation}
\Big[ - i V_{7b} \Big]_{AA}^{\Delta\alpha} =
	(\kappa \lambda)^2
	\delta^D(x\!-\!y) \delta^D(x'\!-\!y')
	(a_xa_{x'})^D
	i \Delta_A(x;x')
	K_{AA}(x';x')
	\!\times\!
	\overline{\nabla}{}_{x'} \!\cdot\! \overline{\nabla}{}_{y'}
	\, .
\label{AA7bRED}
\end{equation}
Nevertheless, this will not be an obstacle in the end, as this contribution will be canceled
by a contribution coming from the external mode function corrections.

\paragraph{Cancellation of non-local contributions.}
Diagram~$0a$, and diagrams from classes~$\overline{\overline{4}}$ and~$\overline{5}$ 
contain non-local contributions, while all the other diagrams contain contributions of 
the same topology as the first diagram on the right-hand side of~(\ref{4ReducedDiagrams}). 
These contributions can rightfully be called non-local, compared to the contributions from 
the other diagrams, that take the form of local vertex corrections. It is therefore natural 
to consider them together. In fact, when taken together their non-local contributions all 
cancel,
\begin{align}
\MoveEqLeft[6]
\Bigl[ - i V_{0a} - i V_{\overline{\overline{4a+b+c+d}}} - i V_{\overline{5a+b+c+d}} 
	\Bigr]_{AA}^{\Delta\alpha}
	=
	-
	(\kappa\lambda)^2
	\delta^D(x\!-\!y) \delta^D(x'\!-\!y')
	(a_xa_{x'})^D
\nonumber \\
&
	\times i\!
	\int\! d^{D\!}z \, d^{D\!}z' \,
	\bigl[ \nabla_z^\mu i \Delta_A(x;z) \bigr]
	\left(\begin{matrix}
	K_{AA} (z;z')
	\\
	\mathcal{D}_z \delta^D(z\!-\!z')
	\end{matrix}\right)
	\bigl[ \nabla^{z'}_\mu i \Delta_A(z';x') \bigr]
	\, .
\label{cancel1}
\end{align}
The remaining contribution can be further reduced, but we first need to migrate derivatives
away from the delta function. We do this by first symmetrizing the derivatives acting on the
delta function,
\begin{equation}
\mathcal{D}_z \delta^D(z\!-\!z')
	=
	-
	\partial_z \! \cdot\! \partial_{z'} \bigl[ a_z^{D-2} \delta^D(z \!-\! z') \bigr]
	\, ,
\label{deltaSym}
\end{equation}
and then integrating by parts each derivative around its vertex. This frees up the 
delta function so that we can integrate over it,
\begin{align}
\MoveEqLeft[3]
\Bigl[ - i V_{0a} - i V_{\overline{\overline{4a+b+c+d}}} - i V_{\overline{5a+b+c+d}} 
	\Bigr]_{AA}^{\Delta\alpha}
	=
	(\kappa\lambda)^2
	\delta^D(x\!-\!y) \delta^D(x'\!-\!y')
	(a_xa_{x'})^D
\nonumber \\
&
	\times
	i\! 
	\int\! d^{D\!}z \, 
	a_z^{D-2}
	\biggl\{
	\tfrac{1}{2}
	\partial_z^\nu\nabla_z^\mu i \Delta_A(x;z)
    \!\times\!
	\partial^{z}_\nu \bigl[
	K_{AA} (z;z) \!\times\! \nabla^{z}_\mu i \Delta_A(z;x')
	\bigr]
\nonumber \\
&	\hspace{3cm}
	+
	\tfrac{1}{2}
	\partial_z^\nu \bigl[ 
	\nabla_z^\mu i \Delta_A(x;z) \!\times\!
	K_{AA} (z;z)
	\bigr]
    \!\times\!
    \partial^{z}_\nu \nabla^{z}_\mu i \Delta_A(z;x')
\nonumber \\
&	\hspace{3cm}
	+
	\bigl[ \nabla_z^\mu i \Delta_A(x;z) \bigr]
	\!\times\!
	\bigl[ \partial_z \! \cdot\! \partial_{z'} K_{AA} (z;z') \bigr]_{z\to z'}
	\!\times\!
	\bigl[ \nabla^{z}_\mu i \Delta_A(z;x') \bigr]
	\biggr\}
	\, ,
\label{AAnlIntermediate}
\end{align}
where we have used~(\ref{CoincidentDerivative}).
Now integrating by parts the derivative in the first two terms 
in braces of~(\ref{AAnlIntermediate}) produces the scalar
d'Alembertian~$\mathcal{D}$ that eliminates a single scalar leg in each term.
This finally produces,
\begin{align}
\MoveEqLeft[3]
\Bigl[ - i V_{0a} - i V_{\overline{\overline{4a+b+c+d}}} - i V_{\overline{5a+b+c+d}} 
	\Bigr]_{AA}^{\Delta\alpha}
	=
	(\kappa\lambda)^2
	\delta^D(x\!-\!y) \delta^D(x'\!-\!y')
	(a_xa_{x'})^D
\nonumber \\
&
	\times
	\biggl\{
	-
	\tfrac{1}{2}
	\nabla_x \!\cdot\!
	\bigl[
	K_{AA} (x;x) \!\times\! \nabla_{x} i \Delta_A(x;x')
	\bigr]
	-
	\tfrac{1}{2}
	\nabla_{x'} \!\cdot\!
	\bigl[
	\nabla_{x'} i \Delta_A(x;x') \!\times\! K_{AA} (x';x')
	\bigr]
\nonumber \\
&	\hspace{1cm}
	+
	i\! 
	\int\! d^{D\!}z \, 
	a_z^{D-2}
	\bigl[ \nabla_z^\mu i \Delta_A(x;z) \bigr]
	\!\times\!
	\bigl[ \partial_z \! \cdot\! \partial_{z'} K_{AA} (z;z') \bigr]_{z' \to z}
	\!\times\!
	\bigl[ \nabla^{z}_\mu i \Delta_A(z;x') \bigr]
	\biggr\}
	\, .
\label{AA045cancel}
\end{align}
%

\subsection{Preliminary reduction of $BB$ parts}
\label{subsec: Preliminary reduction of BB parts}

The tensor structure of the~$BB$ part of the graviton propagator factorizes,
and consequently so do the contractions needed for the diagrams of 
Fig.~\ref{4ptDiagrams}. There are three contractions we need, the first one with the 
C-vertex,
\begin{align}
\MoveEqLeft[5]
a_z^{D-2} \Bigl[
	\overline{\partial}{}_z^\mu \partial_z^\nu
	\!-\!
	\tfrac{1}{2} \eta^{\mu\nu} \overline{\partial}_z \!\cdot\! \partial_z
	\Bigr]
	\!\times\!
	2\Bigl[ \delta^0_{(\mu} \widetilde{\partial}{}^z_{\nu)} \!-\! \eta_{\mu\nu} Ha_z \Bigr]
\nonumber \\
={}&
	a_z^{D-2} \Bigl[
	-
	\bigl( \partial_z \!\cdot\! \widetilde{\partial}{}_z \bigr) \overline{\partial}{}^z_0
	-
	\bigl( \overline{\partial}{}_z \!\cdot\! \widetilde{\partial}{}_z \bigr) \partial^z_0
	+
	\bigl( \overline{\partial}_z \!\cdot\! \partial_z \bigr)
	\Bigl( \widetilde{\partial}{}^z_0 \!+\! (D\!-\!2) Ha_z \Bigr)
	\Bigr]
\nonumber \\
\longrightarrow{}&
	a_z^{D-2} \Bigl[
	-
	\partial_z \!\cdot\! \bigl( \widetilde{\partial}{}_z \!+\! \overline{\partial}_z \bigr) 
		\overline{\partial}{}^z_0
	-
	\overline{\partial}{}_z \!\cdot\! 
	\bigl( \widetilde{\partial}{}_z \!+\! \partial_z \bigr) \partial^z_0
	\Bigr]
\longrightarrow
	\mathcal{D}_z \overline{\partial}{}^z_0
	+
	\overline{\mathcal{D}}_z  \partial^z_0
	\, .
\label{BBfirstContraction}
\end{align}
Here we first expanded the expression and contracted indices, followed by partially
integrating the uncontracted derivative in the last term using the first vertex
identity~(\ref{VertexIdentity1}), to then apply the second vertex 
identity~(\ref{VertexIdentity2}) to the contracted derivatives.
The second contraction we need is with the~$\overline{\rm B}$-vertex,
\begin{align}
\MoveEqLeft[5]
\Bigl[
	a_z^{D-2} \overline{\partial}{}_z^\mu \partial_z^\nu
	\!-\! \tfrac{1}{4} \eta^{\mu\nu} \widetilde{\mathcal{D}}_z
		\Bigr]
	\!\times\!
	2\Bigl[ \delta^0_{(\mu} \widetilde{\partial}{}^z_{\nu)} \!-\! \eta_{\mu\nu} Ha_z \Bigr]
\nonumber \\
={}&
	-
	a_z^{D-2}
	\Bigl[
	\bigl( \partial_z \!\cdot\! \widetilde{\partial}{}_z \bigr)
	\overline{\partial}{}^z_0 
	+
	\bigl( \overline{\partial}{}_z \!\cdot\! \widetilde{\partial}{}_z \bigr)
	\partial^z_0
	+
	2Ha_z \bigl( \overline{\partial}{}_z \!\cdot\! \partial_z\bigr)
	\Bigr]
	+
	\tfrac{1}{2} \widetilde{\mathcal{D}}_z
	\bigl( \widetilde{\partial}{}^z_{0} \!+\! D Ha_z \bigr)
\nonumber \\
\longrightarrow{}&
	\tfrac{1}{2} \bigl( \mathcal{D}_z \!-\! \overline{\mathcal{D}}_z \bigr)
		\bigl( \overline{\partial}{}^z_0 \!-\! \partial^z_0 \bigr)
	+
	H a_z \bigl( \mathcal{D}_z \!+\! \overline{\mathcal{D}}_z \bigr)
	+
	\tfrac{1}{2} \widetilde{\mathcal{D}}_z
	\Bigl[ \widetilde{\partial}{}^z_{0} \!+\! 
		\overline{\partial}{}^z_0 \!+\! \partial^z_0 \!+\! (D\!-\!2) Ha_z \Bigr]
\nonumber \\
\longrightarrow{}&
	\tfrac{1}{2} \Bigl[ \overline{\partial}{}^z_0 \!-\! \partial_0^z \!+\! DH a_z \Bigr]
		\bigl( \mathcal{D}_z \!-\! \overline{\mathcal{D}}_z \bigr)
	+
	D Ha_z \overline{\mathcal{D}}_z
	+
	\tfrac{1}{2} 
	\bigl( \widetilde{\partial}^z_0 \!+\! \overline{\partial}{}^z_0 \!+\! \partial^z_0 \bigr)
	\widetilde{\mathcal{D}}_z
\nonumber \\
\longrightarrow{}&
	\tfrac{1}{2} \Bigl[ 2 \overline{\partial}{}^z_0
			\!+\! \widetilde{\partial}_0^z \!+\! DH a_z \Bigr]
		\bigl( \mathcal{D}_z \!-\! \overline{\mathcal{D}}_z \bigr)
	+
	D Ha_z \overline{\mathcal{D}}_z
	\, .
\label{BBsecondContraction}
\end{align}
Deriving the final expression first required explicitly contracting derivatives,
and then applying the second vertex identity~(\ref{VertexIdentity2}) to the
contracted derivatives. Getting to the penultimate line required the use of commutation
identities~(\ref{CommutationIdentities}), followed by integrating
derivatives by parts using the first vector identity~(\ref{VertexIdentity1}) to get to the 
last expression.
Lastly, we need the contraction with the B-vertex, that is obtained by the
same sequence of steps,
\begin{align}
\MoveEqLeft[5]
a_z^{D-2}
	\Bigl[
		\overline{\partial}{}_z^{(\mu} \partial_z^{\nu)}
		-
		\tfrac{1}{2} \eta^{\mu\nu} 
			\bigl( \overline{\partial}_z \!\cdot\! \partial_z \!+\! a_z^2 m^2 \bigr) 
		\Bigr]
	\!\times\!
	2\Bigl[ \delta^0_{(\mu} \widetilde{\partial}{}^z_{\nu)} \!-\! \eta_{\mu\nu} Ha_z \Bigr]
\nonumber \\
={}&
	a_z^{D-2} \Big[
		( \overline{\partial}_z \!\cdot\! \partial_z )
		\big( \widetilde{\partial}_z^0 + (D\!-\!2) Ha_z \big)
	-
	(\partial_z \!\cdot\! \widetilde{\partial}_z) \overline{\partial}_0^z
	-
	( \overline{\partial}_z \!\cdot\! \widetilde{\partial}_z ) \partial_0^z
	+
	a_z^2m^2 \big( \widetilde{\partial}{}_0^z + DHa_z \big)
	\Big]
\nonumber \\
\longrightarrow{}&
	\tfrac{1}{2} \widetilde{\mathcal{D}}_z
		\big( \widetilde{\partial}{}_0^z + \overline{\partial}{}_0^z + \partial{}_0^z \big)
	-
	\tfrac{1}{2} \overline{\mathcal{D}}_z
		\big( \widetilde{\partial}{}_0^z + \overline{\partial}{}_0^z - \partial{}_0^z \big)
	-
	\tfrac{1}{2} \mathcal{D}_z
		\big( \widetilde{\partial}{}_0^z - \overline{\partial}{}_0^z + \partial{}_0^z \big)
\nonumber \\
&
	+
	\tfrac{1}{2} (D\!-\!2) H a_z \big( \widetilde{\mathcal{D}}_z
		- \overline{\mathcal{D}}_z - \mathcal{D}_z \big)
	-
	(D\!-\!2) H^2 a_z^D \widetilde{\partial}{}_0^z
	-
	m^2 a_z^{D} \big( \widetilde{\partial}{}_0^z + DHa_z \big)
\nonumber \\
\longrightarrow{}&
	\big( \mathcal{D}_z \!-\! a_z^Dm^2 \big) \overline{\partial}{}_0^z
	+
	\big( \overline{\mathcal{D}}_z \!-\! a_z^Dm^2 \big) \partial_0^z
	\, .
\label{BBthirdContraction}
\end{align}
%

\paragraph{Diagram class 0.}

The diagram in~(\ref{defV0a}) requires two contractions in~(\ref{BBfirstContraction}),
one for each vertex. The operators in these contractions pinch the~$A$-type
propagators they act on, reducing the diagram to
\begin{align}
\Bigl[ - i V_{0a} \Bigr]_{BB}^{\Delta\alpha}
={}&
	(\kappa\lambda)^2
	\delta^D(x\!-\!y) \delta^D(x'\!-\!y')
	(a_xa_{x'})^D
	\biggl\{
	-
	K_{BB}(x;x') \!\times\! \partial^x_0 \partial^{x'}_0 i \Delta_A(x;x')
\nonumber \\
&
	-
	\partial^y_0 \Bigl[
	K_{BB}(x;y)
	\!\times\!
	\partial^{y}_0 i \Delta_A(y;x')
	\Bigr]
	-
	\partial^{y'}_0 
	\Bigl[ \partial^{y'}_0 i \Delta_A(x;y')
	\!\times\!
	K_{BB}(y';x')
	\Bigr]
\nonumber \\
&
	+
	i\!
	\int\! d^{D\!}z \, d^{D\!}z' \,
	\bigl[ \partial^z_0 i \Delta_A(x;z) \bigr]
	\left(\begin{matrix}
	K_{BB}(z;z')
	\\
	\mathcal{D}_z
	\delta^D(z\!-\!z')
	\end{matrix}\right)
	\bigl[ \partial^{z'}_0 i \Delta_A(z';x') \bigr]
	\biggr\}
	\, .
\label{V0aBBresult}
\end{align}
For the $0b$ diagram in~(\ref{defV0b}) we perform all the contractions, and 
use that any odd number of spatial derivatives acting on the graviton taken at coincidence
vanishes,
\begin{align}
\MoveEqLeft[2]
\Bigl[ - i V_{0b} \Bigr]_{BB}^{\Delta\alpha}
    =
	(\kappa\lambda)^2 \delta^D(x\!-\!y) \delta^D(x'\!-\!y')
	(a_x a_{x'})^D
	\times i\!\int\! d^{D\!}z \,  a_z^{D-2} \,
\nonumber \\
&
    \times\!
    \bigl[\partial^z_\mu i \Delta_A(x;z) \bigr]
    \!\times\!
	\Bigl\{
    \Bigl[
    \tfrac{1}{2}
    \eta^{\mu\nu}
    \bigl( \partial{}_z \!\cdot\! \partial{}_{z'}
    + 2 (D\!-\!4) H a_z \partial{}_0^z
    +
    (D\!-\!2)(D\!-\!4) H^2 a_z^2
    \bigr)
\nonumber \\
&
    +
    \delta^\mu_0\delta^\nu_0
    \bigl(
    \partial{}_z \!\cdot\! \partial{}_{z'}
    +
    2(D\!-\!4) Ha_z \partial{}_0^z
    \bigr)
    -
    \partial{}_z^\mu \partial{}_{z'}^\nu
    \Bigr]
	K_{BB}(z;z') 
    \Bigr\}_{z' \to z}
    \!\times\!
	\bigl[\partial_\nu^z i \Delta_A(z;x') \bigr]
    \, .
\label{BB0bRES}
\end{align}
%

\paragraph{Diagram class~$\boldsymbol{\overline{1}}$.}
For the two diagrams in this class we need the simple contraction
\begin{align}
\MoveEqLeft[7]
\Bigl[
	\eta^{\mu(\rho} \eta^{\sigma)\nu}
	-
	\tfrac{1}{2} \eta^{\mu\nu} \eta^{\rho\sigma}
	\Bigr]
	\!\times\!
	4 \Bigl[ \delta^0_{(\mu} \widetilde{\partial}{}^z_{\nu)} 
		- \eta_{\mu\nu} Ha_z \Bigr]
	\Bigl[ \delta^0_{(\rho} \widetilde{\partial}{}^{z'}_{\sigma)}
		- \eta_{\rho\sigma} Ha_{z'} \Bigr]
\nonumber \\
={}&
	-
	2 ( \widetilde{\partial}{}_z \!\cdot\! \widetilde{\partial}{}_{z'} )
	-
	2 (D\!-\!2) \Big[ Ha_z \widetilde{\partial}{}_0^{z'}
		+ Ha_{z'} \widetilde{\partial}{}_0^z
		+ D H^2a_za_{z'} \Big]
	\, .
\end{align}
Applied to diagrams in~(\ref{defV1b}) and~(\ref{defV1c}) this produces
\begin{align}
\MoveEqLeft[4]
\Big[ - i V_{1b} \Big]_{BB}^{\Delta\alpha} =
	- \tfrac{1}{2} (\kappa\lambda)^2
	\delta^D(x\!-\!y) \delta^D(x' \!-\! y')
	(a_x a_{x'})^D
\label{BB1bRES}
\\
&
	\times\!
	\Big\{\Big[ \partial_x \!\cdot\! \partial_{y}
	+ (D\!-\!2) \big( Ha_x \partial_0^y + Ha_y \partial_0^x + D H^2a_xa_y \big) \Big]
	K_{BB}(x;y) \Big\}
	\!\times\!
	i \Delta_A(x;x') 
	\, ,
\nonumber 
\\
\MoveEqLeft[4]
\Big[ - i V_{1c} \Big]_{BB}^{\Delta\alpha} =
	-
	\tfrac{1}{2} (\kappa\lambda)^2
	\delta^D(x\!-\!y) \delta^D(x' \!-\! y')
	(a_x a_{x'})^D
\label{BB1cRES}
\\
&
	\times\!
	i \Delta_A(x;x') 
	\!\times\!
	\Big\{\Big[ \partial_{x'} \!\cdot\! \partial_{y'}
		+ (D\!-\!2) \big( Ha_{x'} \partial_0^{y'}
		+ Ha_{y'} \partial_0^{x'} + D H^2a_{x'} a_{y'} \big) \Big]
	K_{BB}(x';y') \Big\}
	\, .
\nonumber 
\end{align}
%

\paragraph{Diagram class~$\boldsymbol{\overline{\overline{4}}}$.}
The~$\overline{\overline{4a}}$ diagram in~(\ref{4barbara})
takes two contractions in~(\ref{BBsecondContraction}),
\begin{align}
\Bigl[ - i V_{\overline{\overline{4a}}} \Bigr]_{BB}^{\Delta\alpha}
	={}&
	-
	\tfrac{1}{4} 
	(\kappa\lambda)^2
	\delta^D(x\!-\!y)
	\delta^D(x' \!-\! y')
	(a_x a_{x'})^D
\nonumber \\
&
	\times
	\Bigl(
		2 \overline{\partial}{}^y_0 
			\!+\! \widetilde{\partial}_0^y \!+\! DH a_y
		\Bigr)
	\Bigl( 2 \overline{\partial}{}^{y'}_0 
			\!+\! \widetilde{\partial}_0^{y'} \!+\! DH a_{y'} \Bigr)
	\left(\begin{matrix}
	K_{BB}(y;y')
	\\
	i \Delta_A(x;x')
	\end{matrix}\right)
	\, ,
\end{align}
By combining this with the remaining three diagrams~(\ref{4barbarb})--(\ref{4barbard})
the derivatives in vertices combine neatly, and can be integrated by parts away from the 
external mode functions,
\begin{align}
\MoveEqLeft[2]
\Bigl[ - i V_{\overline{\overline{4a+b+c+d}}} \Bigr]_{BB}^{\Delta\alpha}
	=
	-
	(\kappa\lambda)^2
	\delta^D(x\!-\!y)
	\delta^D(x' \!-\! y')
	(a_x a_{x'})^D
	K_{BB}(x;x')
	\!\times\!
	\partial_0^{x} \partial_0^{x'} i \Delta_A(x;x')
	\, .
\label{V4abcdBB}
\end{align}
%

\paragraph{Diagram class~$\boldsymbol{\overline{5}}$.}
Diagram~$\overline{5a}$ from~(\ref{defV5bara}) requires the contraction 
in~(\ref{BBsecondContraction})
for the~$y$-vertex and contraction~(\ref{BBfirstContraction}) for the~$z$-vertex,
\begin{align}
\Bigl[ - i V_{\overline{5a}} \Bigr]_{BB}^{\Delta\alpha}
	={}&
	-
	\tfrac{1}{2} 
	(\kappa\lambda)^2
	\delta^D(x\!-\!y)
	\delta^D(x'\!-\!y')
	(a_x a_{x'})^D
\nonumber \\
&
	\times\!
	\Bigl( 2 \overline{\partial}{}^y_0
			\!+\! \widetilde{\partial}_0^y \!+\! DH a_y \Bigr)
	\biggl[ 
	\left(\begin{matrix}
	K_{BB}(y;x')
	\\
	\partial^{x'}_0 i \Delta_A(x;x')
	\end{matrix} \right)
	+
	\left(\begin{matrix}
	K_{BB}(y;x)
	\\
	\partial^{x}_0 i \Delta_A(x;x')
	\end{matrix}\right)
	\biggr]
	\, .
\end{align}
Adding to it the~$\overline{5b}$ contribution, obtained by reflecting coordinates,
the derivatives in the second line combine
into a vertex identity, and are integrated by parts to obtain
\begin{align}
\Bigl[ - i V_{\overline{5a+b}} \Bigr]_{BB}^{\Delta\alpha}
	={}&
	(\kappa\lambda)^2
	\delta^D(x\!-\!y)
	\delta^D(x'\!-\!y')
	(a_x a_{x'})^D
\nonumber \\
&
	\times\!
	\Bigl\{
	K_{BB}(x;x') \!\times\! \partial_0^x\partial^{x'}_0 i \Delta_A(x;x')
	+
	\partial_0^y \Bigl[ K_{BB}(x;y) \!\times\! \partial^{y}_0 i \Delta_A(y;x') \Bigr]
	\Bigr\}
	\, .
\label{V5abBB}
\end{align}
The two remaining contributions~$\overline{5c}$ and~$\overline{5d}$ are obtained
by interchanging pairs of coordinates,
\begin{align}
\Bigl[ - i V_{\overline{5c+d}} \Bigr]_{BB}^{\Delta\alpha}
	={}&
	(\kappa\lambda)^2
	\delta^D(x\!-\!y)
	\delta^D(x'\!-\!y')
	(a_x a_{x'})^D
\nonumber \\
&
	\times\!
	\Bigl\{
	K_{BB}(x;x') \!\times\! \partial^{x}_0 \partial_0^{x'} i \Delta_A(x;x')
	+
	\partial_0^{y'}
	\Bigl[
	K_{BB}(x';y') \!\times\! \partial^{y'}_0 i \Delta_A(x;y')
	\Bigr]
	\Bigr\}
	\, .
\label{V5cdBB}
\end{align}
%

\paragraph{Diagram class 6.}
For this class of diagrams, given in~(\ref{defV6a})--(\ref{defV6d}),
we need the contraction in~(\ref{BBthirdContraction}), and also
to take the trace of one of the tensor structures in the~$BB$ part of graviton 
propagator,
\begin{align}
\MoveEqLeft[3]
\Big[ - i V_{6a} \Big]_{BB}^{\Delta\alpha} =
	i (\kappa \lambda)^2
	\delta^D(x'\!-\!y')
	(a_xa_{x'})^D
\nonumber \\
&
	\times\!
	\Bigl[
	\big( \mathcal{D}_y \!-\! a_y^Dm^2 \big) \overline{\partial}{}_0^y
	+
	\big( \overline{\mathcal{D}}_y \!-\! a_y^Dm^2 \big) \partial_0^y
	\Big]
	\big( \widetilde{\partial}{}_0^x \!+\! D Ha_x \big)
	\begin{pmatrix}
	K_{BB}(x;y)
	\\
	i \Delta_m(x;y) 
	\end{pmatrix}
	i \Delta_A(x;x')
	\, .
\end{align}
Applying equations of motion we have
\begin{equation}
\Big[ - i V_{6a} \Big]_{BB}^{\Delta\alpha} =
	-
	\tfrac{1}{2}
	(\kappa \lambda)^2
	\delta^D(x\!-\!y) \delta^D(x'\!-\!y')
	(a_xa_{x'})^D
	\overline{\partial}{}_0^y
	\Big[\big( \partial_0^x \!+\! 2D Ha_{x} \big) K_{BB}(x;x) \Big]
	i \Delta_A(x;x')
	\, ,
\end{equation}
where we used identity~(\ref{CoincidentDerivative}) applied to~$K_{BB}$.
Adding to this the~$6b$ diagram, we can move the temporal derivatives inside,
away from the external mode functions,
\begin{align}
\Big[ - i V_{6a+b} \Big]_{BB}^{\Delta\alpha}
	={}&
	\tfrac{1}{2} (\kappa \lambda)^2
	\delta^D(x\!-\!y) \delta^D(x'\!-\!y')
	(a_xa_{x'})^D
\nonumber \\
&	\hspace{1cm}
	\times\!
	\big( \partial_0^x \!+\! DHa_x \big)
	\Big[ \big( \partial_0^x \!+\! 2D Ha_{x} \big) K_{BB}(x;x)
	\!\times\!
	i \Delta_A(x;x')
	\Big]
	\, .
\label{BB6a+bRES}
\end{align}
The two remaining contributions are inferred by reflecting coordinates,
\begin{align}
\Big[ - i V_{6c+d} \Big]_{BB}^{\Delta\alpha}
	={}&
	\tfrac{1}{2} (\kappa \lambda)^2
	\delta^D(x\!-\!y) \delta^D(x'\!-\!y')
	(a_xa_{x'})^D
\nonumber \\
&	\hspace{1cm}
	\times\!
	\big( \partial_0^{x'} \!+\! DHa_{x'} \big)
	\Big[
	i \Delta_A(x;x')
	\!\times\!
	\big( \partial_0^{x'} \!+\! 2D Ha_{x'} \big) K_{BB}(x';x')
	\Big]
	\, .
\label{BB6c+dRES}
\end{align}
%

\paragraph{Diagram class 7.}
Here we just need two contractions~(\ref{BBthirdContraction})
applied to~(\ref{defV7a}) and~(\ref{defV7b})
that immediately yield
\begin{align}
\Big[ - i V_{7a} \Big]_{BB}^{\Delta\alpha}
	={}&
	-
	(\kappa \lambda)^2
	\delta^D(x\!-\!y) \delta^D(x'\!-\!y')
	(a_xa_{x'})^D
		\overline{\partial}{}_0^x
		 \overline{\partial}{}_0^y
	\!\times\!
	K_{BB} (x;x)
	i\Delta_A(x;x')
	\, ,
\label{BB7aRED}
\\
\Big[ - i V_{7b} \Big]_{BB}^{\Delta\alpha}
	={}&
	-
	(\kappa \lambda)^2
	\delta^D(x\!-\!y) \delta^D(x'\!-\!y')
	(a_xa_{x'})^D
	i \Delta_A(x;x')
	K_{BB} (x';x')
	\!\times\!
		\overline{\partial}{}_0^{x'}
		 \overline{\partial}{}_0^{y'}
	\, .
\label{BB7bRED}
\end{align}
%

\paragraph{Cancellation of non-local contributions.}
Just as for the~$AA$ part, it is natural to group together 
contributions~(\ref{V0aBBresult}),~(\ref{V4abcdBB}),~(\ref{V5abBB}) and~(\ref{V5cdBB}),
as the non-local contributions cancel, and we are left with
\begin{align}
\MoveEqLeft[7]
\Bigl[ - i V_{0a} - i V_{\overline{\overline{4a+b+c+d}}} 
	- i V_{\overline{5a+b+c+d}} \Bigr]_{BB}^{\Delta\alpha}
\nonumber \\
={}&
	i
	\int\! d^{D\!}z \, d^{D\!}z' \,
	\bigl[ \partial^z_0 i \Delta_A(x;z) \bigr]
	\left(\begin{matrix}
	K_{BB}(z;z')
	\\
	\mathcal{D}_z
	\delta^D(z\!-\!z')
	\end{matrix}\right)
	\bigl[ \partial^{z'}_0 i \Delta_A(z';x') \bigr]
	\, .
\end{align}
Applying here the same sequence of operations as we applied to Eq.~(\ref{cancel1}) 
at end of the~$AA$ section, we get the remaining term to reduce to
\begin{align}
\MoveEqLeft[1]
\Bigl[ - i V_{0a} - i V_{\overline{\overline{4a+b+c+d}}} 
	- i V_{\overline{5a+b+c+d}} \Bigr]_{BB}^{\Delta\alpha}
	=
	(\kappa\lambda)^2
	\delta^D(x\!-\!y) \delta^D(x'\!-\!y')
	(a_xa_{x'})^D
\nonumber \\
&	\hspace{0.3cm}
	\times\!
	\biggl\{
	-
	i\! 
	\int\! d^{D\!}z \, 
	a_z^{D-2}
	\bigl[ \partial_0^z i \Delta_A(x;z) \bigr]
	\!\!\times\!\!
	\Bigl[ \bigl( \partial_z \! \cdot\! \partial_{z'} \!-\! (D\!-\!2)H^2 a_z^2 \bigr) K_{BB} (z;z') \Bigr]_{z\to z'}
	\!\!\times\!\!
	\bigl[ \partial_0^{z} i \Delta_A(z;x') \bigr]
\nonumber \\
&	\hspace{1.5cm}
	+
	\tfrac{1}{2}
	K_{BB} (x;x) \!\times\! 
	\bigl[ \partial_0^x \!+\! (D\!-\!2) H a_x \bigr]
	\partial_0^x i \Delta_A(x;x')
\nonumber \\
&	\hspace{3cm}
	+
	\tfrac{1}{2}
	\bigl[ \partial_0^{x'} \! \!+\! (D\!-\!2) H a_{x'} \bigr]
	\partial_0^{x'} \! i \Delta_A(x;x')
    \!\times\!
    K_{BB} (x';x')
\nonumber \\
&	\hspace{1.5cm}
	+ \tfrac{1}{2} \partial_0^x K_{BB}(x;x) \times \partial_0^x i \Delta_A(x;x')
	+ \tfrac{1}{2} \partial_0^{x'} K_{BB}(x';x') \times \partial_0^{x'} i \Delta_A(x;x')
	\biggr\}
	\, .
\label{BB0a45cancel}
\end{align}
%

\subsection{Final reduction}
\label{subsec: Final reduction 1}

Having exploited derivative interactions to reduce the~$AA$ and~$BB$ parts
of diagram classes 0--7, we managed to get all contributions to the form that require
only the coincidence limits of integrated propagators~$K_{AA}$ and~$K_{BB}$ and their
first two derivatives, that are given in~(\ref{KAAcoincidence})--(\ref{ddKBBcoincidence}).
Now we substitute for these, keeping only the relevant logarithm
terms, and bring the contributions from both parts together.
All the final contributions are summarized in the first three rows of 
Table~\ref{FinalTable} in the concluding section.

\paragraph{Diagram classes 0--5.}
It is convenient to consider the first six diagram classes together. 
The total contribution from the~$0b$ diagram comes from adding~(\ref{AA0bRES})
and~(\ref{BB0bRES}),
\begin{equation}
\Bigl[ - i V_{0b} \Bigr]^{\Delta\alpha}
	=
	k (\kappa\lambda)^2 \delta^D(x\!-\!y) \delta^D(x'\!-\!y')
	(a_x a_{x'})^D
	\Big[
	\tfrac{1}{2} D(D\!-\!1) Q_{AA}(x;x')
	-
	2D L_{AA}(x;x')
	\Big]
	\, ,
\end{equation}
where
\begin{align}
Q_{AA}(x;x')
	={}&
	- i \!\int\! d^{D\!}z \, a_z^{D-2}
	\ln(a_z)
	\partial^\mu_z i \Delta_A(x;z)
	\!\times\! \partial_\mu^z i \Delta_A(z;x')
	\, ,
\label{QAAdef}
\\
L_{AA}(x;x')
	={}&
	- i \!\int\! d^{D\!}z \, a_z^{D-2}
	\ln(a_z)
	\nabla^\mu_z i \Delta_A(x;z)
	\!\times\! \nabla_\mu^z i \Delta_A(z;x')
    \, ,
\label{LAAdef}
\end{align}
are special cases of integrated propagators defined in~(\ref{Qdef})
and~(\ref{Ldef}) in Appendix~\ref{app:Integrated propagators}.
Contributions from diagrams in class~$\overline{1}$, given in~(\ref{AA1bRES}),
(\ref{AA1cRES}),~(\ref{BB1bRES}), and~(\ref{BB1cRES}), simply give
\begin{equation}
\Bigl[ - i V_{\overline{1b+c}} \Bigr]^{\Delta\alpha} =
	k (\kappa\lambda)^2
	\delta^D(x\!-\!y) \delta^D(x' \!-\! y')
	(a_x a_{x'})^D
	\!\times\!
	\tfrac{D(D-1)}{2} \ln(a_x a_{x'})
	i \Delta_A(x;x') 
	\, ,
\end{equation}
while the contributions from diagram~$0a$ and classes~$\overline{\overline{4}}$
and~$\overline{5}$, given in~(\ref{AA045cancel}) and~(\ref{BB0a45cancel}), 
combine into
\begin{align}
\MoveEqLeft[2]
\Bigl[ - i V_{0a} - i V_{\overline{\overline{4}}} 
	- i V_{\overline{5}} \Bigr]^{\Delta\alpha}
	=
	k (\kappa\lambda)^2
	\delta^D(x\!-\!y) \delta^D(x'\!-\!y')
	(a_xa_{x'})^D
\nonumber \\
&
	\times\!
	\Bigl\{
	-
	\tfrac{D-1}{D-3} Q_{AA}(x;x')
	+
	\tfrac{(D-1)(D-2)}{D-3} L_{AA}(x;x')
	+
	\Big[
	\tfrac{1}{D-3}
	\Big(
		\tfrac{\ln(a_x)}{Ha_x} \partial_0^x
	+
	\tfrac{\ln(a_{x'})}{Ha_{x'}} \partial_0^{x'} 
	\Big)
\nonumber \\
&	\hspace{0.6cm}
	+
	\tfrac{D-4}{2(D-3)} \Big(
	\tfrac{\ln(a_x)}{H^2a_x^2} \nabla_{x}^2 
	+
	\tfrac{\ln(a_{x'})}{H^2a_{x'}^2} \nabla_{x'}^2 
	\Big)
	+
	\tfrac{1}{2(D-3)}
	\Big(
	\tfrac{\ln(a_x)}{H^2a_x^D} \mathcal{D}_x
	+
	\tfrac{\ln(a_{x'})}{H^2a_{x'}^D} \mathcal{D}_{x'}
	\Big)
	\Big] i \Delta_A(x;x')
	\Bigr\}
	\, .
\end{align}
Upon using the identity that follows from the solutions~(\ref{Qresult}) and~(\ref{Lresult})
for integrated propagators,
\begin{equation}
(D\!-\!1) Q_{AA}(x;x') -2 L_{AA}(x;x')
	\longrightarrow
	\Big( \tfrac{ \ln(a_x) }{ H a_x } \partial_0^x
	+
	\tfrac{ \ln(a_{x'}) }{ Ha_{x'} } \partial_0^{x'}
	\Big)
	i \Delta_A(x;x')
	\, ,
\end{equation}
where we kept only the logarithm terms,
the total contribution from diagrams 0--5 reads:
\begin{align}
\MoveEqLeft[3]
\Bigl[ - i V_{\text{0--5}} \Bigr]^{\Delta\alpha}
	=
	k (\kappa\lambda)^2
	\delta^D(x\!-\!y) \delta^D(x'\!-\!y')
	(a_xa_{x'})^D
	\Bigl[
	\tfrac{D(D-1)}{2} \ln(a_x a_{x'})
	+
	\tfrac{D}{2}
	\Big( \tfrac{ \ln(a_x) }{ H a_x } \partial_0^x
	+
	\tfrac{ \ln(a_{x'}) }{ Ha_{x'} } \partial_0^{x'}
	\Big)
\nonumber \\
&
	+
	\tfrac{D-4}{2(D-3)} \Big(
	\tfrac{\ln(a_x)}{H^2a_x^2} \nabla_{x}^2 
	+
	\tfrac{\ln(a_{x'})}{H^2a_{x'}^2} \nabla_{x'}^2 
	\Big)
	+
	\tfrac{1}{2(D-3)}
	\Big(
	\tfrac{\ln(a_x)}{H^2a_x^D} \mathcal{D}_x
	+
	\tfrac{\ln(a_{x'})}{H^2a_{x'}^D} \mathcal{D}_{x'}
	\Big)
	\Bigr] i \Delta_A(x;x')
	\, .
\end{align}
%

\paragraph{Diagram classes 6--7.}
The~$AA$ part of class 6 in~(\ref{6a+bRED}) and~(\ref{6c+dRED}) does not
contribute anything on account of the spatial derivative acting on the coincidence limit of 
the integrated propagator. The entire result comes from the~$BB$ part in~(\ref{BB6a+bRES})
and~(\ref{BB6c+dRES}),
\begin{align}
\Big[ - i V_{6} \Big]^{\Delta\alpha}
	={}&
	k (\kappa \lambda)^2
	\delta^D(x\!-\!y) \delta^D(x'\!-\!y')
	(a_xa_{x'})^D
\nonumber \\
&
	\times\!
	\Big[
	-
	\tfrac{(D-1)^2 }{D-3} \ln(a_x a_{x'})
	- \tfrac{D-1 }{D-3}
	\Big( \tfrac{\ln(a_x)}{Ha_x} \partial_0^x
		+ \tfrac{\ln(a_{x'})}{Ha_{x'}} \partial_0^{x'} \Big)
	\Big]
	i \Delta_A(x;x')
	\, .
\end{align}
The total contribution of class 7 is also easily obtained by substituting in the appropriate
coincidence limits of integrated propagators 
into~(\ref{AA7aRED}),~(\ref{AA7bRED}),~(\ref{BB7aRED}) and~(\ref{BB7bRED}),
\begin{align}
\MoveEqLeft[1.5]
\Big[ - i V_{7} \Big]^{\Delta\alpha} =
	k (\kappa \lambda)^2
	\delta^D(x\!-\!y) \delta^D(x'\!-\!y')
	(a_xa_{x'})^D
\\
&
	\times\!
	\Big[
	- \tfrac{D-4}{D-3}
	\Big(
	\tfrac{\ln(a_x)}{H^2 a_x^2} 
	\overline{\nabla}{}_x \!\cdot\! \overline{\nabla}{}_y
	+
	\tfrac{\ln(a_{x'})}{H^2 a_{x'}^2} 
	\overline{\nabla}{}_{x'} \!\cdot\! \overline{\nabla}{}_{y'}
	\Big)
	-
	\tfrac{1}{D-3}
	\Big(
	\tfrac{\ln(a_x)}{H^2a_x^2}
		\overline{\partial}{}_x \!\cdot\! \overline{\partial}{}_y
	+
	\tfrac{\ln(a_{x'})}{H^2a_{x'}^2}
		\overline{\partial}{}_{x'} \!\cdot\!  \overline{\partial}{}_{y'}
		\Big)
	\Big]
	i\Delta_A(x;x')
	\, .
\nonumber 
\end{align}
%

\section{Reducing mode function corrections}
\label{sec: Mode function corrections}

In addition to the eight classes of diagrams considered in the preceding section,
here we work out the contributions from the corrections to external mode
functions. In what follows we will compute the two contributions~(\ref{defMI}) 
and~(\ref{defMII}) to the self-mass of the massive scalar, assuming that the tree-level
mode function is connected to its~$y$-leg, as appears in Eq.~(\ref{deltaUNeq}).

\subsection{Preliminary reduction of~$AA$ parts}

Reducing diagram $I$ requires the same 
contraction~(\ref{7AAcontraction}) already worked out for 
diagram class 7. Dropping immediately the terms containing kinetic operators
that annihilate the external mode function, we get
\begin{align}
\Big[ -i\mathcal{M}^2_I \Big]_{AA}^{\Delta\alpha}
	=
	\kappa^2
	\Big[
	\big( \mathcal{D}_x \!-\! a_x^Dm^2 \big)
		( \overline{\nabla}_x \!\cdot\! \overline{\nabla}_y )
	+
	\big( \overline{\mathcal{D}}_x \!-\! a_x^Dm^2 \big)
		( \nabla_x \!\cdot\! \overline{\nabla}_y )
	\Big]
	\big( \mathcal{D}_y \!-\! a_y^Dm^2 \big)
	\begin{pmatrix}
	K_{AA}(x;y)
	\\
	i\Delta_m(x;y)
	\end{pmatrix}
	.
\end{align}
The kinetic operator factored out acts on the massive propagator in the loop to 
create a delta function,
\begin{equation}
\Big[ -i\mathcal{M}^2_I \Big]_{AA}^{\Delta\alpha}
	=
	i
	\kappa^2
	\Big[
	\big( \mathcal{D}_x \!-\! a_x^Dm^2 \big)
		( \overline{\nabla}_x \!\cdot\! \overline{\nabla}_y )
	+
	\big( \overline{\mathcal{D}}_x \!-\! a_x^Dm^2 \big)
		( \nabla_x \!\cdot\! \overline{\nabla}_y )
	\Big]
	\begin{pmatrix}
	K_{AA}(x;y)
	\\
	\delta^D(x\!-\!y)
	\end{pmatrix}
	\, .
\end{equation}
We then symmetrize the derivatives acting on the delta function in the first term 
using~(\ref{deltaSym}), and in the
second term move the derivative away from the delta function,
\begin{align}
\MoveEqLeft[13]
\Big[ -i\mathcal{M}^2_I \Big]_{AA}^{\Delta\alpha}
	=
	-i\kappa^2
	\bigg[
	( \overline{\nabla}_x \!\cdot\! \overline{\nabla}_y )
	\big( \partial_x \!\cdot\! \partial_y \!+\! a_x^2m^2 \big)
		\begin{pmatrix}
	K_{AA}(x;y)
	\\
	a_x^{D-2} \delta^D(x\!-\!y)
	\end{pmatrix}
\nonumber \\
&
	+
	\big( \overline{\mathcal{D}}_x \!-\! a_x^Dm^2 \big)
		 ( \overline{\nabla}{}_x \!+\! \widetilde{\nabla}{}_x ) \!\cdot\! \overline{\nabla}_y
		\begin{pmatrix}
	K_{AA}(x;y)
	\\
	\delta^D(x\!-\!y)
	\end{pmatrix}
	\bigg]
	\, .
\end{align}
Next, in the first term we integrate by parts the two derivatives away from the delta
function, and in the second term we drop the spatial derivative of~$K_{AA}$ at coincidence,
\begin{align}
\Big[ -i\mathcal{M}^2_I \Big]^{\Delta\alpha}_{AA}
	=
	-i
	\kappa^2
	( \overline{\nabla}{}_x\!\cdot\! \overline{\nabla}_y )
	\Big\{
	a_x^{D-2} 
	\Big[ (\overline{\partial}_x \!+\! \widetilde{\partial}_x) \!\cdot\! 
		(\overline{\partial}_y \!+\! \widetilde{\partial}_y) + a_x^2m^2 \Big]
	+
	\big( \overline{\mathcal{D}}_x \!-\! a_x^Dm^2 \big)
	\Big\}
	\begin{pmatrix}
	K_{AA}(x;y)
	\\
	\delta^D(x\!-\!y)
	\end{pmatrix}
	.
\label{AApreliminaryI}
\end{align}

For diagram~$I\!I$ we need to contract the F-vertex into the tensor structure of 
the~$AA$ part of the graviton propagator, which produces the expression
\begin{align}
\Big[ -i\mathcal{M}^2_{I\!I} \Big]_{AA}^{\Delta\alpha}
	={}&
	i \kappa^2 a_x^{D-2}
	\Bigl[
	(D\!-\!1) ( \overline{\partial}{}_x \!\cdot\! \widetilde{\partial}{}_x )
		( \overline{\partial}{}_y \!\cdot\! \widetilde{\partial}{}_y )
	+
	(\overline{\nabla}{}_x \!\cdot\! \overline{\nabla}{}_y)
		( \widetilde{\partial}{}_x \!\cdot\! \widetilde{\partial}{}_y )
\nonumber \\
&	\hspace{3cm}
	-
	\tfrac{1}{2} (D\!-\!1)
		\big( \overline{\partial}{}_x \!\cdot\! \overline{\partial}{}_y 
			\!+\! a_x^2m^2 \big)
			( \widetilde{\partial}{}_x \!\cdot\! \widetilde{\partial}{}_y )
	\Big]
	\begin{pmatrix}
	K_{AA}(x;y)
	\\
	\delta^D(x\!-\!y)
	\end{pmatrix}
	\, ,
\label{AApreliminaryII}
\end{align}
that has the same structure as the expression~(\ref{AApreliminaryI}) obtained for 
diagram~$I$.
Adding the two contributions~(\ref{AApreliminaryI}) and~(\ref{AApreliminaryII}) 
and substituting in the appropriate coincidence 
limits~(\ref{KAAcoincidence})--(\ref{ddKAAcoincidence}) then produces:
\begin{align}
\Big[ -i\mathcal{M}^2_{I+I\!I} \Big]_{AA}^{\Delta\alpha}
	={}&
	i k \kappa^2
	\delta^D(x\!-\!y)
	\ln(a_x)
	a_x^{D-2} 
	\Big\{
	\Big[
	\tfrac{\overline{\nabla}{}_x \cdot \overline{\nabla}_y}{H^2a_x^2}
		+
		\tfrac{(D-1)^2}{2}
		\Big]
	\big( \overline{\partial}_x \!\cdot\! \overline{\partial}_y \!+\! a_x^2m^2 \big)
\nonumber \\
&
	+
	\tfrac{\overline{\nabla}{}_x \cdot \overline{\nabla}_y }{H^2a_x^2}
	\Big[
	a_x^{2-D} \big( \overline{\mathcal{D}}_x \!-\! a_x^Dm^2 \big)
	+
	Ha_x \big( \overline{\partial}_0^x \!+\! \overline{\partial}_0^y \big)
	-
	(D\!-\!1) H^2 a_x^2
	\Big]
	\Big\}
	.
\label{AApreliminaryI+II}
\end{align}
%

\subsection{Preliminary reduction of~$BB$ parts}

For diagram~$I$ we require two contractions~(\ref{BBthirdContraction}),
\begin{equation}
\Big[ -i\mathcal{M}^2_I \Big]_{BB}^{\Delta\alpha}
	=
	- \kappa^2
	\Big[
	\big( \mathcal{D}_x \!-\! a_x^Dm^2 \big) \overline{\partial}{}_0^x
	+
	\big( \overline{\mathcal{D}}_x \!-\! a_x^Dm^2 \big) \partial_0^x
	\Big]
	\big( \mathcal{D}_y \!-\! a_y^Dm^2 \big) \overline{\partial}{}_0^y
	\begin{pmatrix}
	K_{BB}(x;y)
	\\
	i \Delta_m(x;y)
	\end{pmatrix}
	\, ,
\end{equation}
where we immediately drop terms that annihilate the mode function.
The factored out kinetic term pinches the massive propagator in the
loop into a delta function,
\begin{equation}
\Big[ -i\mathcal{M}^2_I \Big]_{BB}^{\Delta\alpha}
	=
	-i\kappa^2
	\Big[
	\overline{\partial}{}_0^x \overline{\partial}{}_0^y
	\big( \mathcal{D}_x \!-\! a_x^Dm^2 \big)
	+
	\partial_0^x \overline{\partial}{}_0^y
	\big( \overline{\mathcal{D}}_x \!-\! a_x^Dm^2 \big) 
	\Big]
	\begin{pmatrix}
	K_{BB}(x;y)
	\\
	\delta^D(x\!-\!y)
	\end{pmatrix}
	\, .
\end{equation}
Using~(\ref{deltaSym}) we symmetrize the derivatives acting on the delta function in
the first term,
\begin{align}
\Big[ -i\mathcal{M}^2_I \Big]_{BB}^{\Delta\alpha}
	={}&
	i
	\kappa^2
	\bigg[
	\overline{\partial}{}_0^x \overline{\partial}{}_0^y
	\big( \partial_x \!\cdot\! \partial_y \!+\! a_x^2m^2 \big)
	\begin{pmatrix}
	K_{BB}(x;y)
	\\
	a_x^{D-2} \delta^D(x\!-\!y)
	\end{pmatrix}
\nonumber \\
&	\hspace{2cm}
	-
	\partial_0^x\overline{\partial}{}_0^y
	\big( \overline{\mathcal{D}}_x \!-\! a_x^Dm^2 \big) 
	\begin{pmatrix}
	K_{BB}(x;y)
	\\
	\delta^D(x\!-\!y)
	\end{pmatrix}
	\bigg]
	\, .
\end{align}
We follow this by integrating by parts the derivatives away from the delta function
in both terms, minding that here the order of operators matters,
\begin{align}
\Big[ -i\mathcal{M}^2_I \Big]_{BB}^{\Delta\alpha}
	={}&
	i
	\kappa^2
	\Big\{
	a_x^{D-2} 
	\overline{\partial}{}_0^x \overline{\partial}{}_0^y
	\Big[ (\overline{\partial}{}_x \!+\! \widetilde{\partial}{}_x) 
		\!\cdot\! (\overline{\partial}{}_y \!+\! \widetilde{\partial}{}_y ) + a_x^2m^2 \Big]
\nonumber \\
&	\hspace{2cm}
	+
	\big(\overline{\partial}{}_0^x \!+\! \widetilde{\partial}{}_0^x \big)
	\overline{\partial}{}_0^y
	\big( \overline{\mathcal{D}}_x \!-\! a_x^Dm^2 \big) 
	\Big\}
	\begin{pmatrix}
	K_{BB}(x;y)
	\\
	\delta^D(x\!-\!y)
	\end{pmatrix}
	\, .
\label{BBI}
\end{align}

For diagram $I\!I$ we need the contraction of the F-vertex into the tensor structure of 
the~$BB$ part of the graviton propagator,
\begin{align}
\MoveEqLeft[4]
\Big[ -i\mathcal{M}^2_{I\!I} \Big]_{BB}^{\Delta\alpha}
	=
	i \kappa^2 a_x^{D-2}
	\Bigl\{
	( \overline{\partial}{}_x \!\cdot\! \widetilde{\partial}{}_x )
	( \overline{\partial}{}_y \!\cdot\! \widetilde{\partial}{}_y)
	-
	\overline{\partial}{}_0^x \overline{\partial}{}_0^y
	( \widetilde{\partial}{}_x \!\cdot\! \widetilde{\partial}{}_y )
	+
	(D\!-\!4) Ha_x \overline{\partial}{}_0^x
		( \overline{\partial}{}_y \!\cdot\! \widetilde{\partial}{}_y)
\nonumber \\
&
	+
	(D\!-\!4) Ha_y \overline{\partial}{}_0^y
		( \overline{\partial}{}_x \!\cdot\! \widetilde{\partial}{}_x)
	+
	2Ha_x 
	\big( \widetilde{\partial}_0^x + (D\!-\!2) Ha_x \big)
	(\overline{\partial}{}_x \!\cdot\! \overline{\partial}{}_y)
\nonumber \\
&
	-
	\tfrac{1}{2} 
	\Big[ (\widetilde{\partial}{}_x \!\cdot\! \widetilde{\partial}{}_y)
		+
		(D\!-\!2) H a_x \big( \widetilde{\partial}_0^x 
			\!+\! \widetilde{\partial}_0^y \!+\! D Ha_x \big)
		\Big]
		\big( \overline{\partial}{}_x \!\cdot\! \overline{\partial}{}_y
			\!+\! a_x^2 m^2 \big)
	\Bigr\}
	\begin{pmatrix}
	K_{BB}(x;y)
	\\
	\delta^D(x\!-\!y)
	\end{pmatrix}
	\, .
\label{BBII}
\end{align}
Collecting both contributions~(\ref{BBI}) and~(\ref{BBII}), and substituting in them the 
appropriate coincidence limits of~$K_{BB}$ then gives
\begin{align}
\MoveEqLeft[1.5]
\Big[ -i\mathcal{M}^2_{I+I\!I} \Big]_{BB}^{\Delta\alpha}
	=
	\tfrac{i k }{D-3}
	\kappa^2
	\delta^D(y\!-\!x)
	\ln(a_x)
	a_x^{D-2} 
	\Big\{
	\Big[
	-
	\tfrac{1}{H^2a_x^2} \overline{\partial}{}_0^x \overline{\partial}{}_0^y
	+
	\tfrac{(D-1)(D-3)}{2}
		\Big]
	\big( \overline{\partial}{}_x \!\cdot\! \overline{\partial}{}_y \!+\! a_x^2m^2 \big)
\nonumber \\
&
	-
	\tfrac{1}{H^2a_x^D}
	\big( \overline{\partial}{}_0^x \!-\! Ha_x \big)
	\overline{\partial}{}_0^y
	\big( \overline{\mathcal{D}}_x \!-\! a_x^Dm^2 \big) 
	-
	\tfrac{1}{Ha_x}
	\big( \overline{\partial}{}_0^x \!+\! \overline{\partial}{}_0^y \!-\! Ha_x \big)
		\overline{\partial}{}_0^x \overline{\partial}{}_0^y
	-
	2(D\!-\!3) \overline{\nabla}{}_x \!\cdot\! \overline{\nabla}{}_y
	\Big\}
	\, .
\label{BBpreliminaryI+II}
\end{align}
%

\subsection{Final reduction}
\label{subsec: Final reduction 2}

Adding the total~$AA$ contribution~(\ref{AApreliminaryI+II}) and the 
total~$BB$ contribution~(\ref{BBpreliminaryI+II}) we have
\begin{align}
\Big[ -i\mathcal{M}^2_{I+I\!I} \Big]^{\Delta\alpha}
	={}&
	\tfrac{i k \kappa^2}{D-3}
	\delta^D(x\!-\!y)
	\ln(a_x)
	a_x^{D-2} 
	\Big\{
	\tfrac{1}{H^2a_x^D}
	\Big[
	\overline{\partial}{}_x \!\cdot\! \overline{\partial}{}_y
	+
	(D\!-\!4)\overline{\nabla}{}_x \!\cdot\! \overline{\nabla}_y
	+
	Ha_x \overline{\partial}{}_0^y
	\Big]
	\big( \overline{\mathcal{D}}_x \!-\! a_x^Dm^2 \big)
\nonumber \\
&
	+
	\tfrac{1}{H^2a_x^2} \Big[
	\overline{\partial}{}_x\!\cdot\! \overline{\partial}{}_y
	+
	(D\!-\!4)\overline{\nabla}{}_x \!\cdot\! \overline{\nabla}_y
	+
	\tfrac{D(D-1)(D-3)}{2} H^2 a_x^2
	\Big]
	\big( \overline{\partial}_x \!\cdot\! \overline{\partial}_y \!+\! a_x^2m^2 \big)
\nonumber \\
&
	+
	\tfrac{1}{Ha_x}
	\Big[
	\overline{\partial}{}_x \!\cdot\! \overline{\partial}{}_y
	+
	(D\!-\!4) \overline{\nabla}{}_x \!\cdot\! \overline{\nabla}_y
	\Big]
	\big( \overline{\partial}_0^x \!+\! \overline{\partial}_0^y \!-\! Ha_x \big)
	-
	D(D\!-\!3) \overline{\nabla}{}_x \!\cdot\! \overline{\nabla}_y
	\Big\}
\, .
\end{align}
Further reduction is possible using the following identities in which we keep
only the logarithm terms,
\begin{align}
&
\delta^D(x\!-\!y) \ln(a_x) a_x^{D-3}
	\big( \overline{\partial}{}_0^x \!+\! \overline{\partial}{}_0^y \big)
	\longrightarrow
	- (D\!-\!3) H \delta^D(x\!-\!y) \ln(a_x) a_x^{D-2}
	\, ,
\label{IdUp}
\\
&
\delta^D(x\!-\!y) \ln(a_x) a_x^{D-2}
	\big( \overline{\partial}{}_x \!\cdot\! \overline{\partial}_y \!+\! a_x^2m^2 \big)
\nonumber \\
&	\hspace{1.5cm}
	\longrightarrow
	-\tfrac{1}{2} 
	\delta^D(x\!-\!y) \ln(a_x)
	\Big[
	\big( \overline{\mathcal{D}}_x \!-\! a_x^D m^2 \big)
	+
	\big( \overline{\mathcal{D}}_y \!-\! a_y^D m^2 \big)
	\Big]
	\, ,
\label{IdMiddle}
\\
&
\delta^D(x\!-\!y) \ln(a_x) a_x^{D-4}
	\big( \overline{\partial}{}_x \!\cdot\! \overline{\partial}_y \!+\! a_x^2m^2 \big)
\nonumber \\
&	\hspace{1.5cm}
	\longrightarrow
	\delta^D(x\!-\!y) \ln(a_x)
	\Big[
	(D\!-\!3)H^2 a_x^{D-2}
	-
	\tfrac{1}{2a_x^2} \big( \overline{\mathcal{D}}_x \!-\! a_x^D m^2 \big)
	-
	\tfrac{1}{2a_y^2} \big( \overline{\mathcal{D}}_y \!-\! a_y^D m^2 \big)
	\Big]
	\, ,
\end{align}
that follow from the second vertex identity~(\ref{VertexIdentity2}).
Being careful about the non-commuting derivatives when applying them produces
\begin{align}
\Big[ -i\mathcal{M}^2_{I+I\!I} \Big]^{\Delta\alpha}
	={}&
	i k \kappa^2 \delta^D(x\!-\!y) \ln(a_x)
	\tfrac{1}{(D-3)H^2a_x^2}
	\Big\{
	-
	\tfrac{1}{2} 
	\Big[
	\big( \overline{\mathcal{D}}_x \!-\! a_x^D m^2 \big)
	+
	\big( \overline{\mathcal{D}}_y \!-\! a_y^D m^2 \big)
	\Big]
	\overline{\partial}{}_x \!\cdot\! \overline{\partial}{}_y
\nonumber \\
&
	+
	\Big[
	\overline{\partial}{}_x \!\cdot\! \overline{\partial}{}_y
	+
	\tfrac{D-4}{2} \overline{\nabla}{}_x \!\cdot\! \overline{\nabla}_y
	+
	Ha_x \overline{\partial}{}_0^y
	-
	\tfrac{D(D-1)(D-3)}{4}
	\Big]
	\big( \overline{\mathcal{D}}_x \!-\! a_x^Dm^2 \big)
\nonumber \\
&
	-
	H^2a_x^{D}
	\Big[
	\overline{\partial}{}_x \!\cdot\! \overline{\partial}{}_y
	+
	(D^2\!-\!2D\!-\!4) \overline{\nabla}{}_x \!\cdot\! \overline{\nabla}_y
	\Big]
	\Big\}
\, .
\end{align}
Next we use the commutation identity~(\ref{CommutationIdentities}) on the term in the 
first line,
\begin{align}
\MoveEqLeft[4]
\Big[ -i\mathcal{M}^2_{I+I\!I} \Big]^{\Delta\alpha}
	=
	i k \kappa^2 \delta^D(x\!-\!y) \ln(a_x)
	\tfrac{1}{(D-3)H^2a_x^2}
	\Big\{
	\tfrac{1}{2} \Big[
	\overline{\partial}{}_x \!\cdot\! \overline{\partial}{}_y
	+
	(D\!-\!4) \overline{\nabla}{}_x \!\cdot\! \overline{\nabla}_y
	-
	(D\!-\!4) H a_x \overline{\partial}{}^y_0 
\nonumber \\
&
	-
	\tfrac{D(D-1)(D-3)}{2} H^2a_x^2
	\Big]
	\big( \overline{\mathcal{D}}_x \!-\! a_x^Dm^2 \big)
	-
	H^2a_x^{D}
	\Big[
	(D\!-\!1)\overline{\partial}{}_x \!\cdot\! \overline{\partial}{}_y
	+
	(D^2\!-\!3D\!-\!2) \overline{\nabla}{}_x \!\cdot\! \overline{\nabla}_y
	\Big]
\nonumber \\
&
	+
	H a_x^{D+1} m^2
		\big( \overline{\partial}{}^x_0 \!+\! \overline{\partial}{}^y_0 \big)
	\Big\}
\, .
\end{align}
This is followed by applying the identity~(\ref{SpatialHalfIdentity}) from 
Appendix~\ref{app:Integrated propagators}, in which we keep only the logarithm terms,
\begin{equation}
\delta^D(x\!-\!y) \ln(a_x) 
a_x^{D-2} \overline{\nabla}_x \!\cdot\! \overline{\nabla}_y
	\longrightarrow
	\delta^D(x\!-\!y) \ln(a_x)
	\!\times\!
	\tfrac{-1}{2Ha_x} 
	\Bigl[ \overline{\partial}{}_0^y + \tfrac{D-1}{2} H a_y \Bigr] 
	\big( \overline{\mathcal{D}}_x \!-\! a_x^Dm^2 \big)
	\, ,
\end{equation}
and by once more using~(\ref{IdUp}) and~(\ref{IdMiddle}), which finally produces
\begin{equation}
\Big[ -i\mathcal{M}^2_{I+I\!I} \Big]^{\Delta\alpha}
	\!\!=
	\delta^D(x\!-\!y) 
	\tfrac{ i k \kappa^2\ln(a_x)  }{2(D-3)H^2a_x^2}
	\Big[
	\overline{\partial}{}_x \!\cdot\! \overline{\partial}{}_y
	+
	(D\!-\!4) \overline{\nabla}{}_x \!\cdot\! \overline{\nabla}_y
	+
	(D^2\!-\!4D\!+\!2) Ha_x \overline{\partial}_0^y
	\Big]
	\big( \overline{\mathcal{D}}_x \!-\! a_x^Dm^2 \big)
\, .
\label{FinalExternalLeg}
\end{equation}
From here we can simply read off the solution for the mode function correction
by stripping the last kinetic operator,
\begin{align}
\delta u_{I+I\!I}(x)
	={}&
	- \tfrac{ k \kappa^2 \ln(a_x)}{2(D-3)H^2a_x^2}
	\Big[
	\overline{\partial}{}_x \!\cdot\! \partial{}_x
	+
	(D\!-\!4) \overline{\nabla}{}_x \!\cdot\! \nabla_x
	+
	(D^2\!-\!4D\!+\!2) Ha_x \partial_0^x
	\Big]
	u(x)
	\, ,
\end{align}
where we dropped the homogeneous parts that do not contain logarithms.
The barred derivatives in this expression are assumed to act on the remainder of the 
diagrams in~(\ref{DiagDeltaUa})--(\ref{DiagDeltaUd}) that the mode function correction 
attaches to.

\subsection{Contribution to effective self-mass}

We now connect the~$\Delta\alpha$ variation of the one-loop mode function correction
to the main body of the diagram according to~(\ref{DiagDeltaUa})--(\ref{DiagDeltaUd}). 
The resulting expressions give local vertex corrections with the topology of middle two 
diagrams on the right-hand side of~(\ref{4ReducedDiagrams}), and can thus be 
written using the same conventions as for the 4-point diagrams introduced 
at the beginning of Sec.~\ref{sec: Feynman diagrams},
\begin{align}
\Big[ -iV_{I+I\!I,a} \Big]^{\Delta\alpha}
	={}&
	k (\kappa\lambda)^2 \delta^D(x\!-\!y)\delta^D(x'\!-\!y') (a_xa_{x'})^D
	\tfrac{\ln(a_x)}{2(D-3)H^2 a_x^2}
	\Big[
	\overline{\partial}{}_y \!\cdot\!
		\big( \overline{\partial}{}_x \!+\! \partial{}_x \big)
\nonumber \\
&	\hspace{1cm}
	+
	(D\!-\!4) \overline{\nabla}{}_y \!\cdot\! \big( \overline{\nabla}{}_x \!+\! \nabla_x \big)
	+
	(D^2 \!-\! 5D \!+\! 2) Ha_x \overline{\partial}_0^y
	\Big]
	i\Delta_A(x;x')
	\, ,
\\
\Big[ -iV_{I+I\!I,b} \Big]^{\Delta\alpha}
	={}&
	k (\kappa\lambda)^2 \delta^D(x\!-\!y)\delta^D(x'\!-\!y') (a_xa_{x'})^D
	\tfrac{\ln(a_x)}{2(D-3)H^2 a_x^2}
	\Big[
	\overline{\partial}{}_x \!\cdot\!
		\big( \overline{\partial}{}_y \!+\! \partial{}_x \big)
\nonumber \\
&	\hspace{1cm}
	+
	(D\!-\!4) \overline{\nabla}{}_x \!\cdot\! \big( \overline{\nabla}{}_y \!+\! \nabla_x \big)
	+
	(D^2 \!-\! 5D \!+\! 2) Ha_x \overline{\partial}_0^x
	\Big]
	i\Delta_A(x;x')
	\, ,
\\
\Big[ -iV_{I+I\!I,c} \Big]^{\Delta\alpha}
	={}&
	k (\kappa\lambda)^2 \delta^D(x\!-\!y)\delta^D(x'\!-\!y') (a_xa_{x'})^D
	\tfrac{ \ln(a_{x'}) }{2(D-3)H^2 a_{x'}^2}
	\Big[
	\overline{\partial}{}_{y'} \!\cdot\!
		\big( \overline{\partial}{}_{x'} \!+\! \partial{}_{x'}  \big)
\nonumber \\
&	\hspace{1cm}
	+
	(D\!-\!4) \overline{\nabla}{}_{y'} \!\cdot\! \big( \overline{\nabla}{}_{x'} \!+\! \nabla_{x'} \big)
	+
	(D^2 \!-\! 5D \!+\! 2) Ha_{x'} \overline{\partial}_0^{y'}
	\Big]
	i\Delta_A(x;x')
	\, ,
\\
\Big[ -iV_{I+I\!I,d} \Big]^{\Delta\alpha}
	={}&
	k (\kappa\lambda)^2 \delta^D(x\!-\!y)\delta^D(x'\!-\!y') (a_xa_{x'})^D
	\tfrac{ \ln(a_{x'}) }{2(D-3)H^2 a_{x'}^2}
	\Big[
	\overline{\partial}{}_{x'} \!\cdot\!
		\big( \overline{\partial}{}_{y'} \!+\! \partial{}_{x'} \big)
\nonumber \\
&	\hspace{1cm}
	+
	(D\!-\!4) \overline{\nabla}{}_{x'} \!\cdot\! \big( \overline{\nabla}{}_{y'} \!+\! \nabla_{x'} \big)
	+
	(D^2 \!-\! 5D \!+\! 2) Ha_{x'} \overline{\partial}_0^{x'}
	\Big]
	i\Delta_A(x;x')
	\, .
\end{align}
Therefore, these contributions will combine with the contributions worked out in 
Sec.~\ref{sec: Reducing 4-point diagrams}, but should be further reduced 
before that.

The reduction can be pursued after combining~$a$ and~$b$ contributions, 
\begin{align}
\MoveEqLeft[2]
\Big[ -iV_{I+I\!I,a+b} \Big]^{\Delta\alpha}
	=
	k (\kappa\lambda)^2 \delta^D(x\!-\!y)\delta^D(x'\!-\!y') (a_xa_{x'})^D
	\tfrac{\ln(a_x)}{(D-3)H^2 a_x^2}
	\Big[
	\overline{\partial}{}_x \!\cdot\! \overline{\partial}{}_y
	+
	(D\!-\!4) \overline{\nabla}{}_x \!\cdot\! \overline{\nabla}{}_y
\nonumber \\
&
	+
	\tfrac{1}{2}\big( \overline{\partial}{}_x \!+\! \overline{\partial}{}_y \big) \!\cdot\! \partial{}_x
	+
	\tfrac{1}{2}(D\!-\!4) \big( \overline{\nabla}{}_x \!+\! \overline{\nabla}{}_y \big) \!\cdot\! \nabla_x 
	+
	\tfrac{D^2 - 5D + 2}{2} Ha_x \big( \overline{\partial}_0^x \!+\! \overline{\partial}_0^y \big)
	\Big]
	i\Delta_A(x;x')
	\, .
\end{align}
This allows us to reflect the derivatives in the second line away from the external 
mode functions, which produces the final result,
\begin{align}
\MoveEqLeft[3]
\Big[ -iV_{I+I\!I,a+b} \Big]^{\Delta\alpha}
	=
	k (\kappa\lambda)^2 \delta^D(x\!-\!y)\delta^D(x'\!-\!y') (a_xa_{x'})^D
	\tfrac{\ln(a_x)}{(D-3)H^2 a_x^2}
	\Big[
	\overline{\partial}{}_x \!\cdot\! \overline{\partial}{}_y
	+
	(D\!-\!4) \overline{\nabla}{}_x \!\cdot\! \overline{\nabla}{}_{y}
\nonumber \\
&
	-
	\tfrac{1}{2a_x^{D-2}} \mathcal{D}_x
	-
	\tfrac{D-4}{2} \nabla_x^2
	-
	\tfrac{D^2 - 5D + 2}{2} Ha_x \partial_0^x
	-
	\tfrac{(D-1)(D^2 - 5D + 2)}{2} H^2a_x^2
	\Big]
	i\Delta_A(x;x')
	\, ,
\end{align}
The~$c$ and~$d$ contributions combine in an analogous way,
\begin{align}
\MoveEqLeft[3]
\Big[ -iV_{I+I\!I,c+d} \Big]^{\Delta\alpha}
	=
	k (\kappa\lambda)^2 \delta^D(x\!-\!y)\delta^D(x'\!-\!y') (a_xa_{x'})^D
	\tfrac{\ln(a_{x'})}{(D-3)H^2 a_{x'}^2}
	\Big[
	\overline{\partial}{}_{x'} \!\cdot\! \overline{\partial}{}_{y'}
	+
	(D\!-\!4) \overline{\nabla}{}_{x'} \!\cdot\! \overline{\nabla}{}_{y'}
\nonumber \\
&
	-
	\tfrac{1}{2a_{x'}^{D-2}} \mathcal{D}_{x'}
	-
	\tfrac{D-4}{2} \nabla_{x'}^2
	-
	\tfrac{D^2 - 5D + 2}{2} Ha_{x'} \partial_0^{x'}
	-
	\tfrac{(D-1)(D^2 - 5D + 2)}{2} H^2a_{x'}^2
	\Big]
	i\Delta_A(x;x')
	\, .
\end{align}
Adding these contributions to the ones worked out in 
Sec.~\ref{sec: Reducing 4-point diagrams}
produces a vanishing result for the~$\Delta\alpha$ variation of the effective self-mass, 
as summarized by Table~\ref{FinalTable} in the following section.

\section{Discussion and conclusions}
\label{sec: Discussion}

In this work we set out to test the gauge-independence of our program for purging gauge dependence from 
effective field equations, which was set up with the goal of obtaining physical predictions for 
quantum-gravitational effects during inflation. The central objection to the standard 
procedure---quantum-correcting effective field equations using the 1PI two-point function---is that the 1PI 
two-point function depends on the graviton gauge. We contend that this dependence arises because one 
typically neglects quantum-gravitational corrections associated with the source and the observer, which in 
our setup are represented by a massive matter field, which acts both as a source and as an observer, in close 
analogy with how gauge-independence is achieved for the flat-space S-matrix.

We consider the system in~(\ref{action}), consisting of a heavy scalar interacting through the
exchange of a massless minimally coupled (MMC) scalar on a de Sitter background, and we
incorporate source and observer corrections by forming the position-space amplitudes
corresponding to $t$-channel scattering. These amplitudes receive contributions from the
1PI two-, three-, and four-point functions, as depicted by the skeleton expansion
in~(\ref{SkeletonExpansion}). In addition, as we uncovered here, one must also include
one-loop corrections to the external mode functions, which are captured by the 1PI
two-point function of the massive field. Our strategy is to reduce the full set of
diagrams to the four topologies shown in~(\ref{4ReducedDiagrams}), exploiting the
derivative structure of gravitational interactions and, when needed, applying the Donoghue
identities~\cite{Donoghue:1993eb,Bjerrum-Bohr:2002aqa,Bjerrum-Bohr:2002gqz,Donoghue:1996mt}
to isolate the relevant terms. Upon inserting the exchange propagators using
Eq.~(\ref{Aeom}), these reduced topologies can be interpreted as corrections to the 1PI
two-point function of the MMC scalar. This effective correction is conjectured to be
gauge-independent and can therefore be used to quantum-correct the field equations and
extract physical predictions for quantum-gravitational effects. In flat space this program
was implemented and the resulting effective 1PI two-point function wasverified to be gauge
independent for one-graviton-loop corrections to an MMC scalar~\cite{Miao:2017feh} and to
electromagnetism~\cite{Katuwal:2021thy}. The present work examines gauge dependence of 
the analogous construction on a de Sitter background.

In order to test gauge dependence in de Sitter, we computed the contribution to the 
effective self-mass induced by the part of the de Sitter-breaking graviton 
propagator~\cite{Glavan:2019msf} that depends on one of the two infinitesimal 
gauge-fixing parameters. This defines a one-parameter subfamily of gauges for which 
the dependence on the gauge-fixing parameter is
linear, and hence can be extended to finite values~\cite{Glavan:2025azq}. We refer to this
component of the graviton propagator defined in Eq.~(\ref{GravitonPropagator}), and to the
contributions it generates, as the $\Delta\alpha$ variations.

In Secs.~\ref{sec: Reducing 4-point diagrams} 
and~\ref{sec: Mode function corrections} we
computed the $\Delta\alpha$ variations of the diagrams listed in 
Sec.~\ref{sec: Feynman diagrams}.
After reducing all contributions to the topologies on the right-hand side of~(\ref{4ReducedDiagrams}), which required no Donoghue identities and was achieved 
by judicious integrations by
parts, the full result takes the form
\begin{align}
\MoveEqLeft[2]
\Bigl[ - i V \Bigr]^{\Delta\alpha}
	=
	k (\kappa\lambda)^2
	\delta^D(x\!-\!y) \delta^D(x'\!-\!y')
	(a_xa_{x'})^D
	\bigg[
	C_1 \ln(a_xa_{x'})
	+
	C_2 \bigg( \frac{\ln(a_x)}{Ha_x} \partial_0^x
		+ \frac{\ln(a_{x'})}{Ha_{x'}} \partial_0^{x'} \bigg)
\nonumber \\
&
	+
	C_3
	\bigg(
	\frac{\ln(a_x)}{H^2a_x^2} \nabla_{x}^2 
	+
	\frac{\ln(a_{x'})}{H^2a_{x'}^2} \nabla_{x'}^2 
	\bigg)
	+
	C_4
	\bigg(
	\frac{\ln(a_x)}{H^2 a_x^2} 
	\overline{\nabla}{}_x \!\cdot\! \overline{\nabla}{}_y
	+
	\frac{\ln(a_{x'})}{H^2 a_{x'}^2} 
	\overline{\nabla}{}_{x'} \!\cdot\! \overline{\nabla}{}_{y'}
	\bigg)
\nonumber \\
&
	+
	C_5
	\bigg(
	\frac{\ln(a_x)}{H^2 a_x^2} 
	\overline{\partial}{}_x \!\cdot\! \overline{\partial}{}_y
	+
	\frac{\ln(a_{x'})}{H^2 a_{x'}^2} 
	\overline{\partial}{}_{x'} \!\cdot\! \overline{\partial}{}_{y'}
	\bigg)
	+
	C_6
	\bigg(
	\frac{\ln(a_x)}{H^2a_x^D} \mathcal{D}_x
	+
	\frac{\ln(a_{x'})}{H^2a_{x'}^D} \mathcal{D}_{x'}
	\bigg)
	\bigg]
	i \Delta_A(x;x')
	\, .
\label{coefficients}
\end{align}
This expression contains five contributions with local corrections to the
vertices, parametrized by the coefficients $C_1$--$C_5$, and one ultra-local contribution, 
parametrized by $C_6$, which involves the kinetic operator that
pinches the MMC scalar propagator. The contributions from the different diagram classes to
these coefficients are summarized in Table~\ref{FinalTable}, which shows that they 
all cancel.
\begin{table}[h!]
\renewcommand{\arraystretch}{1.8}
\centering
\begin{tabular}{ w{c}{1cm} w{c}{2cm} w{c}{2cm} w{c}{2cm} w{c}{2cm} w{c}{2cm} w{c}{2cm} } 
\hline
	&
	$C_1$
	&
	$C_2$
	&
	$C_3$
	&
	$C_4$
	&
	$C_5$
	&
	$C_6$
\\
\hline\hline
	{\bf 0--5}
	&
	$\frac{D(D-1)}{2}$
	&
	$\frac{D}{2}$
	&
	$\frac{D-4}{2(D-3)}$
	&
	0
	&
	0
	&
	$\frac{1}{2(D-3)}$
\\
	{\bf 6}
	&
	$- \frac{(D-1)^2}{D-3}$
	&
	$- \frac{D-1}{D-3}$
	&
	0
	&
	0
	&
	0
	&
	0
\\
	\bf{7}
	&
	0
	&
	0
	&
	0
	&
	$-\frac{D-4}{D-3}$
	&
	$-\frac{1}{D-3}$
	&
	0
\\
	$\boldsymbol{\delta u}$
	&
	\hspace{0.2cm}$\frac{(D-1)^2}{D-3} \!-\! \frac{D(D-1)}{2}$
	&
	\hspace{0.1cm}$\, \, \frac{D-1}{D-3}\!-\!\frac{D}{2}$
	&
	$-\frac{D-4}{2(D-3)}$
	&
	$\frac{D-4}{D-3}$
	&
	$\frac{1}{D-3}$
	&
	$- \frac{1}{2(D-3)}$
\\
\hline \hline
	{\bf Total} \
	&
	0
	&
	0
	&
	0
	&
	0
	&
	0
	&
	0
\\
\hline
\end{tabular}
\caption{Contributions to coefficients in the final expression~(\ref{coefficients})
for the~$\Delta\alpha$ gauge variation of the effective self-mass, coming from different
diagrams worked out in Secs.~\ref{sec: Reducing 4-point diagrams} and~\ref{sec: Mode function corrections}.}
\label{FinalTable}
\end{table}
This is the main result of our paper: the $\Delta\alpha$ variation drops out from the final
one-graviton-loop correction to the effective scalar self-mass. The cancellation is highly
non-trivial and occurs only after including contributions from {\it all} diagram classes
and~{\it all} external mode function corrections. 
This situation is considerably more intricate than in flat space~\cite{Miao:2017feh}, where
diagrams from classes~6 and~7 vanish identically, and where diagrams from classes~$I$ 
and~$I\!I$ do not contribute to field-strength renormalization.

The cancellation of the~$\Delta\alpha$ variation reported here is compelling and strengthens confidence in 
our program for constructing gauge-independent quantum-corrected effective field equations in a
cosmological setting. At the same time, it completes only half of the overall gauge test: the graviton 
propagator of~\cite{Glavan:2019msf} also depends on a second  gauge-fixing parameter~$\beta$, 
whose infinitesimal variation around unity we denote by $\delta\beta$. Demonstrating that 
the~$\delta\beta$ variation also cancels between all the diagrams should establish 
gauge-independence of the effective self-mass at the same level as was accomplished in flat 
space~\cite{Miao:2017feh}. We also hope that by performing the computation of the~$\delta\beta$ 
variation we will be able to test our de Sitter generalizations of the Donoghue identities 
made in~\cite{Glavan:2024elz}, and also to understand whether any of the additional diagrams collected in 
Appendix~\ref{app:Additional diagrams} can contribute to the MMC scalar exchange potential. 

Finally, it will be necessary to update the computation of the tentative gauge-independent contribution to 
the MMC scalar self-mass reported in~\cite{Glavan:2024elz}. The key lesson of the present work is that the 
additional diagram classes~6 and~7, together with the external mode-function corrections represented by 
classes $I$ and $I\!I$, are essential for canceling gauge dependence, and therefore must also be included 
when extracting the gauge-independent part. Only once this is done will we have a definitive answer as to 
whether the large quantum-gravitational logarithms persist in the exchange potential mediated by the MMC 
scalar.

\section*{Acknowledgments}
\addcontentsline{toc}{section}{\protect\numberline{}Acknowledgments} 

The authors acknowledge a generous support by the Delta ITP consortium, a program of the Netherlands Organisation for
Scientific Research (NWO) that is funded by the Dutch Ministry of Education, Culture and
Science (OCW) - NWO project number 24.001.027. 
DG was supported by the Czech Science Foundation (GA\v{C}R) grant 24-13079S. 
SPM was partially supported by Taiwan NSTC grants 113-2112-M-006-013 and 114-2112-M-006-020.
TP is supported by the NWA ORC 2023 consortium grant: Cosmic emergence: from abstract simplicity to complex diversity (Kosmische
emergentie: van abstracte eenvoud naar complexe diversiteit) and by The Magnetic Universe NWO grant OCENW.XL.23.147.
RPW was partially supported by NSF grant PHY-2207514 and by the Institute for 
Fundamental Theory at the University of Florida.

\appendix
\section{Integrated propagators}
\label{app:Integrated propagators}

In Sec.~\ref{subsec: Final reduction 1} we need expressions for two 
integrated propagators,
\begin{align}
Q_{\nu\nu}(x;x') ={}& \!- i \!\int\! d^{D\!}z \, a_z^{D-2}
	\ln(a_z)
	\Bigl[
	\partial^\mu_z i \Delta_\nu(x;z)
	\partial_\mu^z i \Delta_\nu(z;x')
	+ a_z^2M_\nu^2 i \Delta_\nu(x;z) i \Delta_\nu(z;x')
	\Bigr]
	\, ,
\label{Qdef}
\\
L_{\nu\nu}(x;x') ={}& \! - i \!\int\! d^{D\!}z \, a_z^{D-2}
	\ln(a_z)
	\nabla^\mu_z i \Delta_\nu(x;z)
	\nabla_\mu^z i \Delta_\nu(z;x')
    \, ,
\label{Ldef}
\end{align}
where the parameter~$\nu$ is related to the mass of the propagators as
\begin{equation}
\nu^2 = \Bigl( \frac{D\!-\!1}{2} \Bigr)^{\!2} - \frac{M_\nu^2}{H^2} \, ,
\end{equation}
so that propagators in the integrands satisfy the equation of motion
\begin{equation}
\bigl( \mathcal{D}_x - a_x^D M_\nu^2 \bigr) i \Delta_\nu(x;x')
    = i \delta^D(x\!-\!x') \, .
\end{equation}
Consequently, the integrated propagators in~(\ref{Qdef}) and~(\ref{Ldef})
satisfy sourced equations of motion,
\begin{align}
\bigl( \mathcal{D}_x \!-\! a_x^D M_\nu^2 \bigr) Q_{\nu\nu}(x;x')
	={}&
	-
	\ln(a_x ) i \delta^D(x\!-\!x')
	+
	Ha_x^{D-1} \partial_0^x i \Delta_\nu(x;x')
    \, ,
\label{Qeom}
\\
\bigl( \mathcal{D}_x \!-\! a_x^D M_\nu^2 \bigr) L_{\nu\nu}(x;x')
    ={}& 
    - \ln(a_x) a_x^{D-2} \nabla_x^2 i \Delta_\nu(x;x')
    \, ,
\label{Leom}
\end{align}
with analogous equations satisfied on the primed coordinate.

The first of the two integrated propagators can be evaluated straightforwardly
using the vertex identity~(\ref{VertexIdentity2}), where the logarithm insertion is 
treated as the third leg of the vertex,
\begin{equation}
Q_{\nu\nu}(x;x')
	=
	-
	\tfrac{1}{2} (D\!-\!1) H^2 I_{\nu\nu}(x;x')
	-
	\tfrac{1}{2} \ln(a_xa_{x'}) i \Delta_\nu(x;x')
	\, ,
\label{Qresult}
\end{equation}
where the result includes an integrated propagator without any derivatives 
or logarithms in the integrand,
\begin{equation}
I_{\nu\nu}(x;x') = 
    - i \int\! d^{D\!}z \, a_z^D
    i \Delta_\nu(x;z)
    i \Delta_\nu(z;x')
    \, .
\end{equation}

Evaluating the second integrated propagator~(\ref{Ldef}) 
is not as straightforward. It is best 
dealt with in momentum space representation, using identities for the 
mode functions (see e.g.~\cite{Glavan:2025azq}). These allow to derive the 
following result,
\begin{align}
L_{\nu\nu}(x;x')
    ={}&
	-
	\nu^2 H^2 I_{\nu\nu}(x;x')
	-
	\frac{ \ln(a_x) \!+\!1 }{ 2 H a_x }
	\Bigl[ \partial_0^x + \tfrac{D-1}{2} H a_x \Bigr] 
	i \Delta_\nu(x;x')
\nonumber \\
&
	-
	\frac{ \ln(a_{x'}) \!+\!1 }{ 2Ha_{x'} }
	\Bigl[ \partial_0^{x'} + \tfrac{D-1}{2} Ha_{x'} \Bigr] 
	i \Delta_\nu(x;x')
	+
	\tfrac{1}{2} i \Delta_\nu(x;x')
    \, .
\label{Lresult}
\end{align}
It is sufficient to check that this identity satisfies the correct equation of motion~(\ref{Leom}), which is 
accomplished by applying the kinetic operator from the left hand-side of~(\ref{Leom}), and commuting it 
over derivatives and functions of the scale factor to obtain the right-hand side of~(\ref{Leom}).

\medskip

In Sec.~\ref{subsec: Final reduction 2} we need another integrated propagator-like expression,
\begin{equation}
\overline{L}_{\nu}(x)
	=
	-
	i \!\int\! d^{D\!}z \, a_z^{D-2}
	\Big[
	\ln(a_z)
	\nabla_\mu^z i \Delta_\nu(x;z)
	\!\times\!
	\nabla^\mu_z u_\nu(z)
	+
	\nu^2 a_z^2H^2 i \Delta_\nu(x;z) u_\nu(z)
	\Big]
	\, ,
\label{Lbardef}
\end{equation}
where~$u_\nu(x)$ is a mode function satisfying a homogeneous equation,
\begin{equation}
\bigl( \mathcal{D}_x - a_x^D M_\nu^2 \bigr) u_\nu(x)
    = 0 \, .
\end{equation}
This integrated quantity satisfies the equation of motion
\begin{equation}
\bigl( \mathcal{D}_x \!-\! a_x^D M_\nu^2 \bigr) \overline{L}_{\nu}(x)
	=
	a_x^{D-2}
    \Big[
    - \ln(a_x) \nabla_x^2
    +
    \nu^2 a_x^2H^2
    \Big]
    u_\nu(x)
    \, .
\label{Lbareom}
\end{equation}
We can guess the form of the solution based on the similarity of equation~(\ref{Lbareom}) with the equation 
in~(\ref{Leom}), whose solution we have found in~(\ref{Lresult}). Up to a homogeneous part that we 
do not need, the solution reads
\begin{equation}
\overline{L}_{\nu}(x)
	=
	-
	\frac{ \ln(a_x) \!+\!1 }{ 2 H a_x }
	\Bigl[ \partial_0^x + \tfrac{D-1}{2} H a_x \Bigr] 
	u_\nu(x)
    \, .
\label{SpatialHalfIdentity}
\end{equation}
 It is sufficient to show that this expression satisfies the equation of motion~(\ref{Lbareom}), which requires
applying the same steps as when proving that~(\ref{Lresult}) satisfies its equation of motion.

\section{Additional diagrams}
\label{app:Additional diagrams}

\begin{figure}[h!]
\vskip-3mm
\hspace{3cm}
\includegraphics[width=3.5cm]{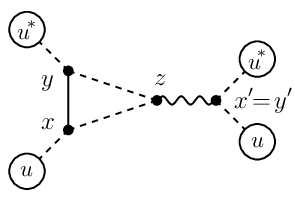}
\hfill
\includegraphics[width=3.5cm]{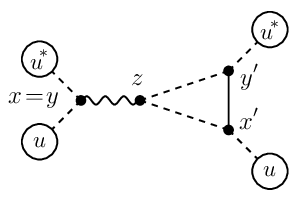}
\hspace{3cm}
\vskip-6mm
\hspace{4.7cm} $\boldsymbol{8a}$ \hfill$\boldsymbol{8b}$ \hspace{4.7cm}
\vskip+3mm
\hspace{3cm}
\includegraphics[width=3.5cm]{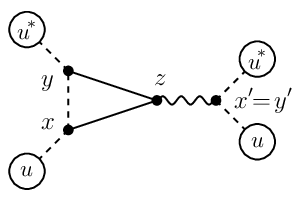}
\hfill
\includegraphics[width=3.5cm]{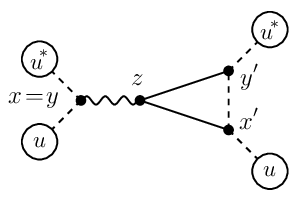}
\hspace{3cm}
\vskip-6mm
\hspace{4.7cm} $\boldsymbol{9a}$ \hfill$\boldsymbol{9b}$ \hspace{4.7cm}
\vskip+3mm
\includegraphics[width=3.5cm]{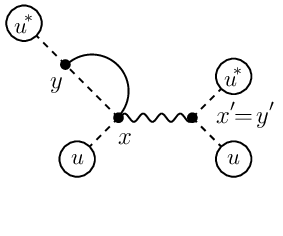}
\hfill
\includegraphics[width=3.5cm]{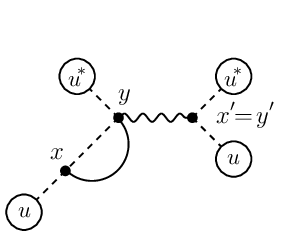}
\hfill
\includegraphics[width=3.5cm]{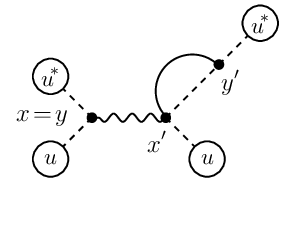}
\hfill
\includegraphics[width=3.5cm]{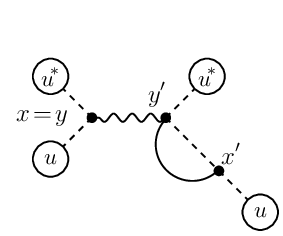}
\vskip-6mm
\hspace{1.4cm} $\boldsymbol{10a}$ \hfill$\boldsymbol{10b}$\hfill $\boldsymbol{10c}$ \ \
\hfill $\boldsymbol{10d}$ \hspace{1.5cm}
\vskip+2mm
\caption{Three additional classes of diagrams that contribute at the same 
order~$(\kappa\lambda)^2$ to the~$t$-channel 4-point function.}
\label{AdditionalDiagrams1}
\end{figure}

In addition to the diagrams in Fig.~\ref{4ptDiagrams} representing the 4-point function
corrections, for which we computed the~$\Delta\alpha$ variation, there are several more 
classes of~$t$-channel diagrams that contribute at the same perturbative 
order~$(\kappa\lambda)^2$; these are classes depicted in Figs.~\ref{AdditionalDiagrams1}.
Even though they naively seem to be corrections to 
the gravitational potential, rather than
the scalar one, we should be careful about making that conclusion prematurely. 
In fact, already in flat
space obtaining the gauge-independent vacuum polarization required an analogue
of one class of these diagrams~\cite{Katuwal:2021thy}. Nonetheless, 
the~$\Delta\alpha$ variations of these additional diagrams all vanish
individually. This is due to the presence of vertex B that connects two external mode 
functions to the graviton mediating the interaction. By virtue of 
contractions~(\ref{contractionAA3}) and~(\ref{BBthirdContraction}), 
and the tree-level equation of motion~(\ref{tree-levelUeom}) for the mode function,
this always gives zero.

\begin{figure}[h!]
\vskip+5mm
\hspace{3.5cm}
$I\!I\!I$
\hspace{5.cm}
$I\!V$
\vskip-4mm
\centering
\hspace{4.cm}
\includegraphics[height=1.7cm]{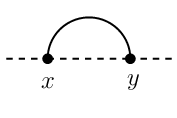}
\hfill
\includegraphics[height=1.7cm]{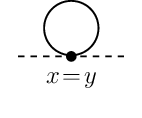}
\hspace{4.cm}
\vskip-6mm
\caption{One-loop diagrams correcting the self-mass of the massive 
scalar from the interaction with the massless scalar: the 3-vertex diagram ($I\!I\!I$) and the 4-vertex diagram~($I\!V$).}
\label{MassiveSelfMassDiagramsIII+IV}
\end{figure}

In addition, we should also list the mode function corrections in diagrams that naively 
look like the~$t$-channel graviton exchange, that contribute at the same order,
\begin{equation}
\vcenter{\hbox{\includegraphics[width=3.1cm]{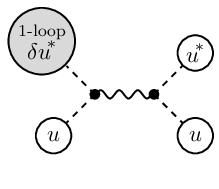}}}
+
\vcenter{\hbox{\includegraphics[width=3.1cm]{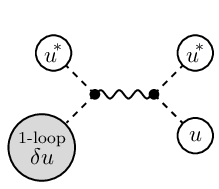}}}
+
\vcenter{\hbox{\includegraphics[width=3.1cm]{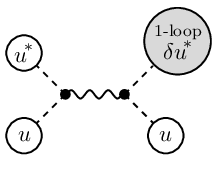}}}
+
\vcenter{\hbox{\includegraphics[width=3.1cm]{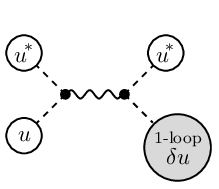}}}
\, .
\end{equation}
The mode function corrections for these diagrams are sourced by the massive scalar 
self-mass with loop corrections engendered by its interactions with the massless scalar, 
given in Fig~\ref{MassiveSelfMassDiagramsIII+IV}. 


\end{document}